# Small but Tubby: A Magnetic Loop Antenna Made from 100 mm Copper Tubing

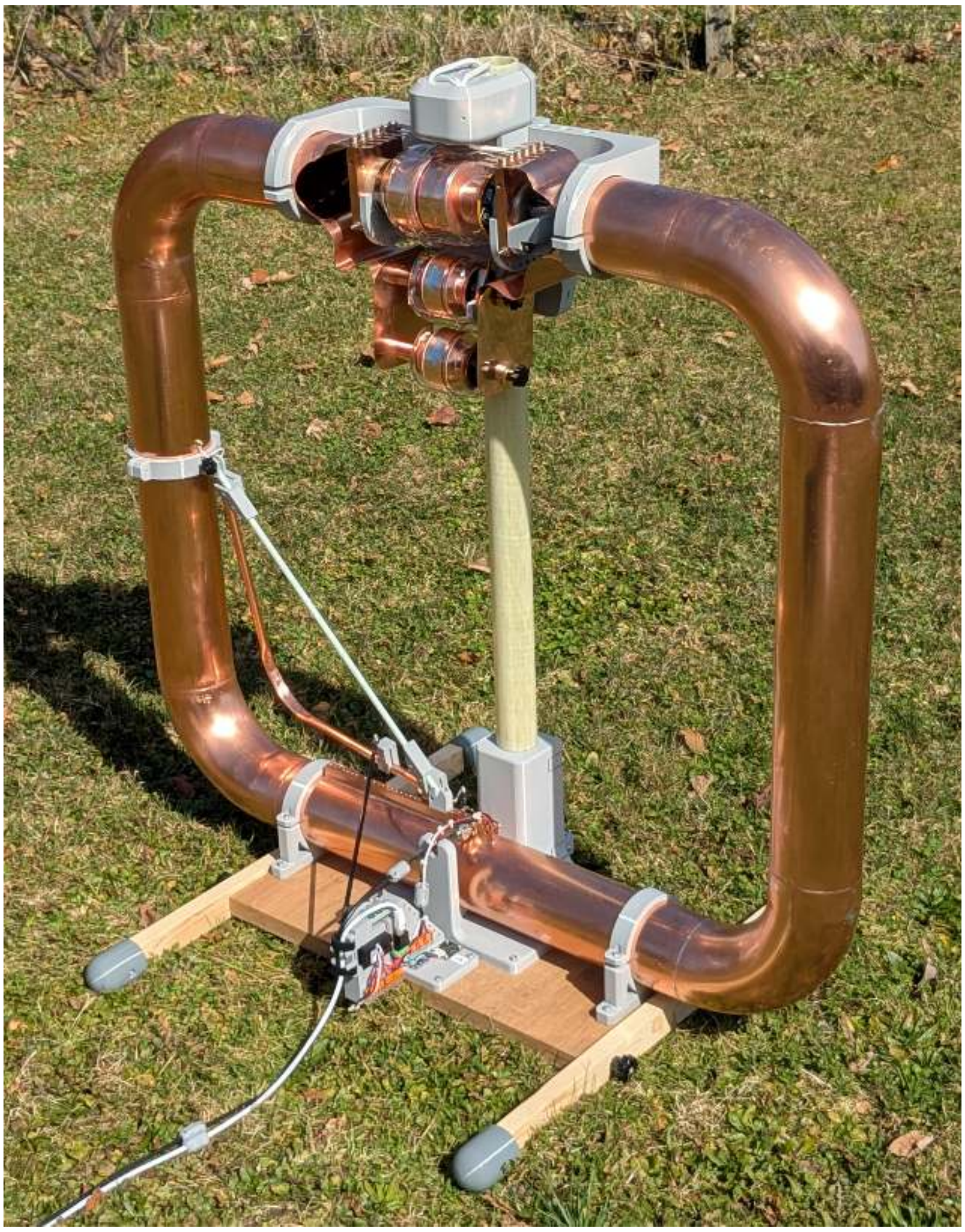

**Author:** HB9ISP Peter Märki, El. Ing. FH (ORCID: 0000-0002-0596-076X), Zelglistrasse 49, 8634 Hombrechtikon, Switzerland
**Co-Author:** Markus Niese, MSc Quantum Engineering, PhD candidate at ETH Zurich (ORCID: 0009-0009-9132-9613)

***Abstract:*** *This paper presents the electrical model, key equations, and practical construction of a small transmitting magnetic loop antenna built from unusually large 100 mm diameter copper tubing. The large conductor surface area and wide-area transitions to the vacuum capacitors were designed to minimize resistive losses. The frequency range from 1.8 MHz to 31 MHz is unusually wide. Frequency, impedance matching, and azimuth are all adjusted automatically by servo motors.*

*Indoor losses originate predominantly from near-field coupling to the environment rather than from the antenna itself. Temperature-rise measurements confirm that the bulk of the dissipated power is absorbed by the environment, not by the antenna components.*

*H-field measurements show good agreement between the measured H-field and the theoretical free-space H-field calculated from the antenna geometry and an estimated loop current. The loop current was estimated from the measured antenna bandwidth and the applied transmit power.*

*The antenna was developed for indoor operation where outdoor installation is not possible.*

***Keywords:*** *magnetic loop antenna, small transmitting loop, vacuum capacitor, KP1-4, copper tubing, antenna efficiency, gamma match, servo tuning, indoor antenna, thermal measurement, building losses*

*Unless otherwise stated, all variables are expressed in SI units.*

# Table of Contents

# 1 Fundamentals

## 1.1 Schematic, resonant circuit

To model the magnetic loop antenna we will successively add elements that represent specific physical effects. Figure 1 shows the magnetic loop together with a capacitor. We introduce a capacitor C and an inductance L, which represents the inductive behaviour of the loop, see Figure 2.

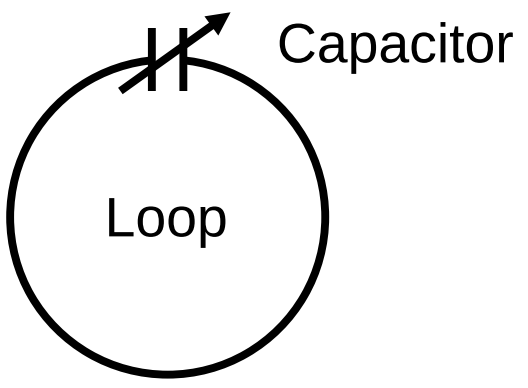

Figure 1: Magnetic loop antenna: main loop and capacitor.

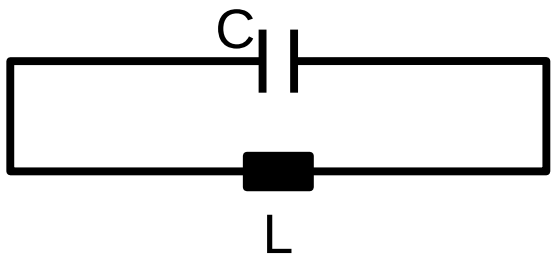

Figure 2: The antenna forms an LC tank circuit, where the loop itself acts as the inductor ($L$).

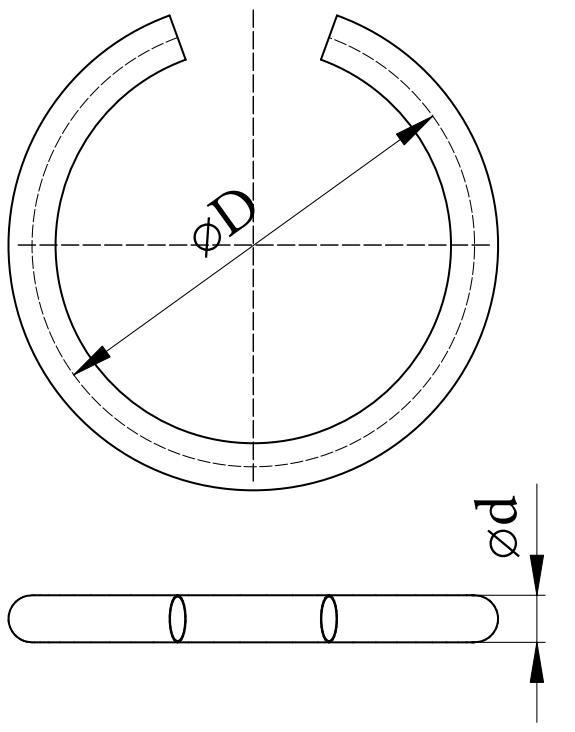

Figure 3: Inductance of a ring-shaped tube.

For a single-turn circular loop with mean diameter $D$ and conductor diameter $d$, the approximate inductance is given by:

$$L = \mu_0 \cdot \frac{D}{2}\left(\ln\left(\frac{8D}{d}\right) - 2\right) \tag{1}$$

where $\mu_0 = 4\pi \cdot 10^{-7} \frac{H}{m}$.

*This approximation assumes a pronounced skin effect where the current flows only on the conductor's surface (e.g., in high-frequency applications or thin-walled tubing).*

The resonance frequency of the tuned loop is given by:

$$f_0 = \frac{1}{2\pi \cdot \sqrt{L \cdot C}} \tag{2}$$

In addition to the inductance, there is also damping in the loop, modeled by $R_T$ (see Figure 4). This damping can be further split up in radiation $R_R$, representing the energy that is radiated away, and losses $R_{\text{loss}}$ in the loop. These losses consist of inductive losses $R_L$, capacitor losses $R_C$ and near-field losses in the environment $R_E$ (see Figure 6).

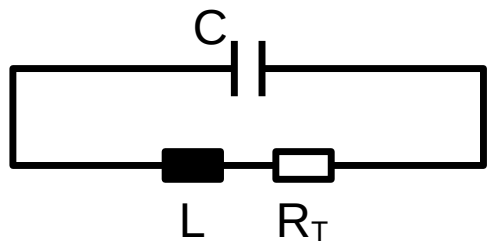

Figure 4: The resonant circuit is damped; this is modeled by inserting a resistance ($R_T$).

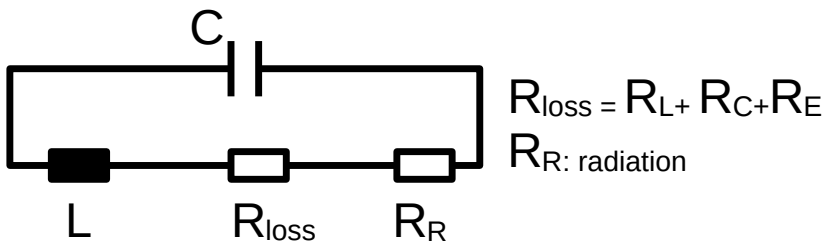

Figure 5: The damping is caused by losses ($R_{\text{loss}}$) and radiation ($R_{\text{R}}$).

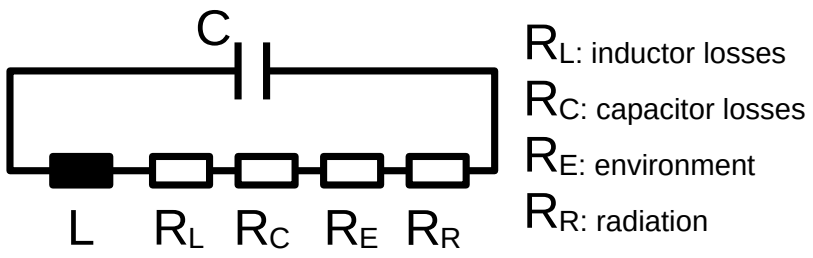

Figure 6: The losses $R_{\text{loss}}$ can be split into inductor losses ($R_L$), capacitor losses ($R_C$), and losses due to energy absorbed in the near vicinity of the antenna ($R_E$).

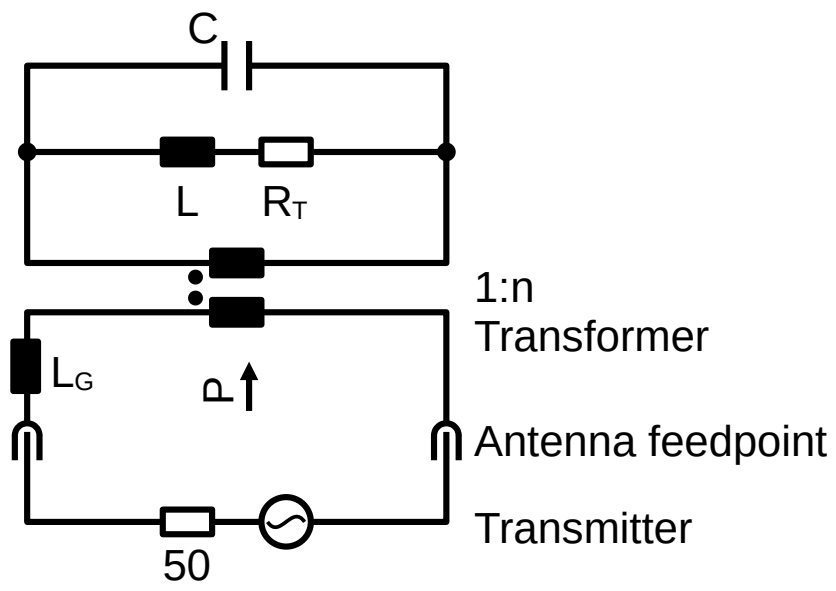

Figure 7: The transmitter, with a 50 Ω internal resistance, delivers the RF power $P$ into the main loop via a matching network (such as a coupling loop, gamma match, or similar transformer).

The series reactance $L_G$ (see Figure 7) models effects of the coupling network, e.g. the inductance of the coupling loop.

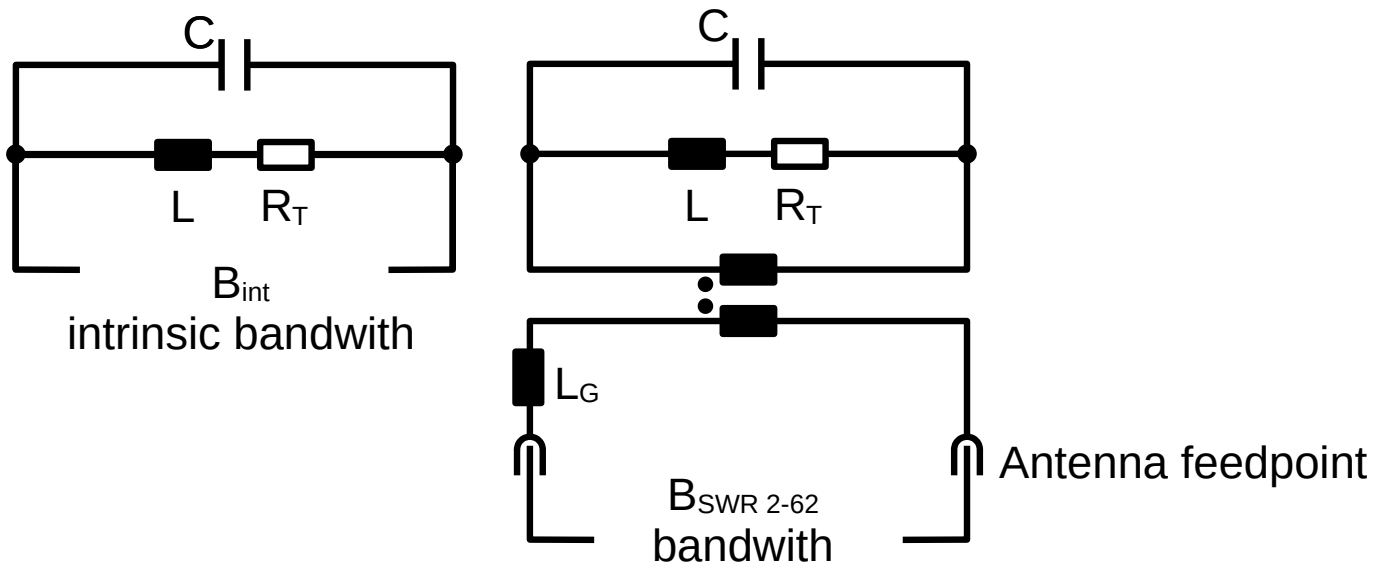

Figure 8: Left: the intrinsic resonant circuit with bandwidth $B_{\text{int}}$; right: the bandwidth $B_{\text{SWR 2.62}}$, which can conveniently be measured at the antenna feed point.

For the calculations that follow, we are interested in the intrinsic bandwidth of the resonant circuit formed by the capacitor and the loop (see Figure 8, left).

From a measurement perspective, however, we have no direct access to the resonant circuit itself.

One option is to determine the S-parameters at the antenna feed point, from which both the series reactance $L_G$ and the intrinsic bandwidth $B_{\text{int}}$ can be extracted.

A simpler measurement is to determine the bandwidth $B_{\text{SWR 2.62}}$ — the bandwidth between the frequencies at which the SWR has risen to 2.62. For a correctly matched antenna with SWR = 1.0 at resonance and negligible series reactance (L_G), this bandwidth agrees well with the intrinsic bandwidth $B_{\text{int}}$ in most cases. For the derivation, see Section 7.1.

$$B_{\text{int}} \approx B_{\text{SWR 2.62}} = f_{\text{upper_SWR_2.62}} - f_{\text{lower_SWR_2.62}} \tag{3}$$

The intrinsic quality factor can be derived as:

$$Q_0 = \frac{f_0}{B_{\text{int}}} \tag{4}$$

From this, the total resistance $R_T$ (including radiation and loss resistance) can be derived using the inductive reactance $X_L$:

$$R_T = \frac{X_L}{Q_0} = 2\pi \cdot f_0 \cdot \frac{L}{Q_0} \tag{5}$$

During transmission, the loop current $I_{\text{main_loop}}$ can be estimated by assuming that the total transmitter power $P$ is delivered to the antenna:

$$I_{\text{main_loop}} = \sqrt{\frac{P}{R_T}} \tag{6}$$

Consequently, the voltage across the capacitor (loop voltage) is given by:

$$U_{\text{loop}} = I_{\text{main_loop}} \cdot X_L = I_{\text{main_loop}} \cdot 2\pi \cdot f_0 \cdot L \tag{7}$$

In both the receiving and the transmitting case, the 3dB bandwidth $\Delta f_{\text{3dB}}$ is determined by the total damping of the system, which defines the loaded quality factor $Q_L$. Under matched conditions, where the receiver load equals the antenna resistance, this bandwidth is given by:

$$\Delta f_{\text{3dB}} = \frac{f_0}{X_L} \cdot 2R_T \tag{8}$$

In this scenario, the total resistance is doubled due to the matched load of the transceiver. This leads to a broader bandwidth than in the unloaded case (antenna without transceiver connected). See also reference [1] and Section 7.2.

## 1.2 Coupling Loop

There are many possible coupling methods. Several variants are shown in references [2] and [3]. This section focuses on two similar methods, both are mechanically and electrically elegant solutions: First, I will discuss the shielded coaxial coupling loop, a small complete coupling loop that is magnetically coupled to the main loop, then I will discuss the gamma match feed system, which only has half a loop in addition to the main loop. There I will also discuss the differences between the two and explain why the gamma match feed system was chosen for the antenna presented in this paper.

### 1.2.1 Shielded Coaxial Coupling Loop

In the following, we describe the shielded coaxial coupling loop. We simplify the coupling loop step by step to explain its function.

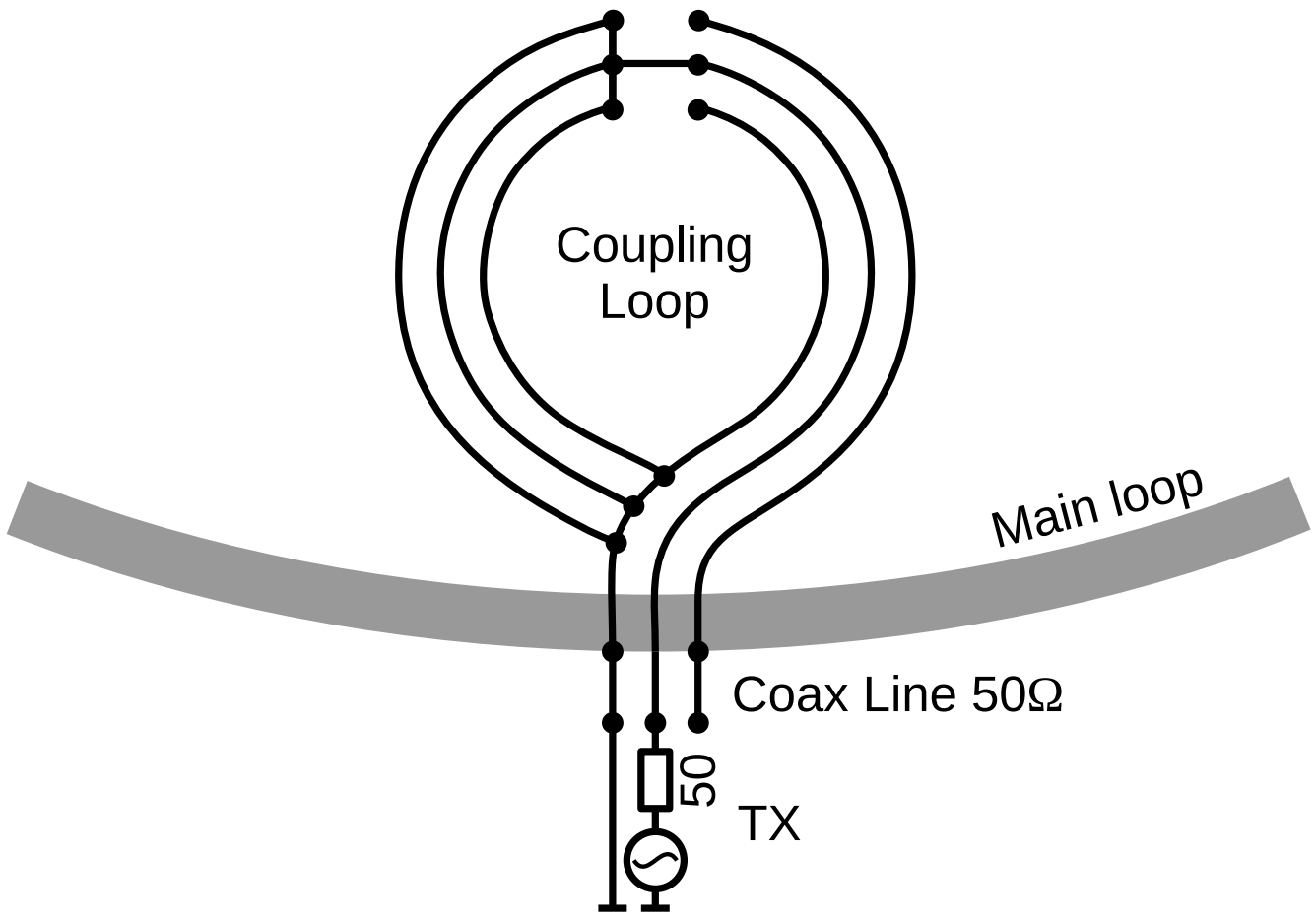

Figure 9: Coupling loop overview: coaxial cable from TX to antenna, with the lower part of the main loop shown in gray; at the top, the shield is opened and the conductors are connected as indicated.

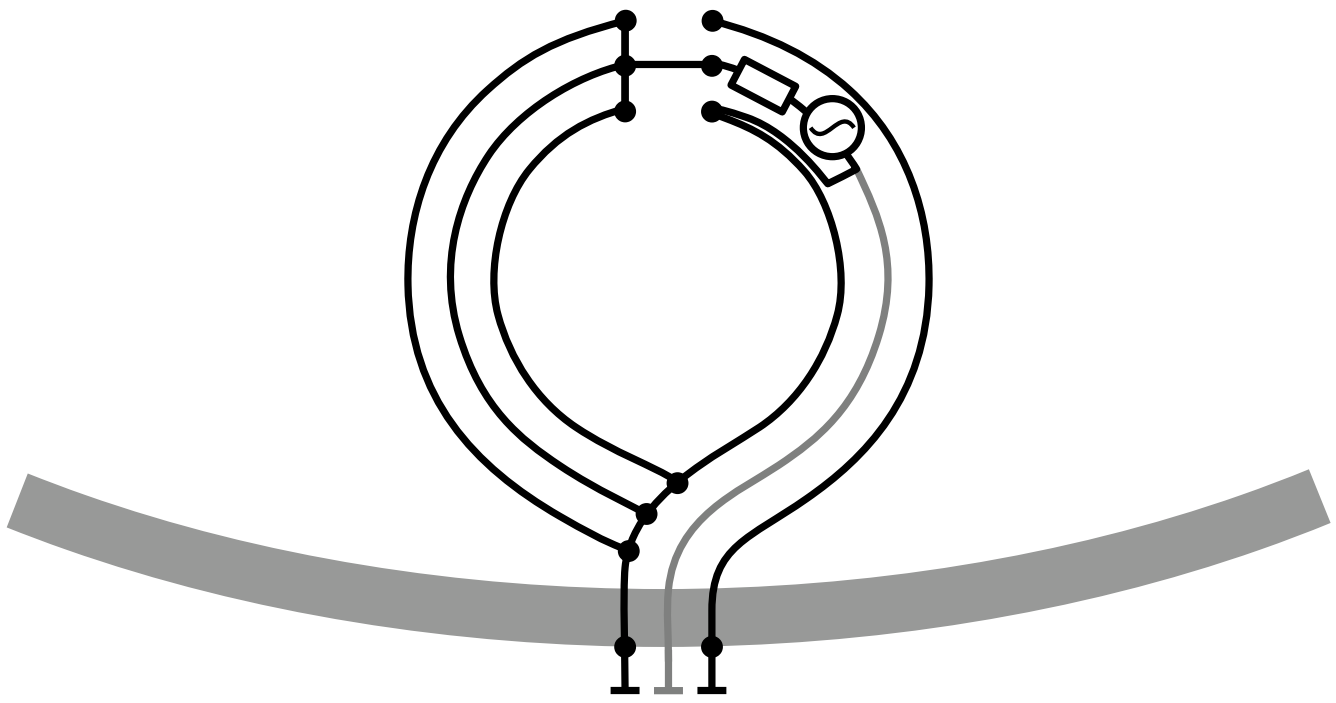

Figure 10: For simplification, the source TX is moved all the way to the top. This simplification is functionally equivalent.

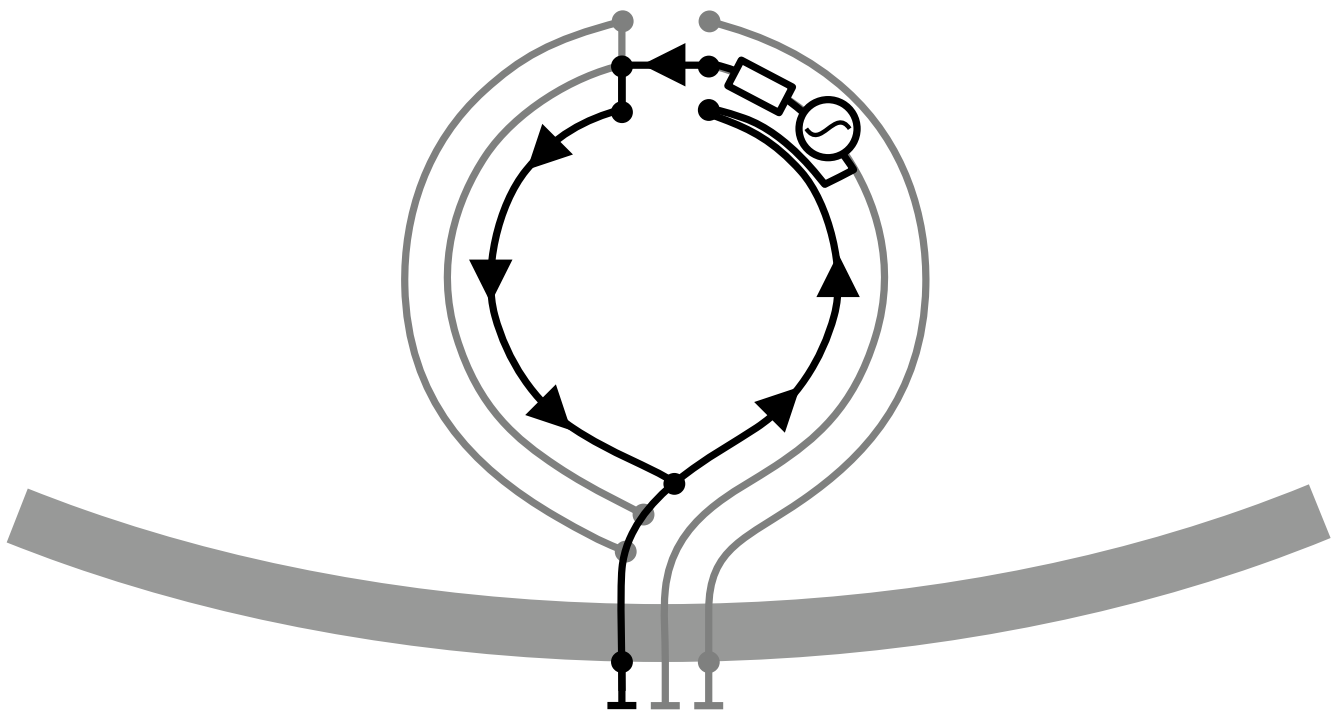

Figure 11: Coupling-loop current flow: Moving the source TX to the top allows us to draw a functionally equivalent picture where the current is modelled to flow only in the one single loop with one conductor, here black, while the other two are grayed out; the source current flows on the outer shield, and symmetry at the GND point helps minimize common-mode feed-line currents toward TX.

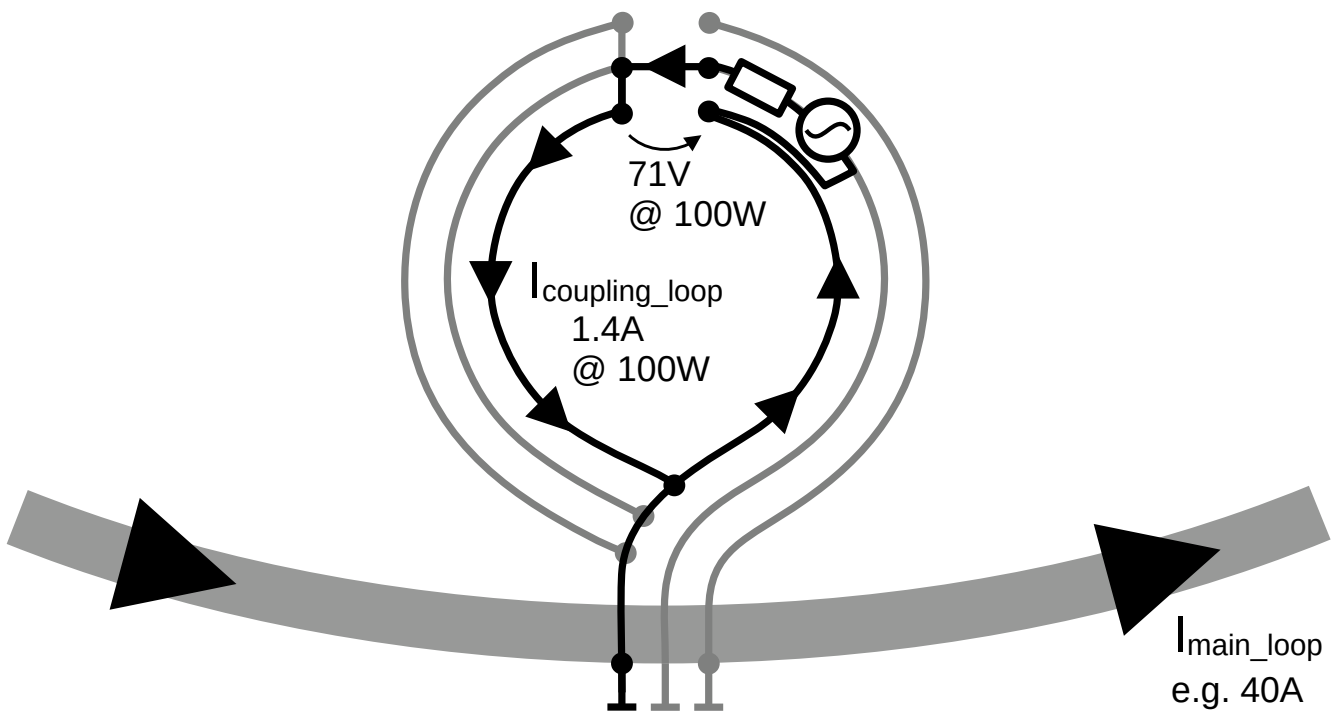

Figure 12: The coupling ratio (equivalent turns ratio) between the coupling loop and the main loop depends on how much magnetic flux links the coupling loop versus the main loop. By changing the position, size, or shape of the coupling loop, the coupling ratio can be adjusted continuously. The adjustment is made so that the input impedance is matched to 50 Ω. Ideally, all power from the transmitter is transferred via the coupling loop into the main loop. Losses in this coupling are usually negligible compared to the other antenna losses. Current and voltage in the coupling loop can therefore be calculated for given power from TX.

Numerical example, consider 100 W into the 50 Ω coupling loop: the voltage is $U = \sqrt{P \cdot R} = \sqrt{100\text{W} \cdot 50\Omega} \approx 71\text{V}$ and the coupling-loop current is $I = \sqrt{P/R} \approx 1.4\text{A}$. Assuming the main loop has $R_T = 0.0625\Omega$, the main-loop current is $I_{\text{main loop}} = \sqrt{P/R_T} = \sqrt{100\text{W}/0.0625\Omega} = 40\text{A}$.

### 1.2.2 The Gamma Match Feed System

An alternative to the coaxial coupling loop is the gamma match feed system. This method also uses a small loop or a shorted stub to magnetically couple to the main loop. In contrast to the coaxial coupling loop, the main loop itself is used as the return path to the coaxial connector.

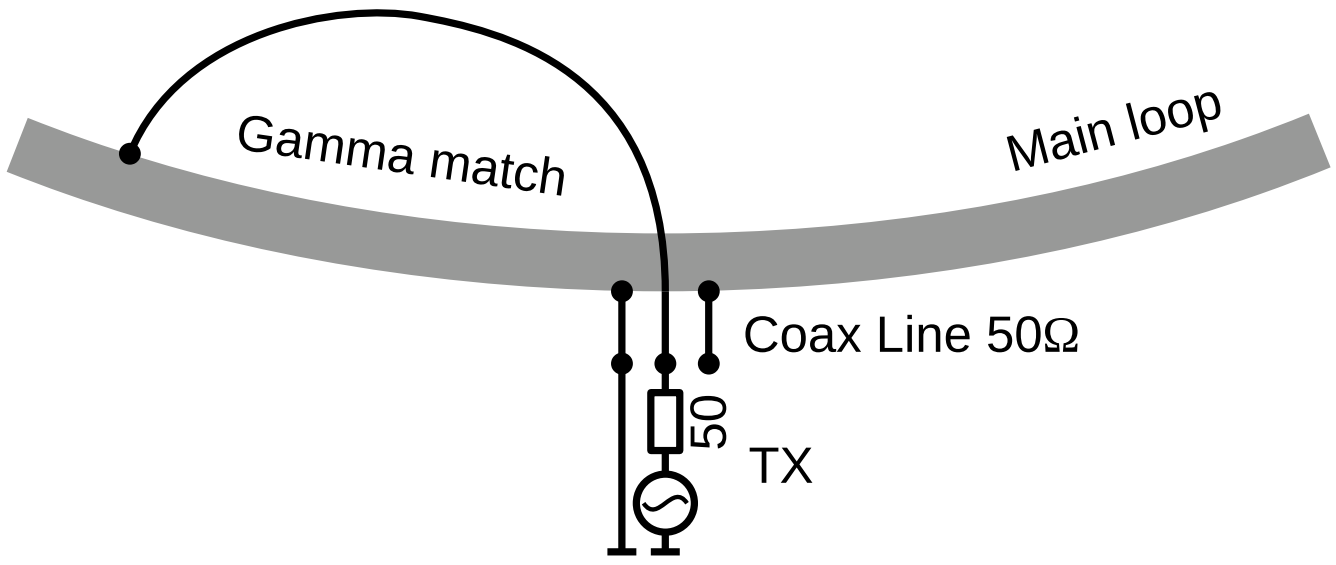

Figure 13: The outer conductor of the coaxial cable is galvanically connected to the main loop at the symmetry point. The inner conductor is run along the main loop at a certain distance and finally also galvanically connected to the left side of the main loop.

In contrast to the previously shown coaxial coupling loop, the gamma match is asymmetric and the conductor is not shielded. In my opinion, however, this is not a relevant disadvantage. The asymmetry is minimal for a main loop with high Q factor because the current in the gamma match is only a small fraction of the current in the main loop. The fact that the gamma match is not shielded is also not a problem. The main loop, much larger and with a larger cross-section, is also not shielded. The small, also unshielded gamma match makes no significant difference.

Gamma matches are usually set to a fixed position or are manually adjustable, possibly with the help of a screwdriver.

Here, I present an approach with a pivoted and thus easily automatically adjustable gamma match. The end point, where the gamma match is connected to the main loop stays fixed, while the arc of the gamma match can be rotated to change the area between the main loop and the gamma match loop.

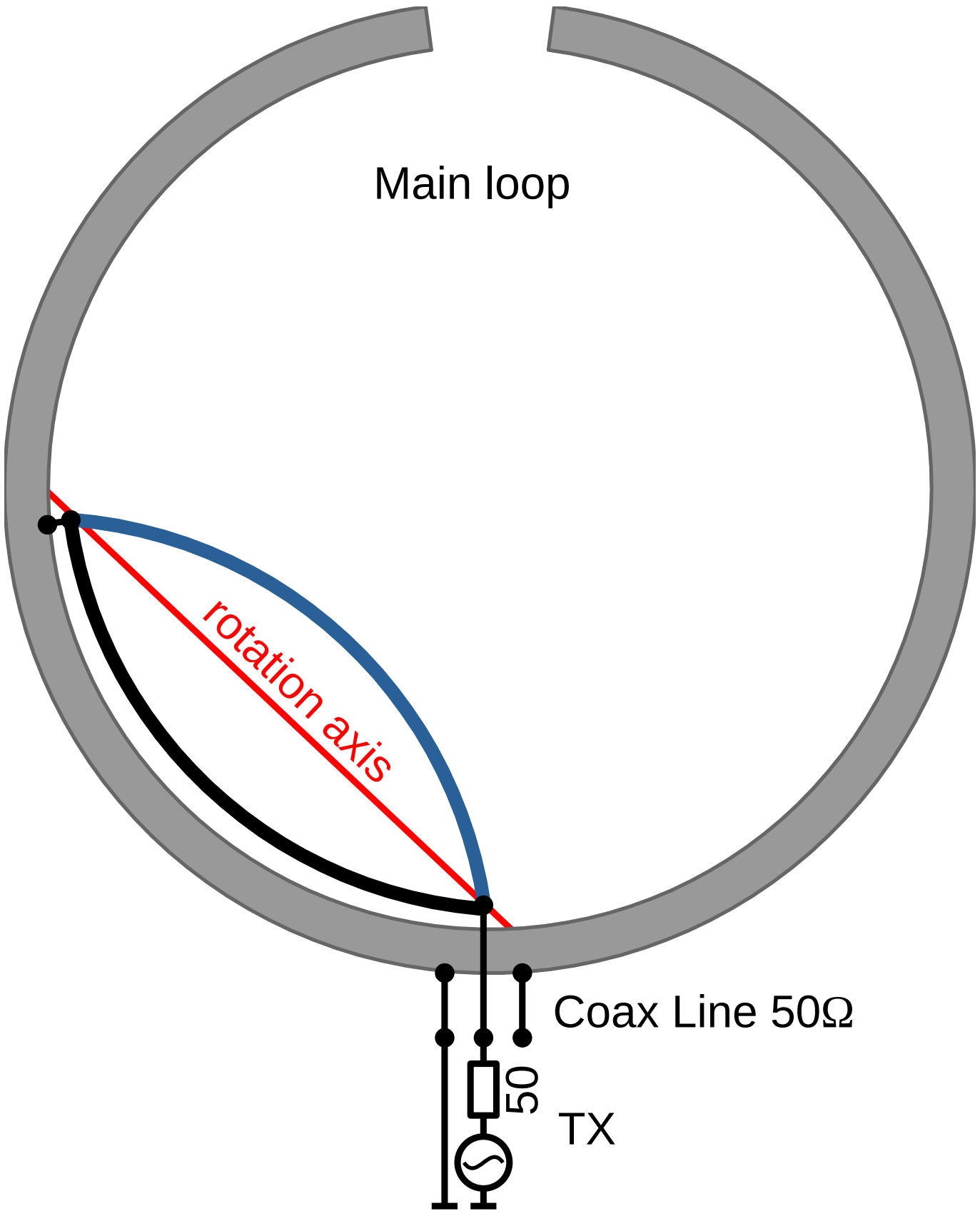

Figure 14: The gamma match is rotatably mounted (red: rotation axis). Position with minimal coupling in black. Maximum coupling in blue.

With minimal coupling, the gamma match conductor is very close to the main loop. The self-inductance of the coupling is very small. In my view, this is a major advantage. A disadvantage is the somewhat complex geometry of the bearing and the rotating contacts, which can be solved using flexible cable sections (see Figure 63).

## 1.3 Radiation Resistance

The radiation resistance $R_\mathrm{R}$ is a crucial parameter determining the antenna's efficiency.

For electrically small loops ($D < \frac{\lambda}{10\pi}$), the radiation resistance is well approximated by the standard Rayleigh formula:

$$R_\mathrm{R} \approx 31171 \cdot \left(\frac{A \cdot f^2}{c^2}\right)^2 \tag{9}$$

For larger loops (up to $D \approx 0.5\lambda$), the small loop approximation loses accuracy and tends to underestimate the radiation resistance. R.W.P. King derived a more general expression for a circular loop with uniform current distribution, which includes a correction factor (second bracket) for larger loops:

$$R_\mathrm{R} \approx 31171 \cdot \left(\frac{A \cdot f^2}{c^2}\right)^2 \cdot \left(1 + 0.5 \cdot \left(\frac{\pi \cdot D \cdot f}{c}\right)^2\right) \tag{10}$$

where:
- $R_\mathrm{R}$: Radiation resistance (Ω)
- $A$: Loop area ($m^2$)
- $f$: Frequency (Hz)
- $c$: Speed of light ($299792458\ \mathrm{m/s}$)
- $D$: Loop diameter (m)

## 1.4 Antenna Efficiency

The antenna efficiency $\eta$ describes how efficiently the antenna radiates power. It is defined as the ratio of radiated power to the total power supplied to the antenna. The efficiency can be calculated as

$$\eta = \frac{R_\mathrm{R}}{R_T} = \frac{R_\mathrm{R}}{R_\mathrm{R} + R_E + R_L + R_C} \tag{11}$$

where $R_T$ is the total resistance, which includes radiation resistance ($R_\mathrm{R}$) and all loss resistances: loss resistance due to conductor losses ($R_L$), losses due to absorption in the environment ($R_E$), and losses in the resonating capacitor ($R_C$).

## 1.5 Magnetic Field Strength near the Loop

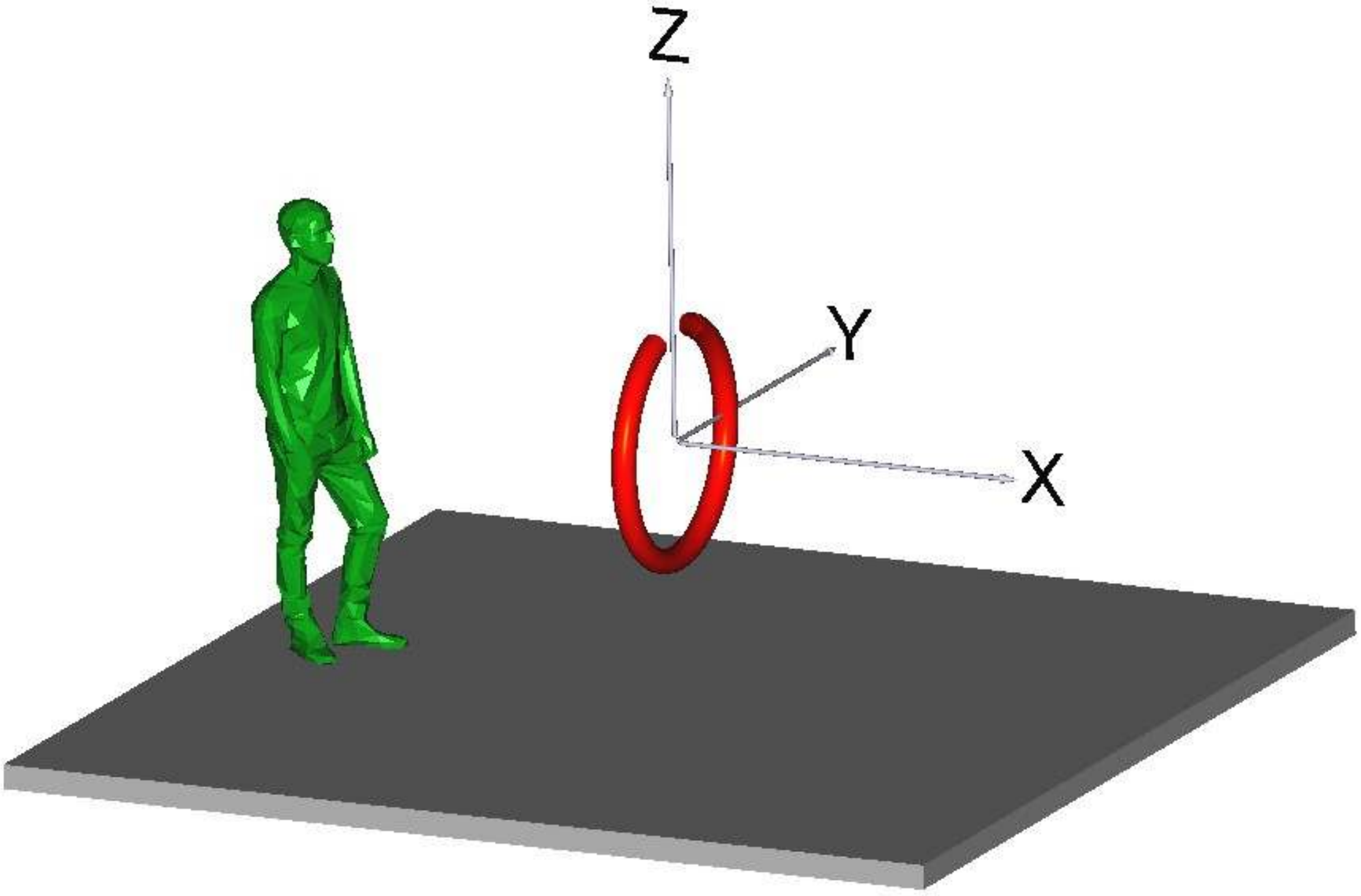

Figure 15: Coordinate system used for the figures in this document. Main loop in red.

It is important to discuss some safety aspects of the antenna. During transmission, the magnetic field close to the antenna can be quite large, requiring people to keep a safety distance.

A simple estimate is obtained by treating the loop as a **magnetic dipole** (valid for $\lambda \gg r$ and $r \gg D$). The magnitude of the magnetic field strength is:

$$H = I_\text{main_loop} \cdot \frac{D^2}{16 \cdot r^3} \cdot \sqrt{1 + 3 \cdot \cos(\varphi)^2} \tag{12}$$

where $D$ is the loop diameter, $r$ is the distance from the loop center to the observation point, and $\varphi$ is the angle measured from the x-axis (loop axis) (see Figure 15).

For a more accurate estimate that is also valid at larger distances, the full retarded magnetic dipole solution is used. With the magnetic dipole moment $m = I_\text{main_loop}\pi\left(\frac{D}{2}\right)^2$ and wavenumber $k = 2\pi\frac{f}{c}$, the H-field magnitude is:

$$H = \frac{m}{4\pi} \cdot \sqrt{4\cos^2(\varphi)\left(\frac{1}{r^6} + \frac{k^2}{r^4}\right) + \sin^2(\varphi)\left(\frac{1}{r^6} - \frac{k^2}{r^4} + \frac{k^4}{r^2}\right)} \tag{13}$$

The first terms ($\frac{1}{r^6}$ inside the square root, i.e., $\propto \frac{1}{r^3}$) dominate in the near field; the last terms ($\frac{k^4}{r^2}$ inside the square root, i.e., $\propto \frac{1}{r}$) dominate in the far field.

The magnetic field strength $H$ around the loop is visualized in the next figure. The figure shows lines of equal magnetic field strength for a loop with a diameter of 1 m and a current of 10.5 A at 14.1 MHz.

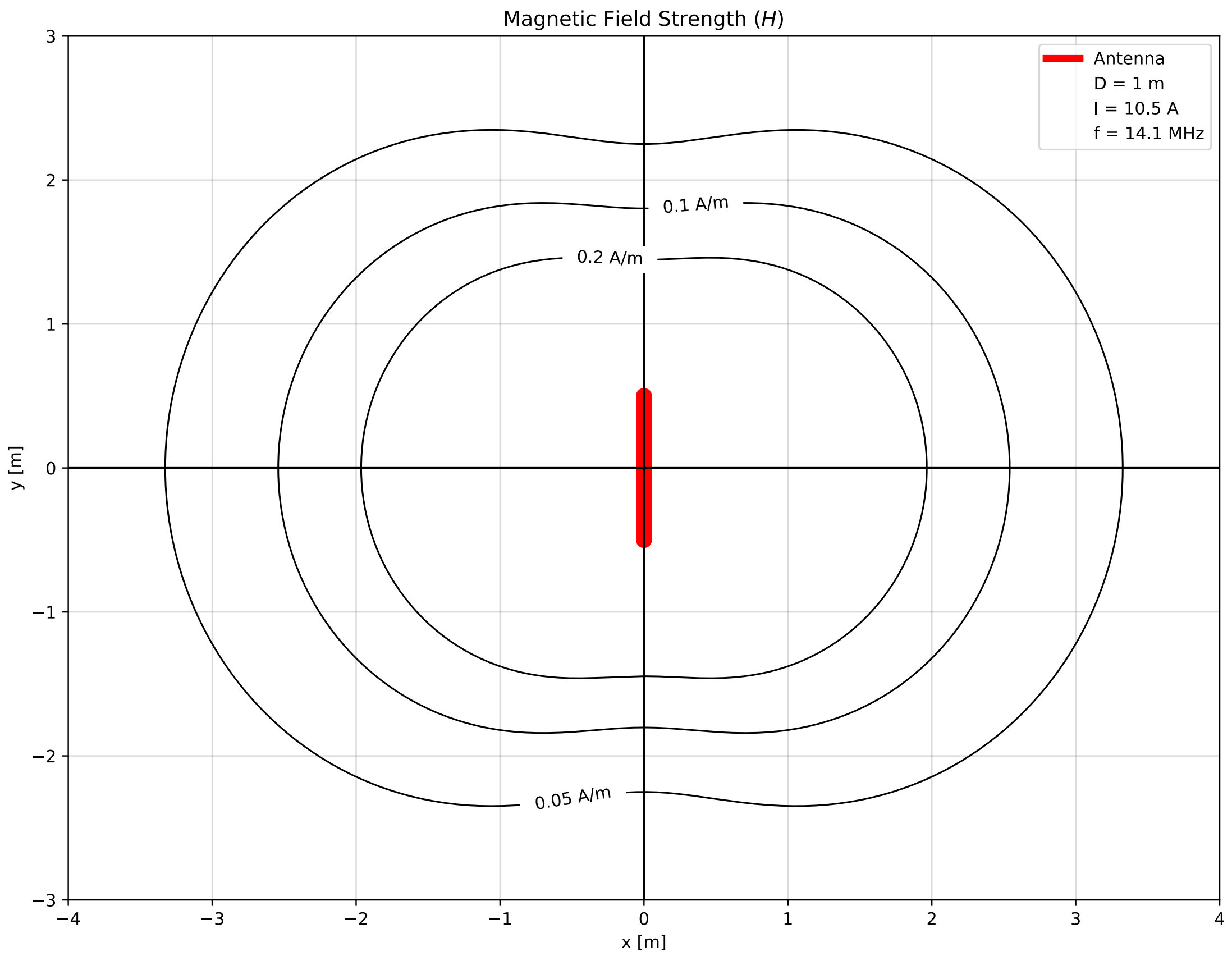

Figure 16: Simulated magnetic field strength (H) around the loop antenna.

A typical safety limit value for the near-field magnetic field strength is 0.073 A/m at 14 MHz (ICNIRP, 1998). In the example shown here, the required safety distance is already quite large.

For indoor operation, safety assessments based solely on theoretical antenna losses may overestimate the magnetic field strength, because additional environmental losses reduce the loop current. Consequently, the actual H-field should be estimated from the measured antenna bandwidth or, preferably, verified directly using an H-field probe.

A wider view highlights the transition toward the far-field pattern: in this cross-sectional view, the field extends predominantly in the loop plane (y-direction) while it narrows along the loop axis (x-direction).

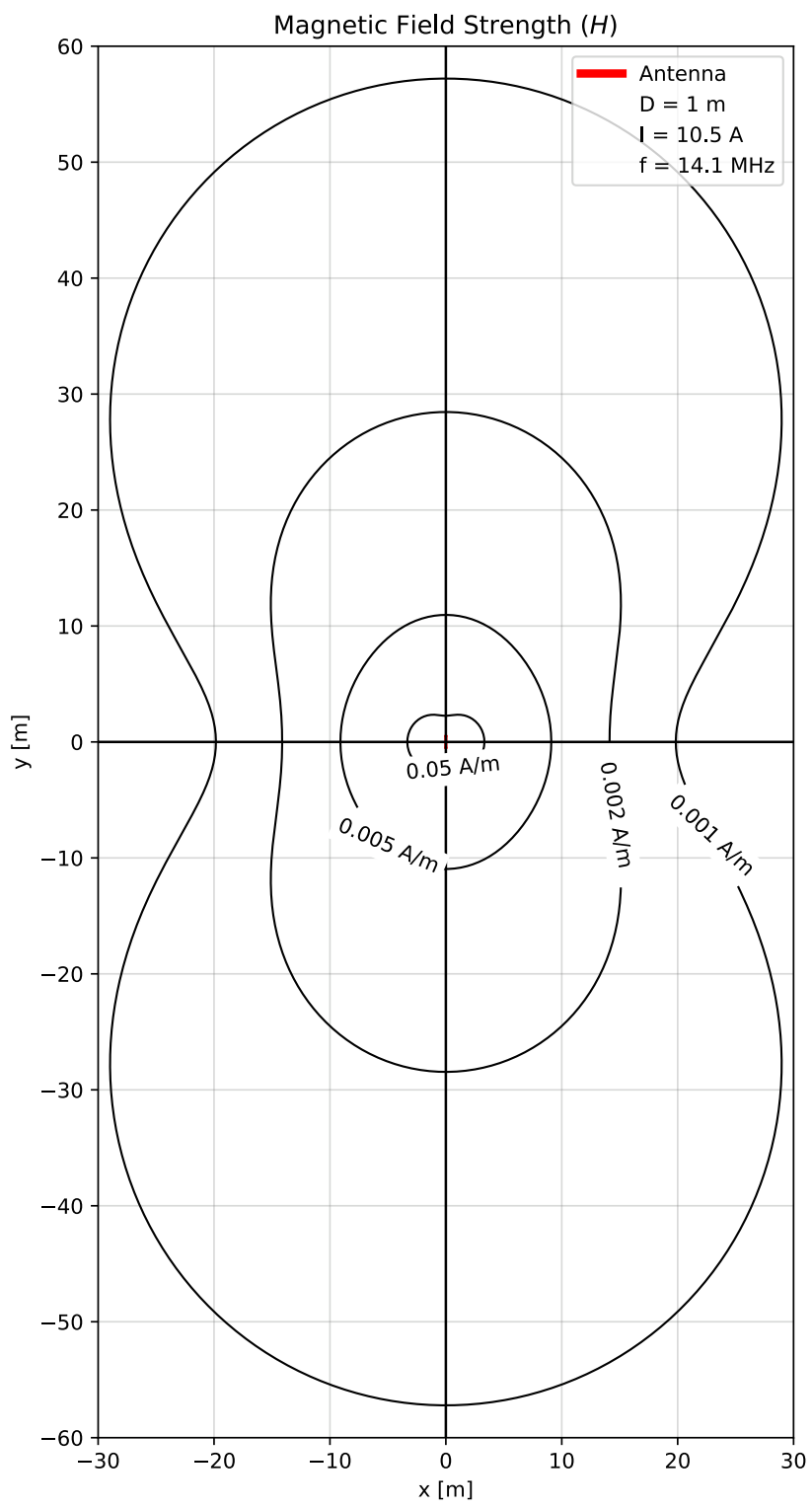

Figure 17: Simulated magnetic field strength (H), wide-view map showing the transition toward the far-field pattern (see next section). The innermost contour here corresponds to the outermost contour in Figure 16.

Examples of H-field measurements can be found in Section 3.7.

## 1.6 Far-Field Radiation Pattern

The following figure shows the normalized linear far-field radiation characteristic.

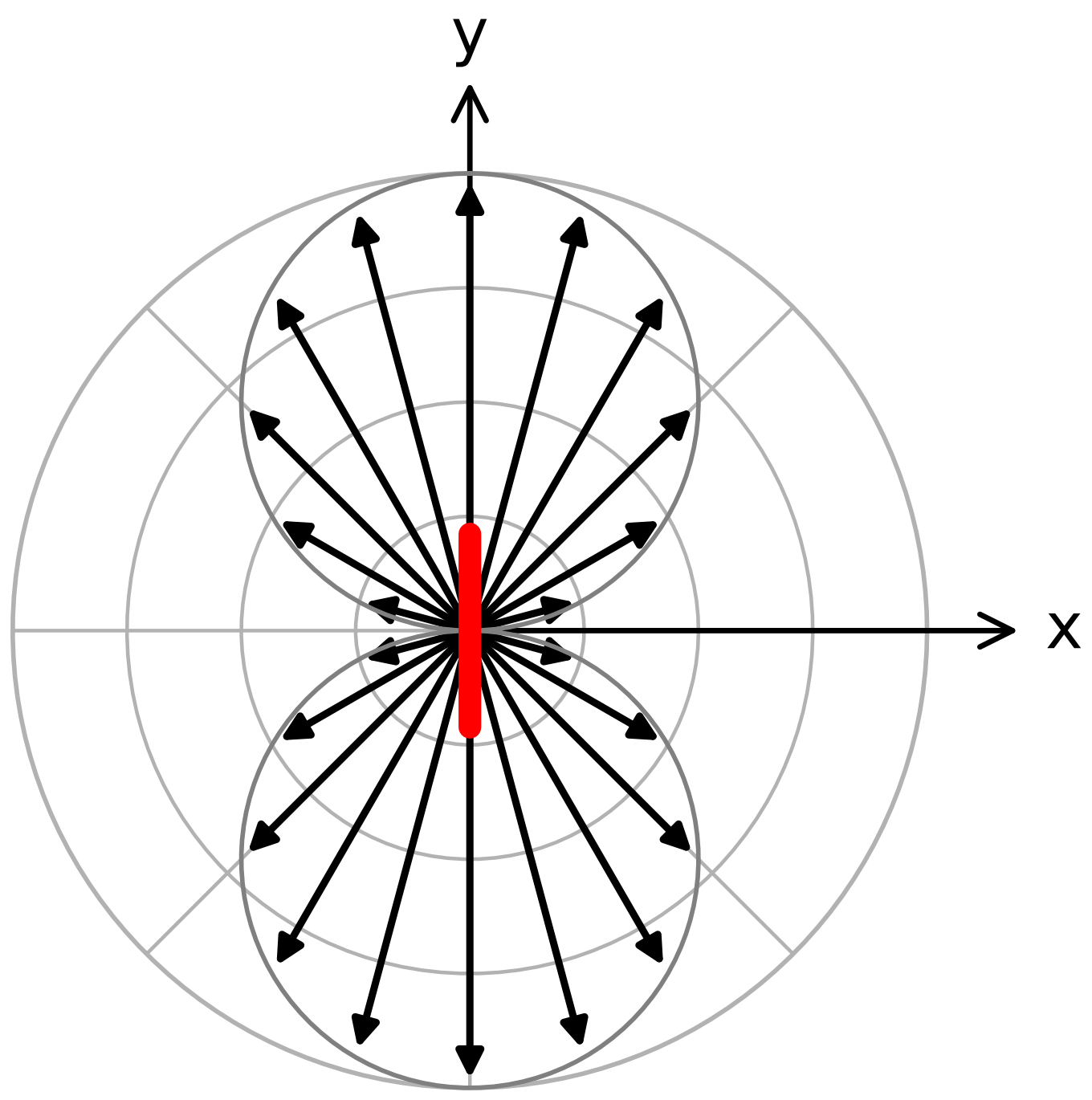

Figure 18: Normalized linear far-field radiation pattern of a loop antenna.

In the far-field region, the magnetic loop antenna exhibits a figure-eight radiation pattern, with maximum radiation occurring in the plane of the loop ($\varphi = 90°$) and nulls along the loop axis ($\varphi = 0°$). This characteristic is typical for small loop antennas.

## 1.7 Environmental Losses ($R_E$): Near-Field Coupling vs. Far-Field Radiation

In discussions about this paper, several people expressed confusion: if the antenna is operated indoors, the power passes through the air and is absorbed by the surrounding walls — so it must be radiated. However, far-field radiation and near-field coupling into nearby objects (as in a transformer) are not the same thing, even though in both cases the power leaves the antenna through the air. The following worked example with concrete numbers is intended to clarify this distinction.

The numerical values in the figures are based on the 20 m band (see Figure 73).

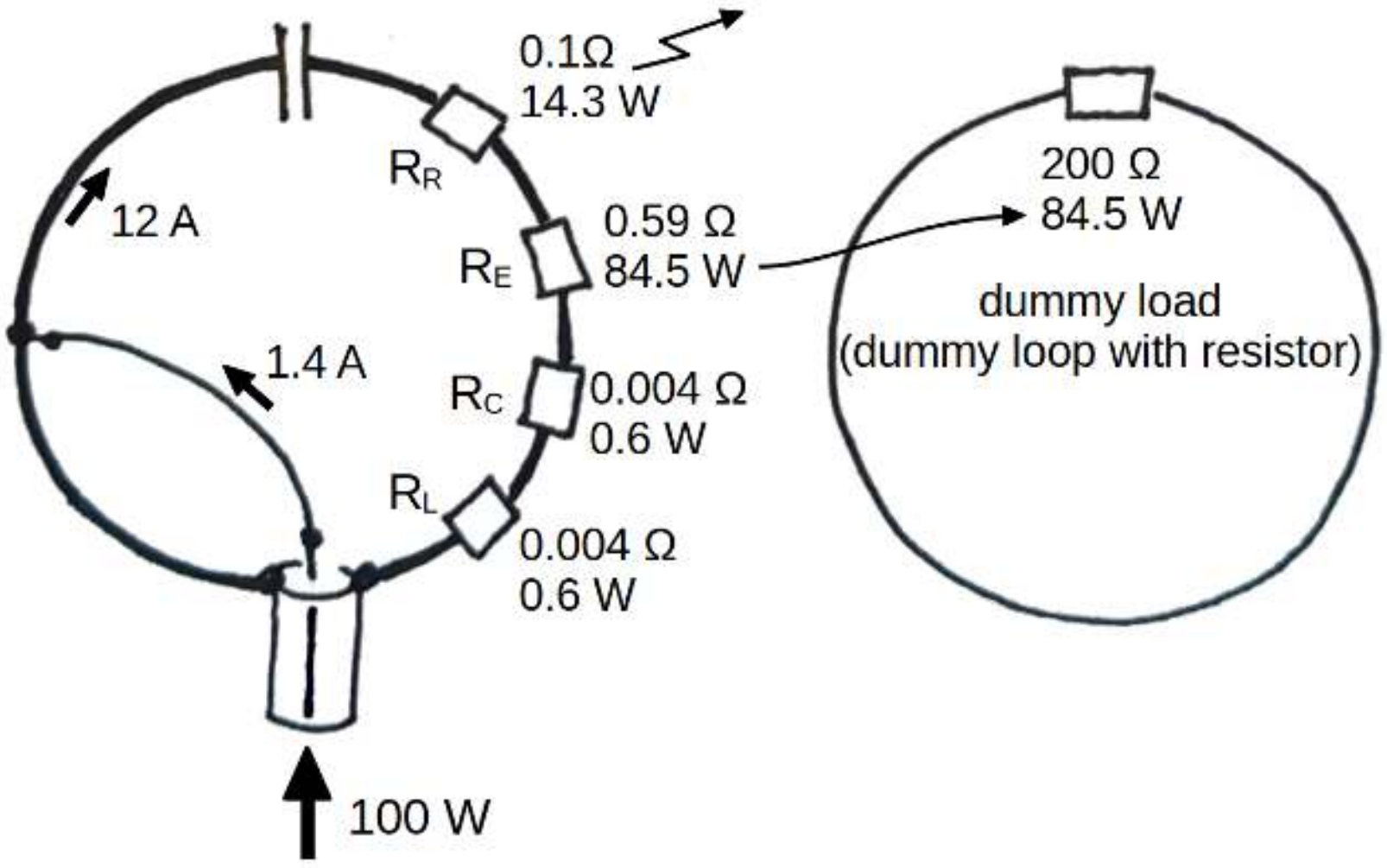

Figure 19: A magnetic loop is shown with its loss resistances. The antenna couples to a dummy load (dummy loop with resistor) placed in close proximity. Most of the energy is coupled from the antenna into this dummy load via the magnetic field, as in a transformer.

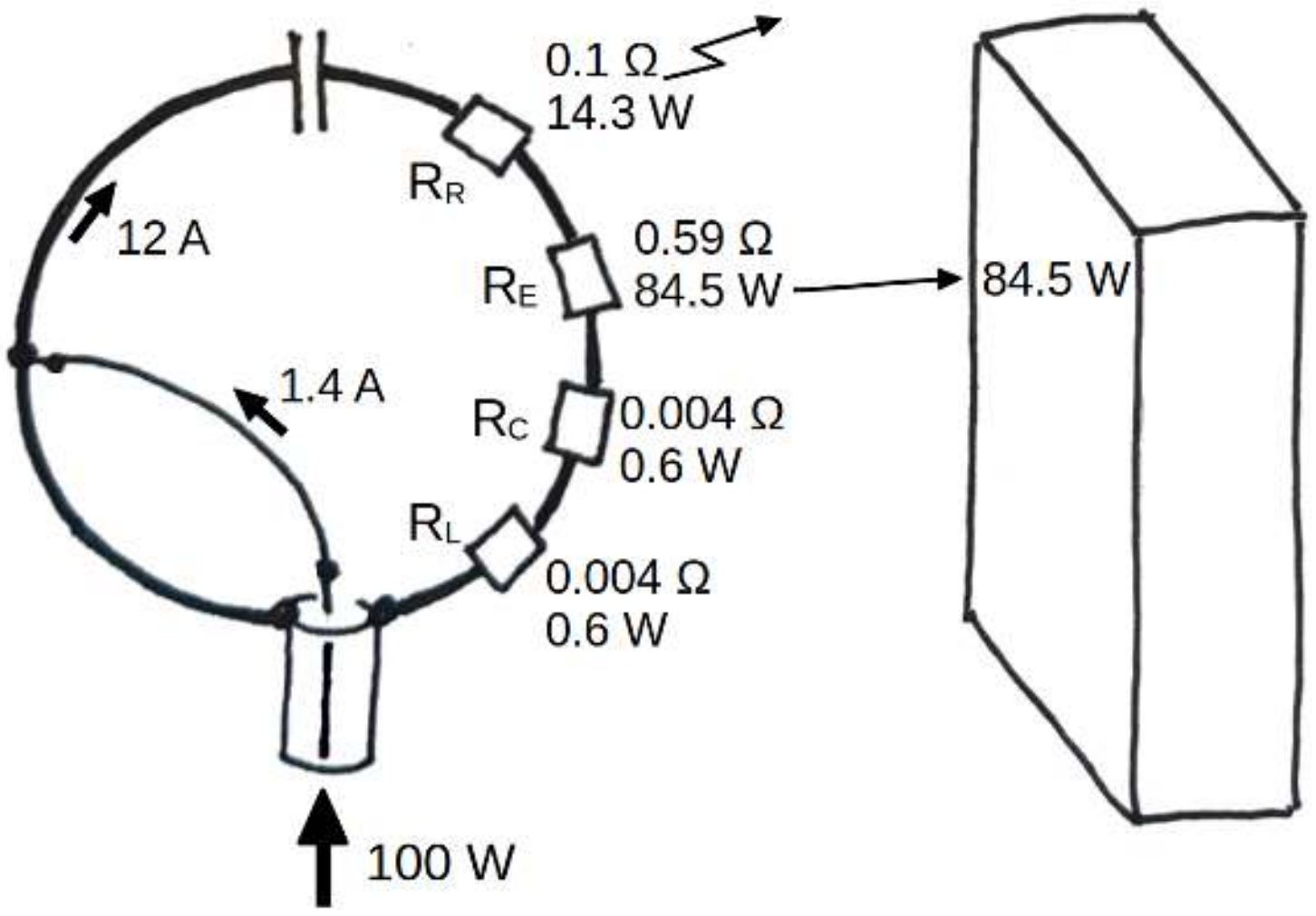

Figure 20: The dummy load has been replaced by a wall. The wall absorbs energy from the antenna's near field in exactly the same way as the dummy load. From the antenna's perspective, the conditions are identical.

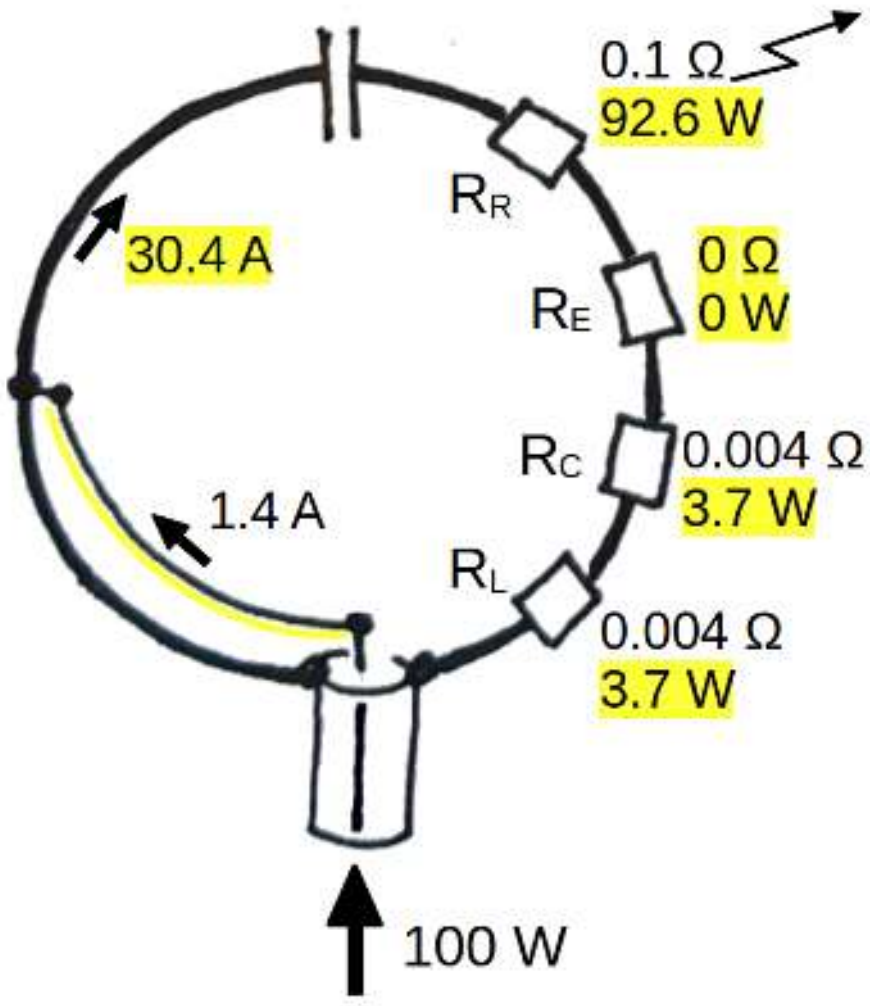

Figure 21: Now the wall is removed and the antenna stands in free space. The gamma match has been re-adjusted so that the feed point still presents 50 Ω. $R_E$ is now 0 Ω. A significantly larger current now flows in the main loop. The power going into radiation is much greater. The losses in $R_L$ and $R_C$ are also larger.

It is somewhat counterintuitive that the antenna in free space, where it has good efficiency, produces more heat loss in the antenna itself than when operated indoors, where the radiation efficiency is worse. Yet this is exactly the case.

# 2 The Small but Tubby Magnetic Loop Antenna

In this chapter I explain how I built the antenna and the choices I made to achieve the optimal design for my use case. In Section 2.1 I describe the construction of the loop, starting with the commercially available copper tubes, to adding mounting and transition parts and assembling the tube. In Section 2.2 I describe the capacitors used. The antenna is constructed in a way such that the capacitors can be exchanged to reach different bands from 10 m to 160 m.

## 2.1 Loop Construction

Commercial copper downspouts were used (see Figure 22).

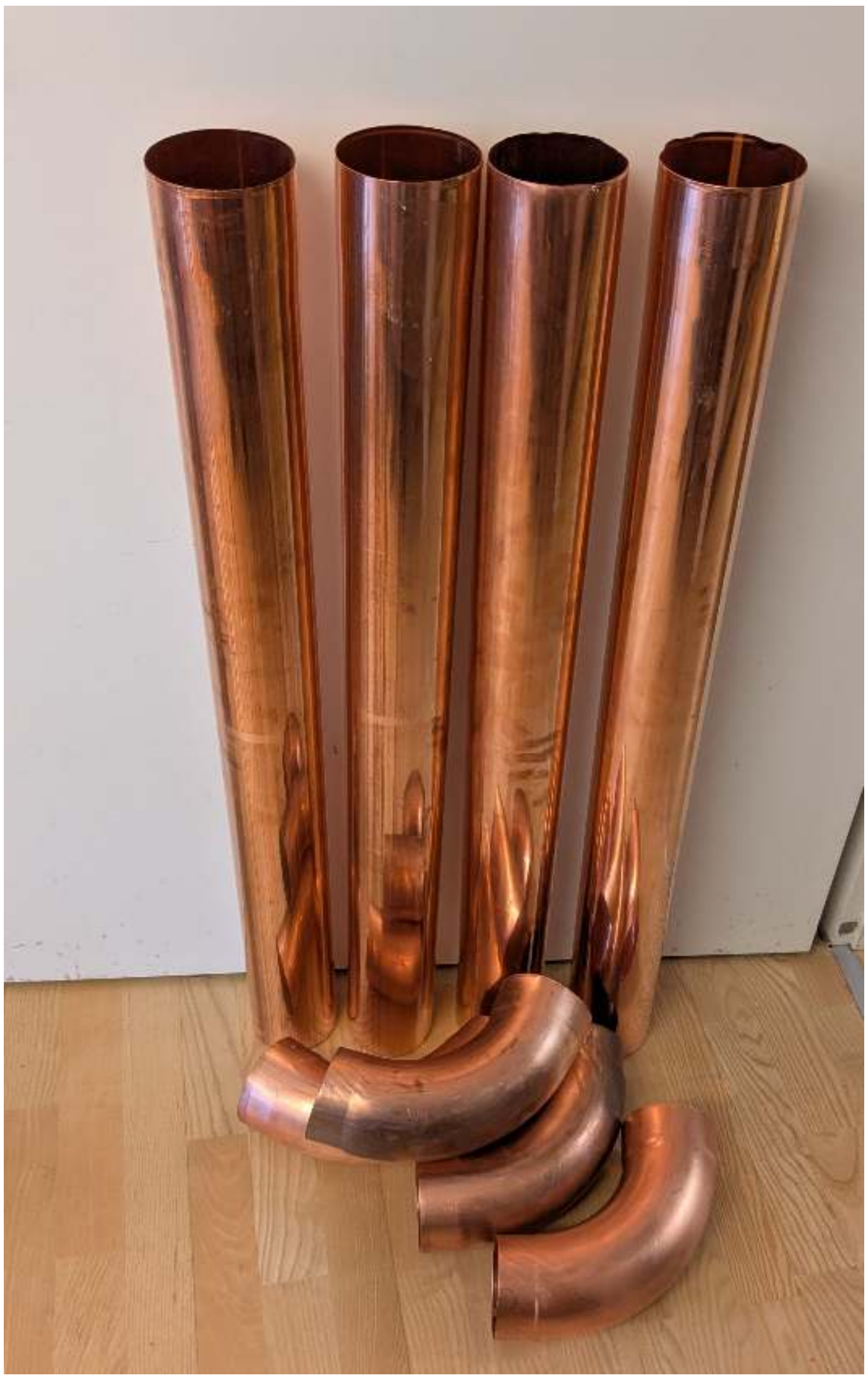

Figure 22: Commercially available copper downspouts with 100 mm diameter.

To mount the coaxial cable to the transmitter and to hold the coupling loop, additional copper parts were attached to the lower part of the loop (see Figure 23).

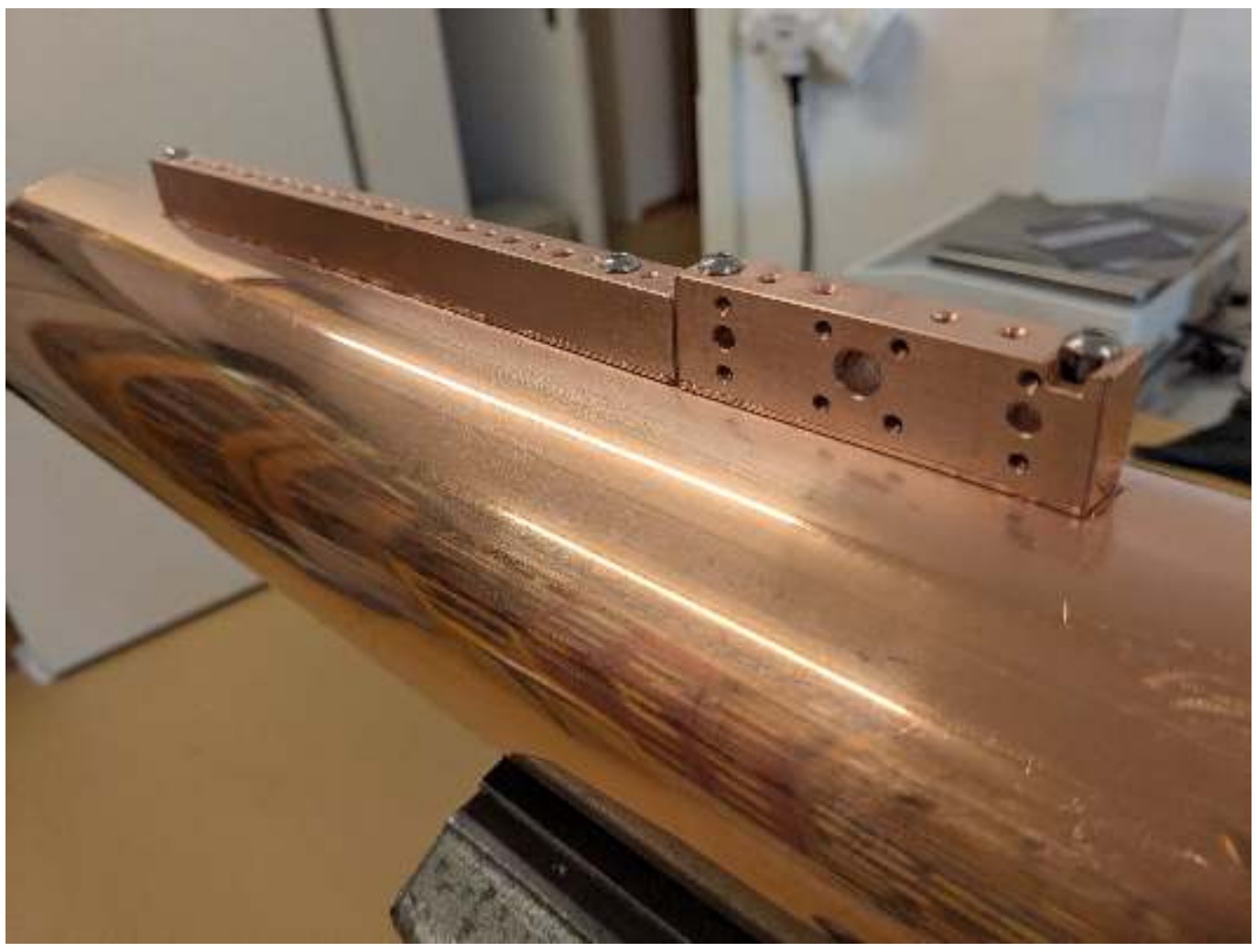

Figure 23: Copper parts for mounting the coupling loop.

For the transitions to the capacitors, I applied cutting patterns using adhesive labels and rough-cut them with tin snips (see Figure 24). The details were then machined using a Dremel cutter.

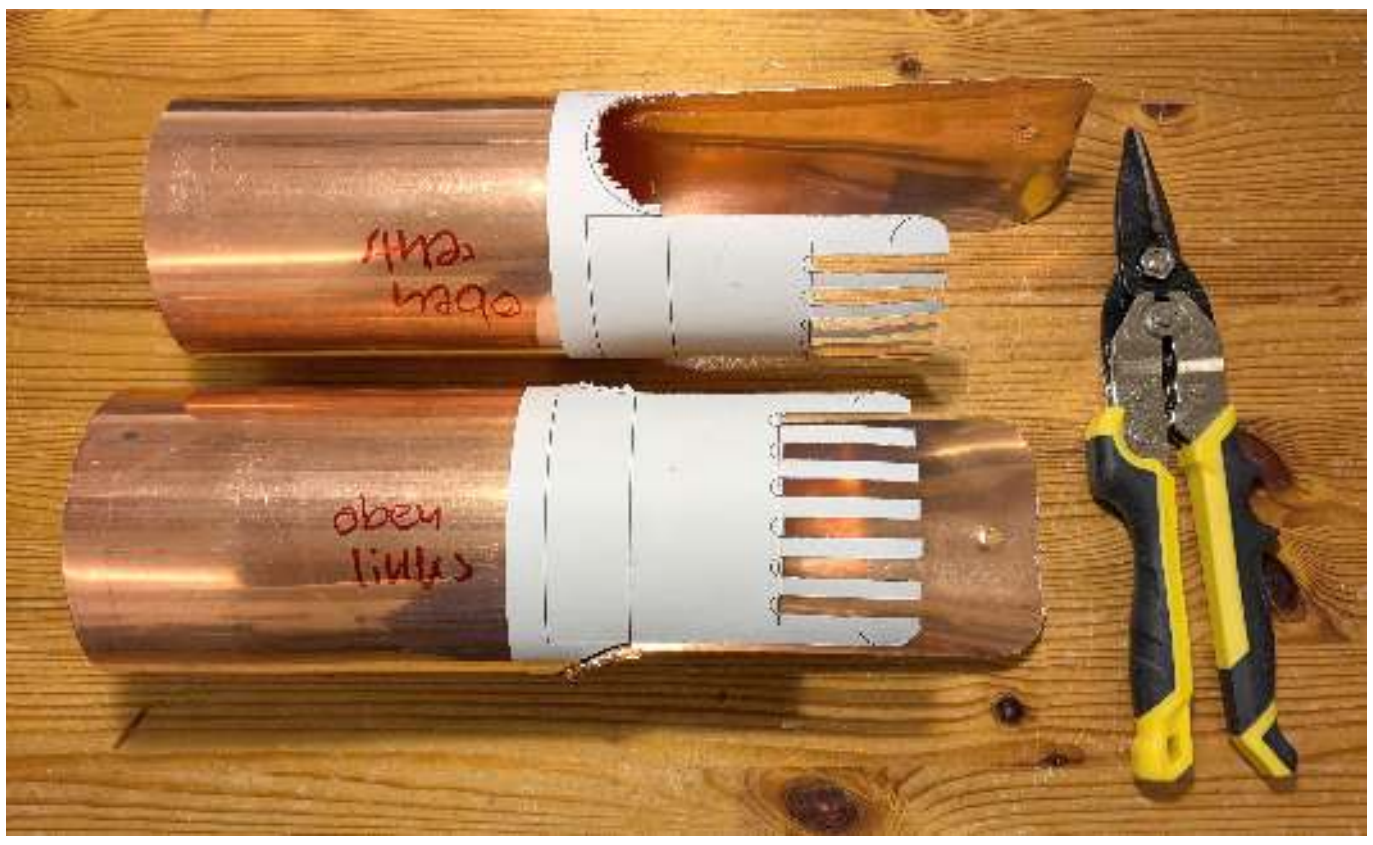

Figure 24: Transitions from pipe to capacitors with glued-on cutting patterns.

The movable parts of the transitions were annealed with a propane torch to make them ductile (see Figure 25).

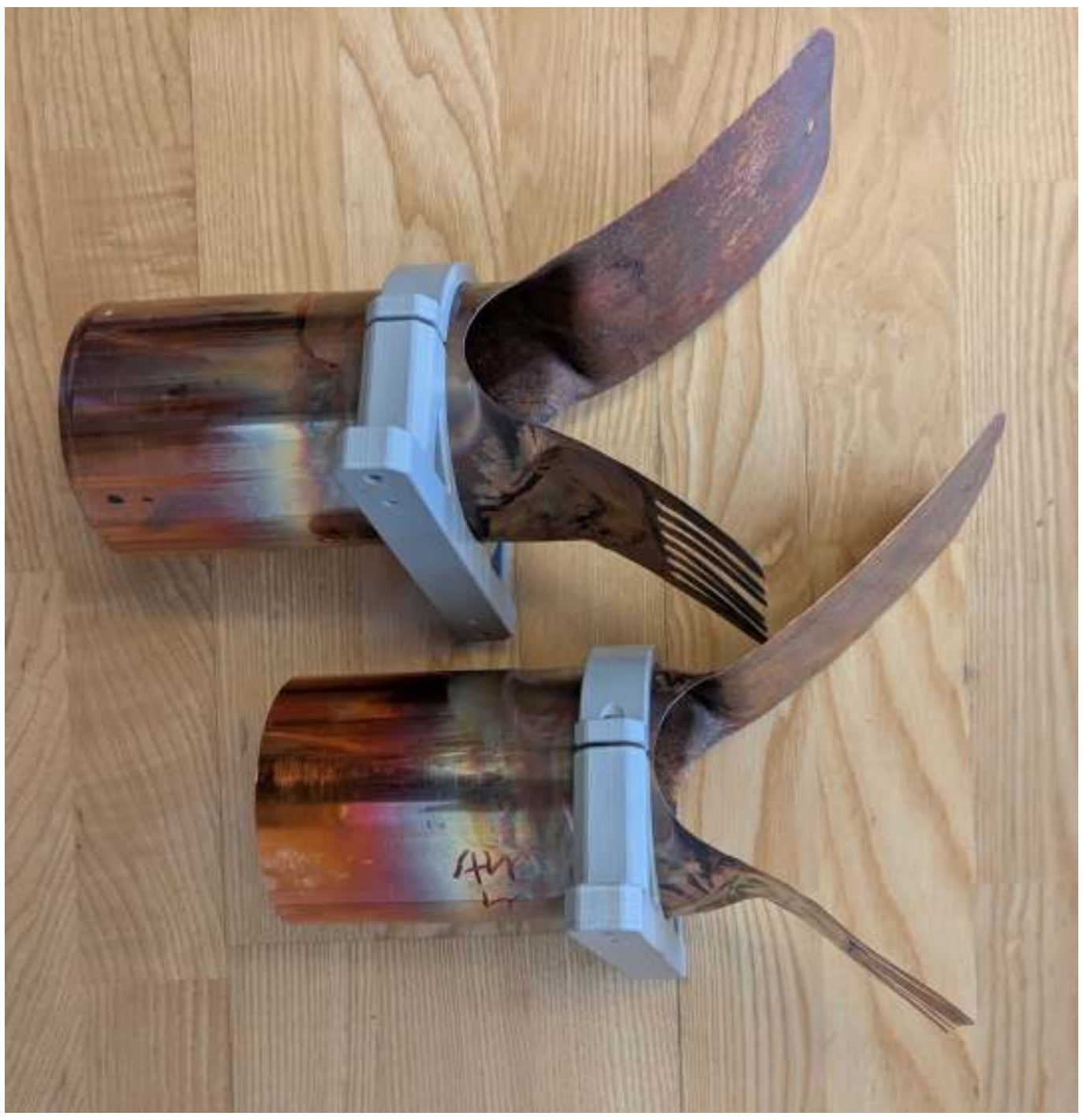

Figure 25: Annealed and bent transitions. The surface is still black from copper oxide.

The commercially available pipe elbows had an angle of 85°. Since I needed 90°, I made a cut on the outside to provide the necessary flexibility. The pipes were cut to length, and the surfaces were cleaned with steel wool.

For the solder joints, the pipes were pressed together and fixed using a strap and screw buckle, ensuring a very narrow solder gap (see Figure 26).

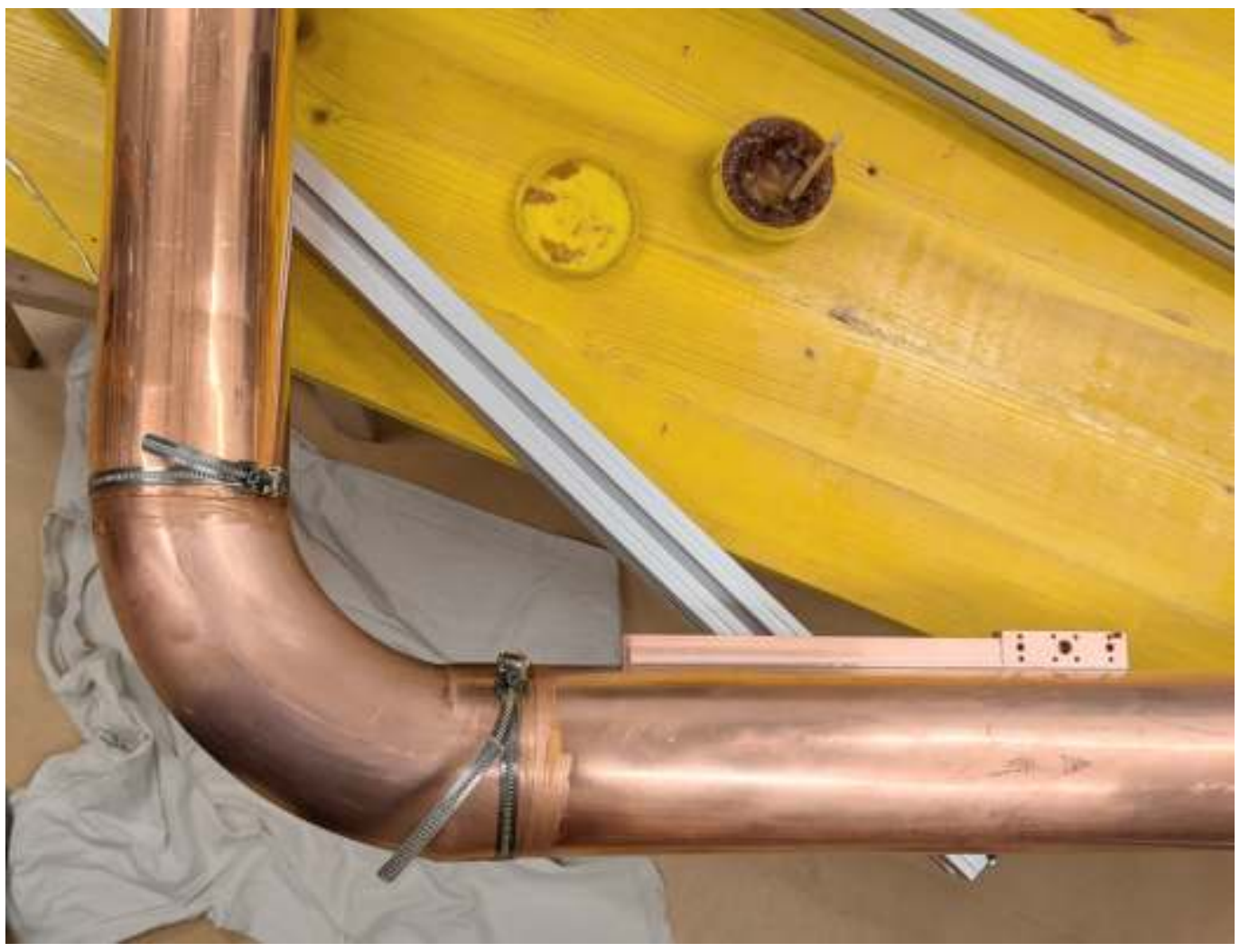

Figure 26: Clamping the copper parts securely before soldering.

Aggressive soldering grease (containing 16% zinc chloride) was used as flux. Soldering was performed with a propane torch and soft solder (Stannol Kristall 611, 1 mm, Sn96.5Ag3Cu0.5), which flows excellently and offers good conductivity.

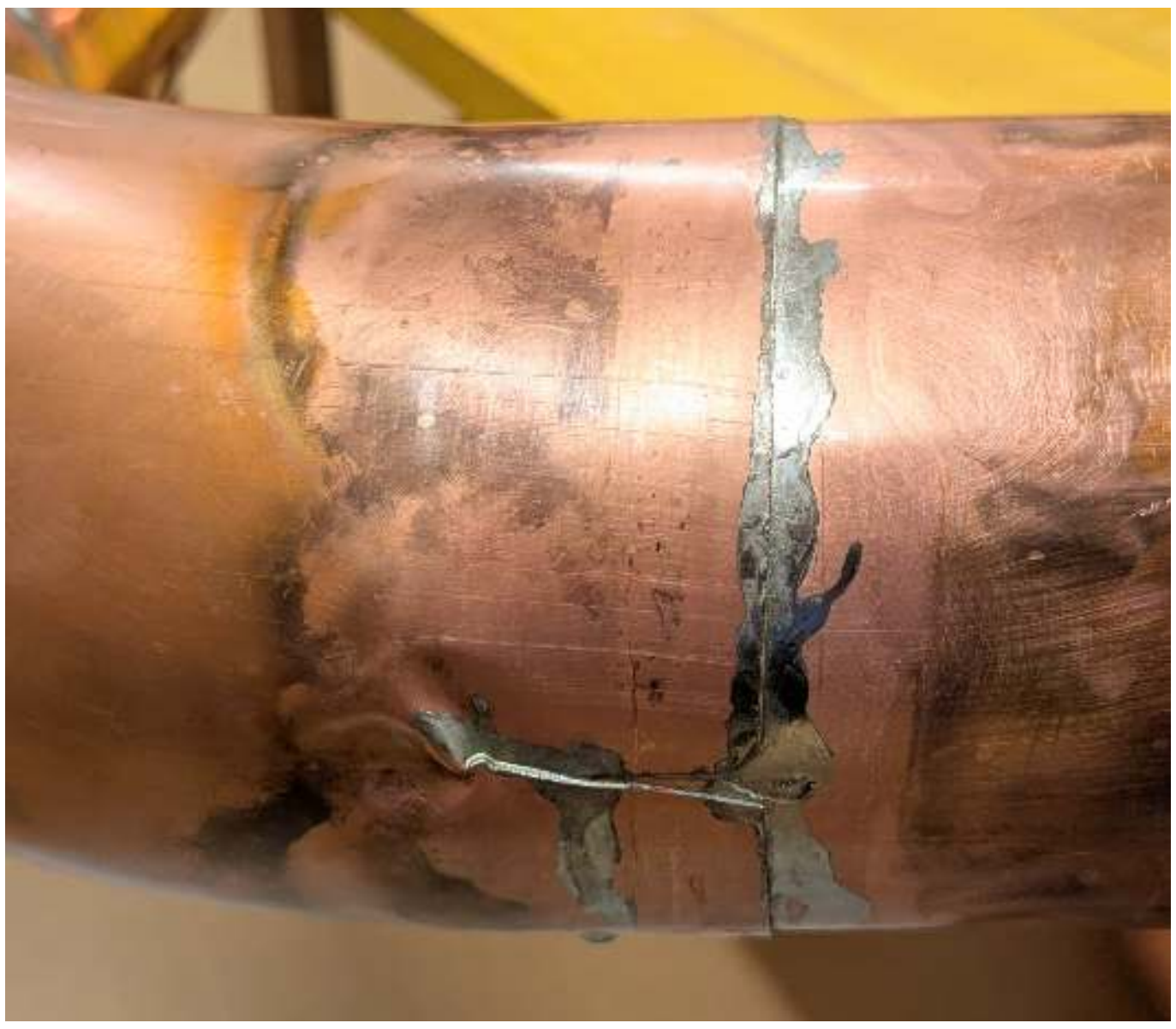

Figure 27: Soldering joint after removing flux residues.

Excess solder was removed using a small cutter on a Dremel tool. The goal was to minimize the path current travels through the solder (see Figure 28).

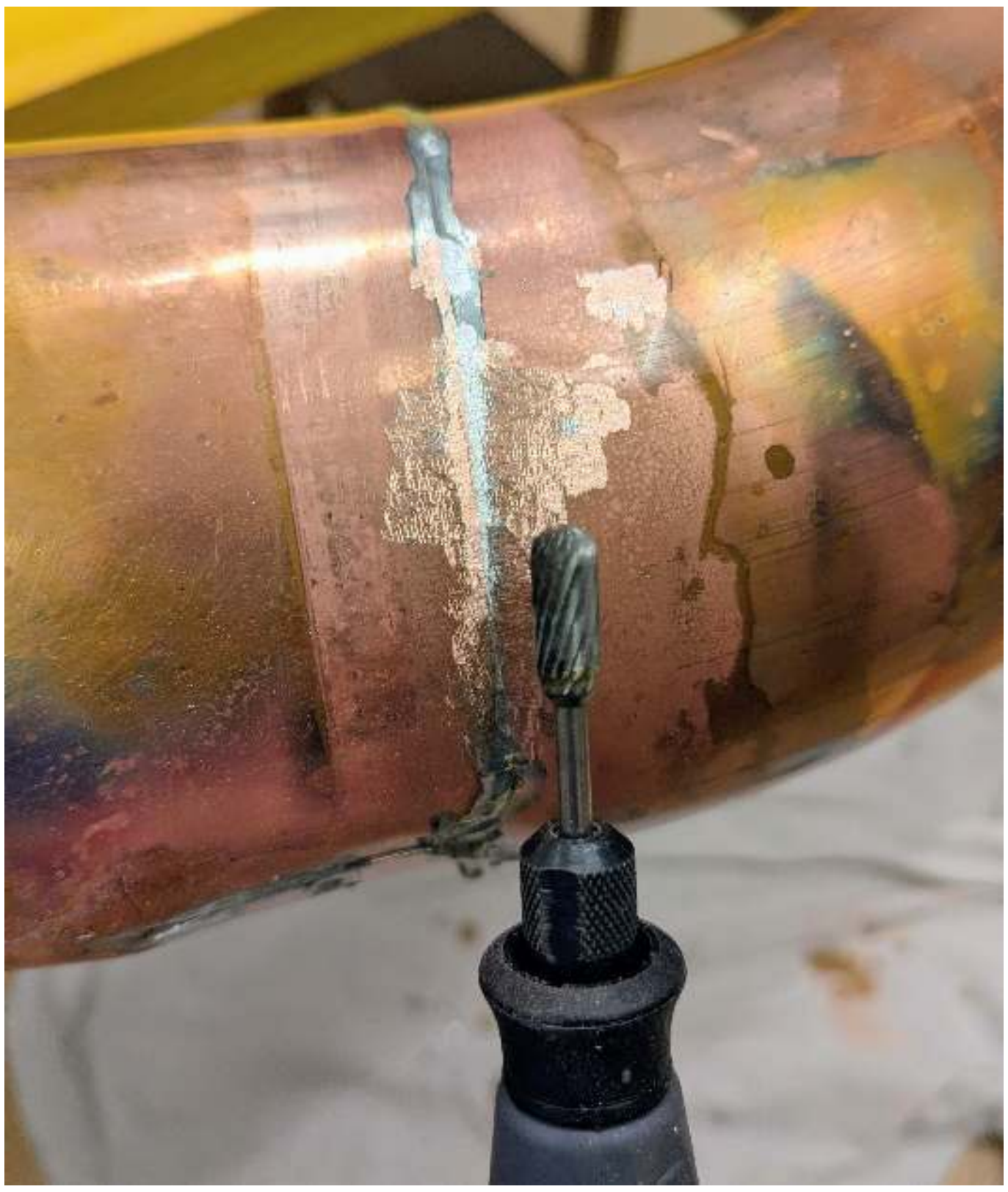

Figure 28: Removing excess solder with a milling cutter.

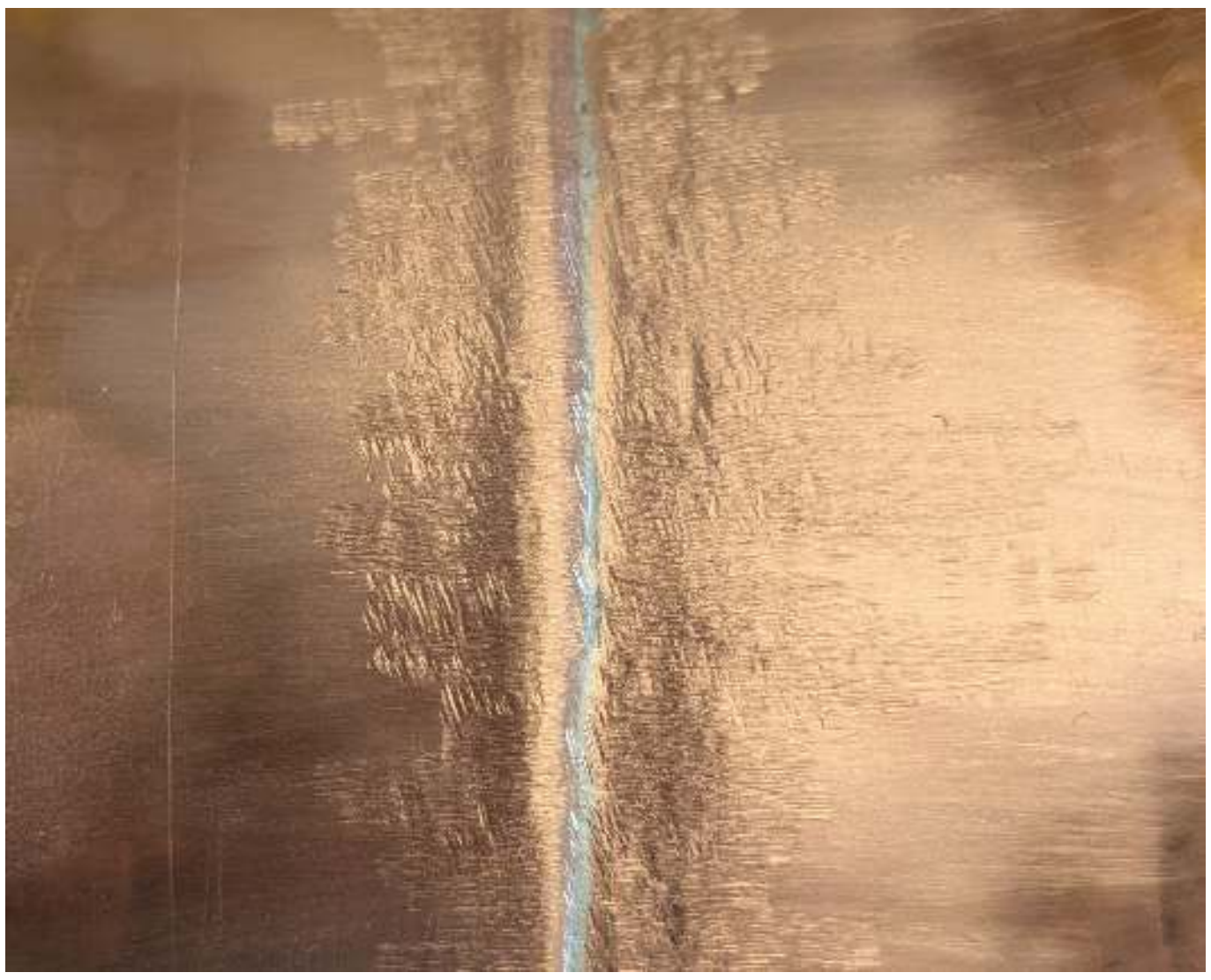

Figure 29: Surface after milling and sanding. The current flows through the solder for only a very short distance.

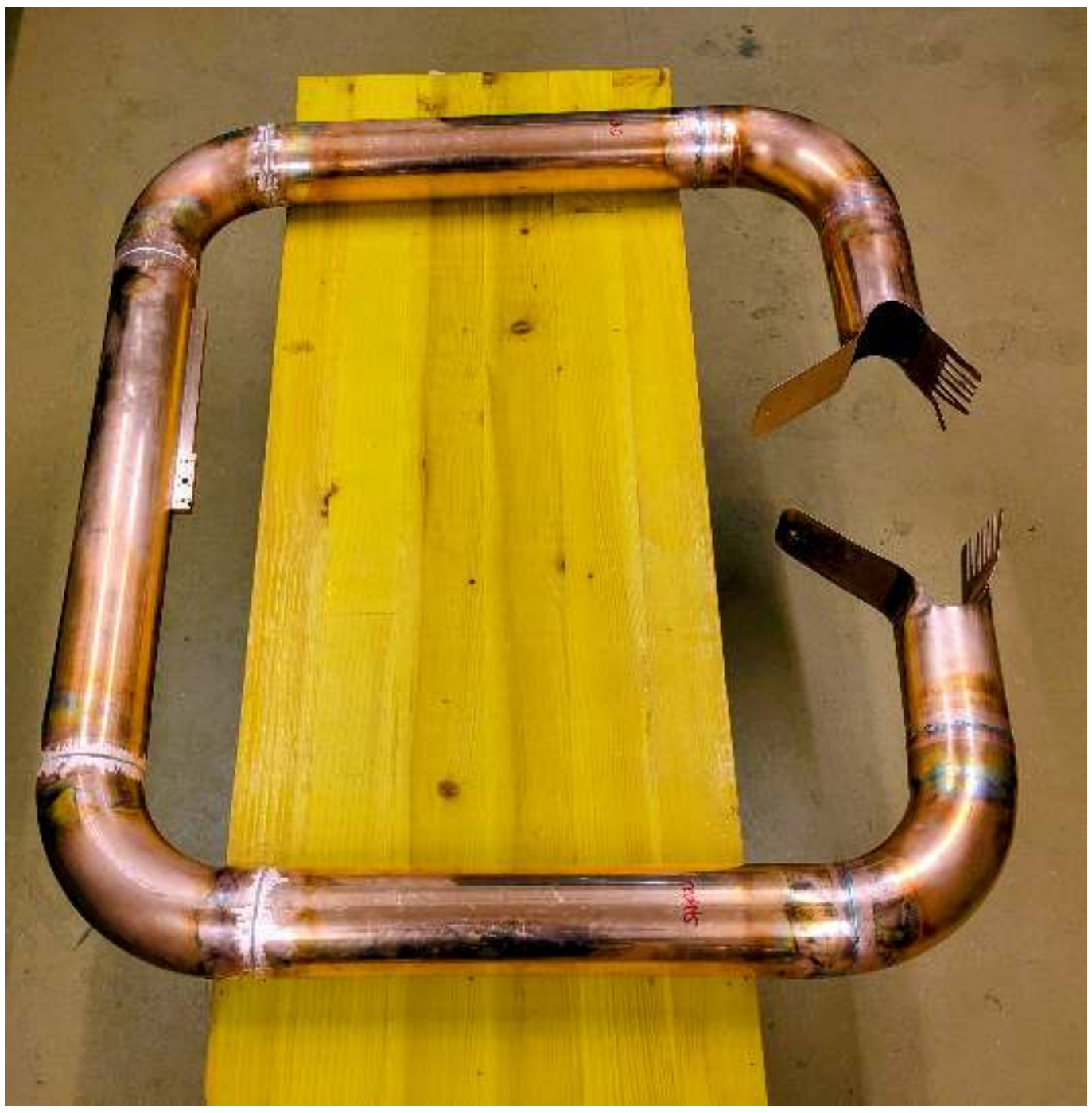

Figure 30: Finished soldered loop.

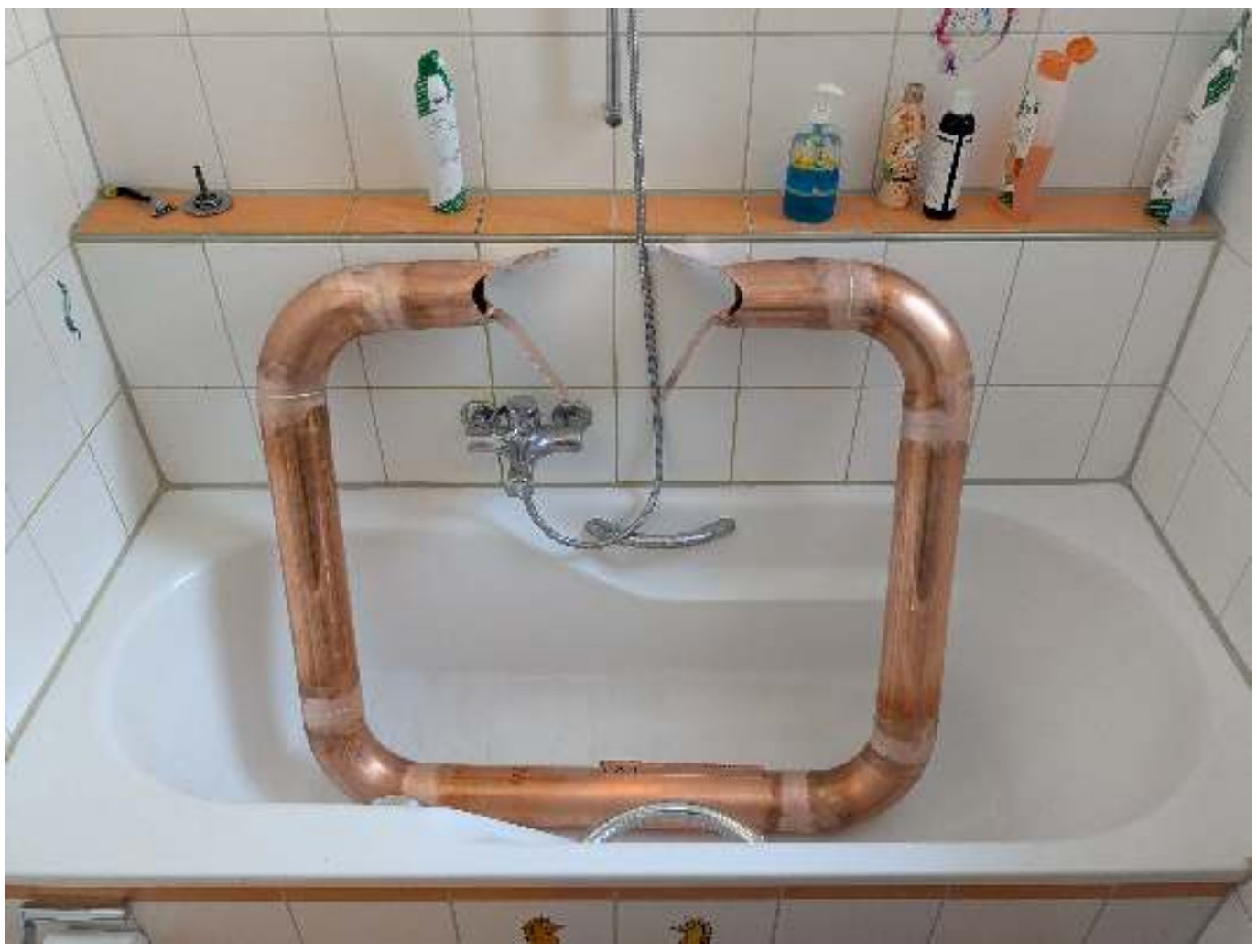

Figure 31: Loop before final cleaning with steel wool.

Finally, the surface was cleaned with sandpaper and steel wool. “Kontakt Chemie PLASTIK 70” varnish was applied for corrosion protection, while contact points were masked with tape (see Figure 32). This was a substantial and laborious task.

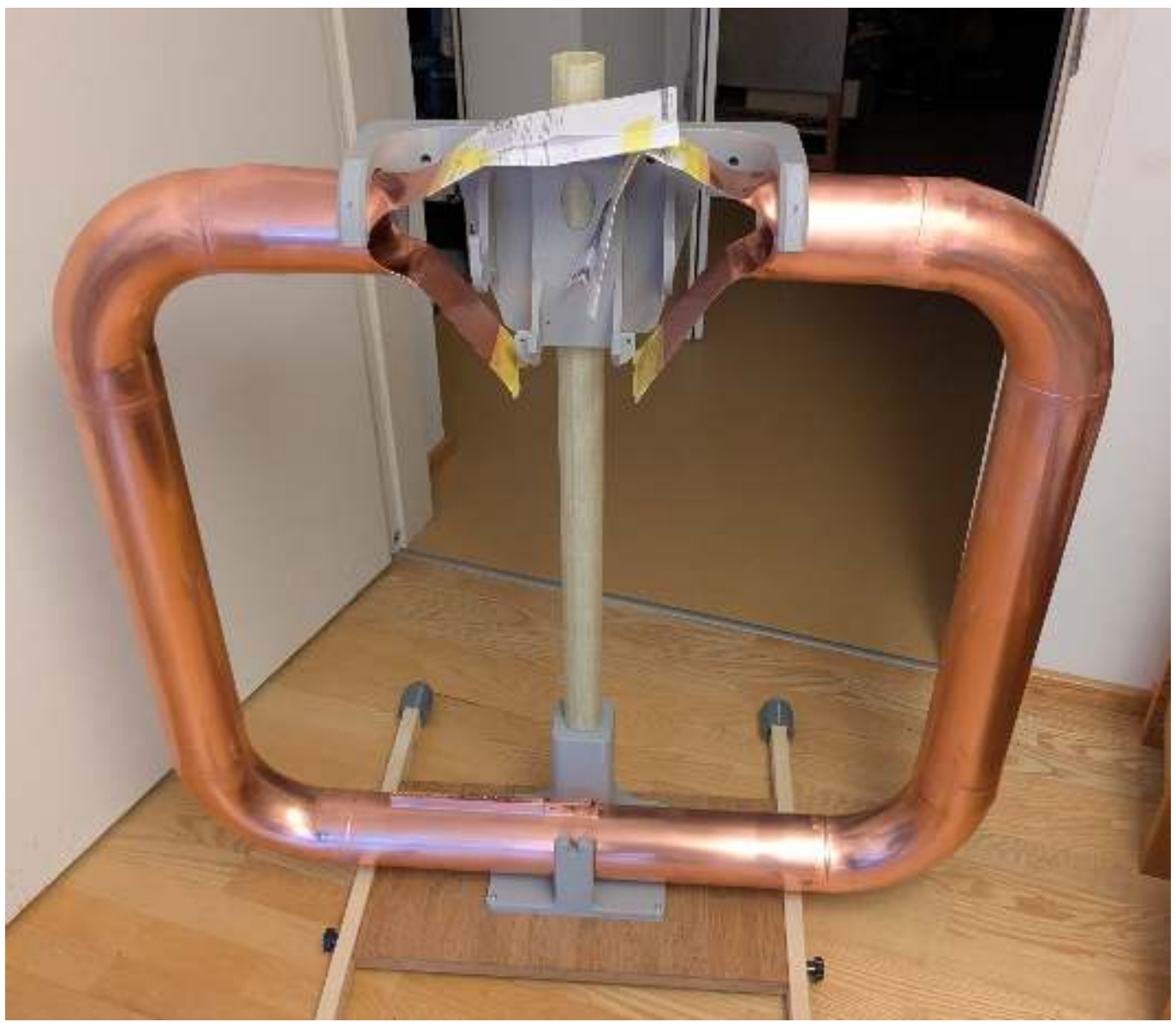

Figure 32: Loop finished and varnished.

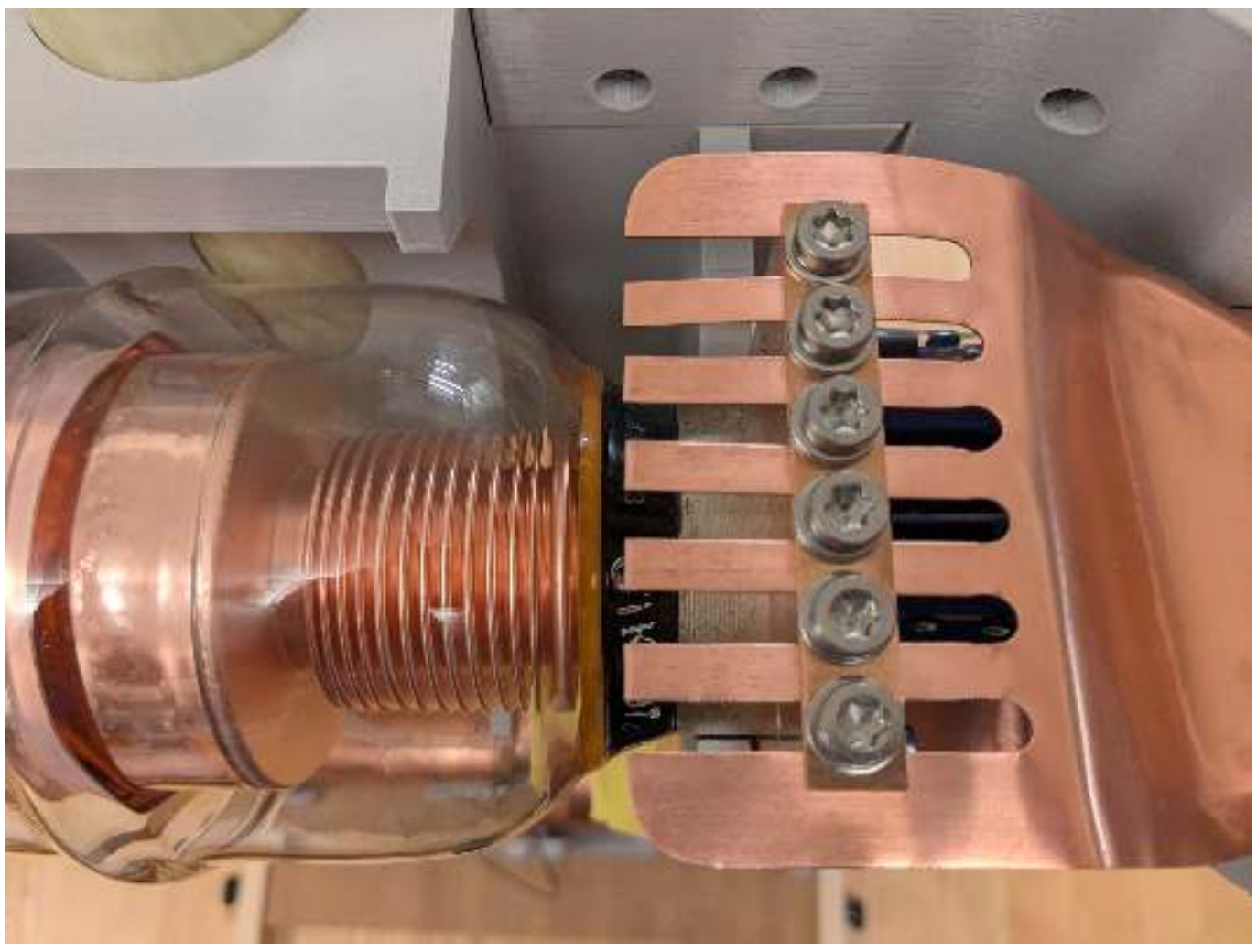

Figure 33: Transition to the variable capacitor. The flattened pipe section is 100 mm wide and is screwed onto the 12 mm thick transition piece to the capacitor using six M6 stainless-steel Torx screws.

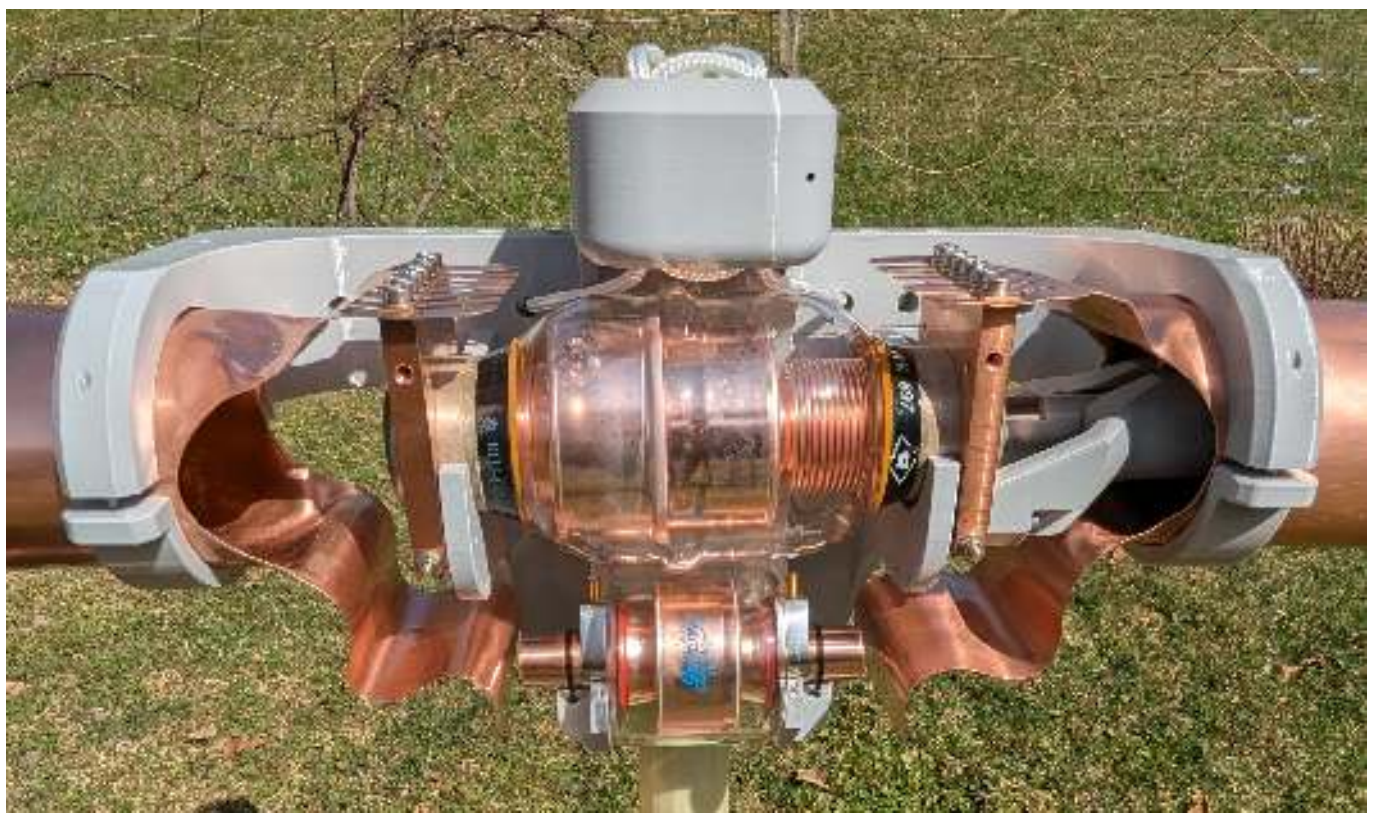

Figure 34: Configuration for 10 m to 40 m: only the variable capacitor is connected. Capacitor A is mechanically mounted, but the side screws are not inserted. The gap between the copper tabs and the capacitor end caps is about 10 mm.

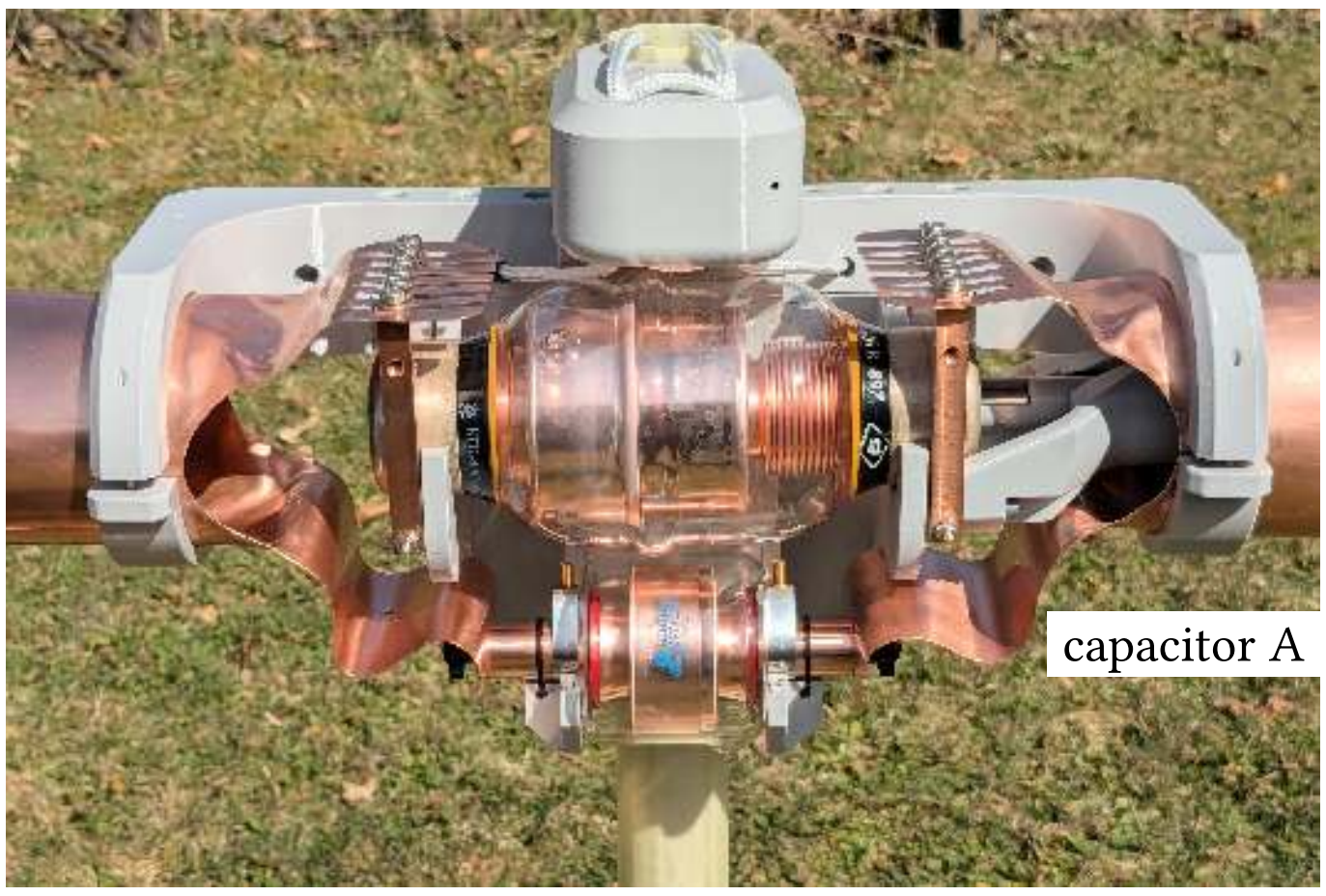

Figure 35: Configuration for 60 m: capacitor A 500 pF is electrically connected. The side screws press the copper tabs onto the capacitor end caps.

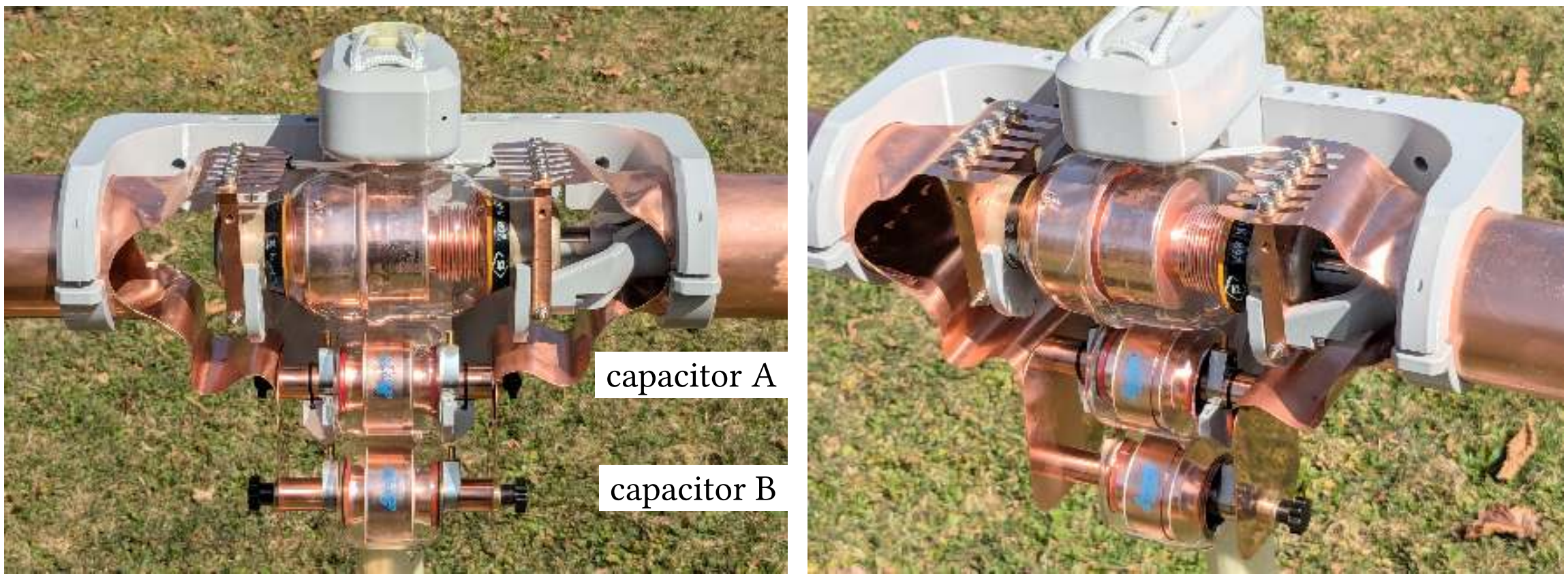

Figure 36: Configuration for 80 m: additional capacitor B 500 pF is installed.

## 2.2 Capacitors

### 2.2.1 Variable Capacitor

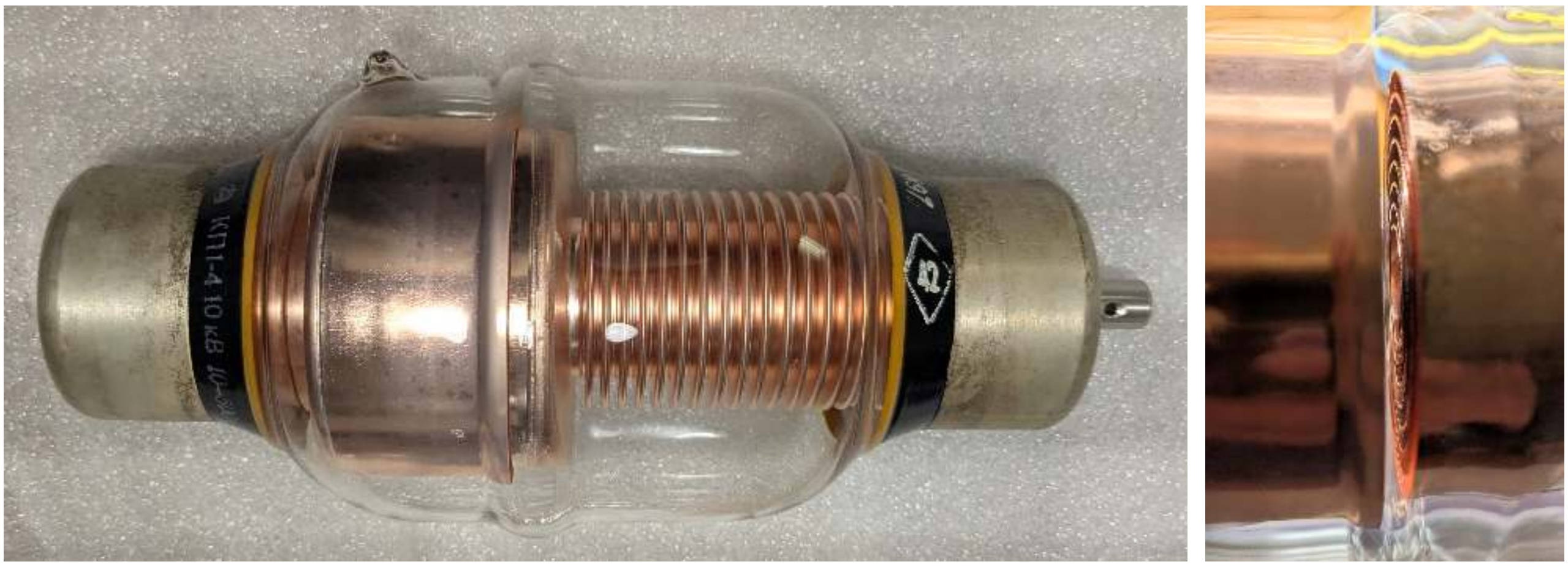

Figure 37: Left: variable vacuum capacitor KP1-4 10–500 pF. Right: when the capacitor is set to minimum capacitance, the concentric copper tubes are visible.

I had long been curious about the internal structure of such a vacuum capacitor, so one had to be sacrificed.

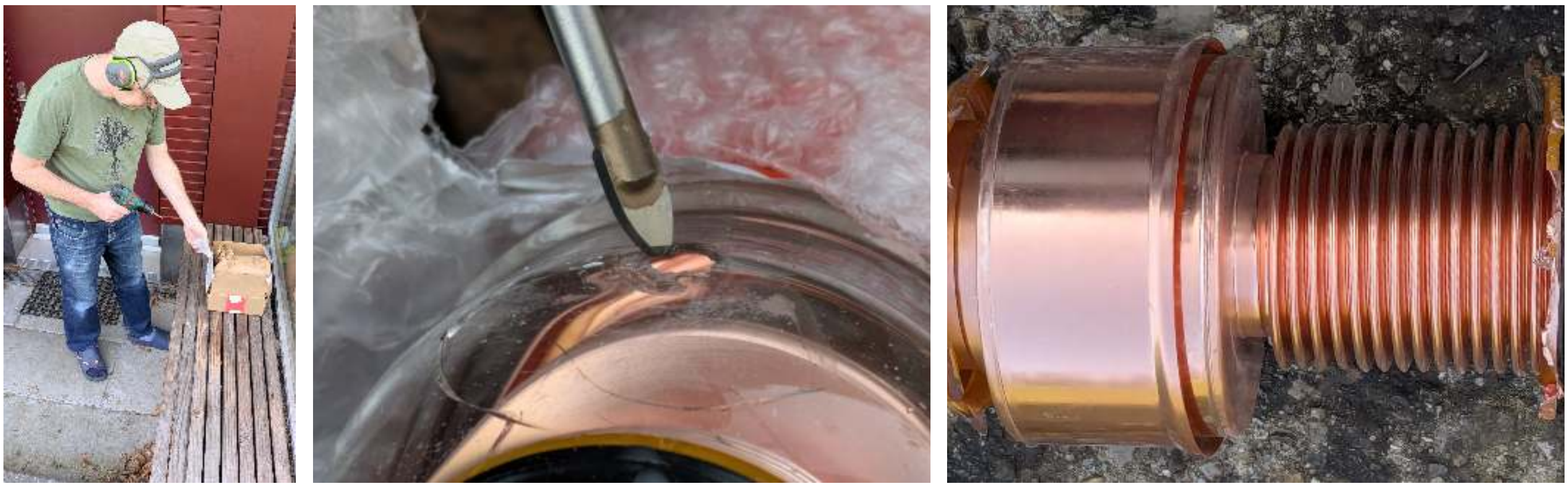

Figure 38: Left: the author and the capacitor wrapped in cardboard as implosion protection. I had no idea whether opening it would be dangerous — it took some courage to drill into the glass envelope. The sound was just a soft blob, not loud at all. Center: glass drill bit; the vacuum is gone. Right: the glass is then removed with a hammer, revealing the beautifully colored copper inside.

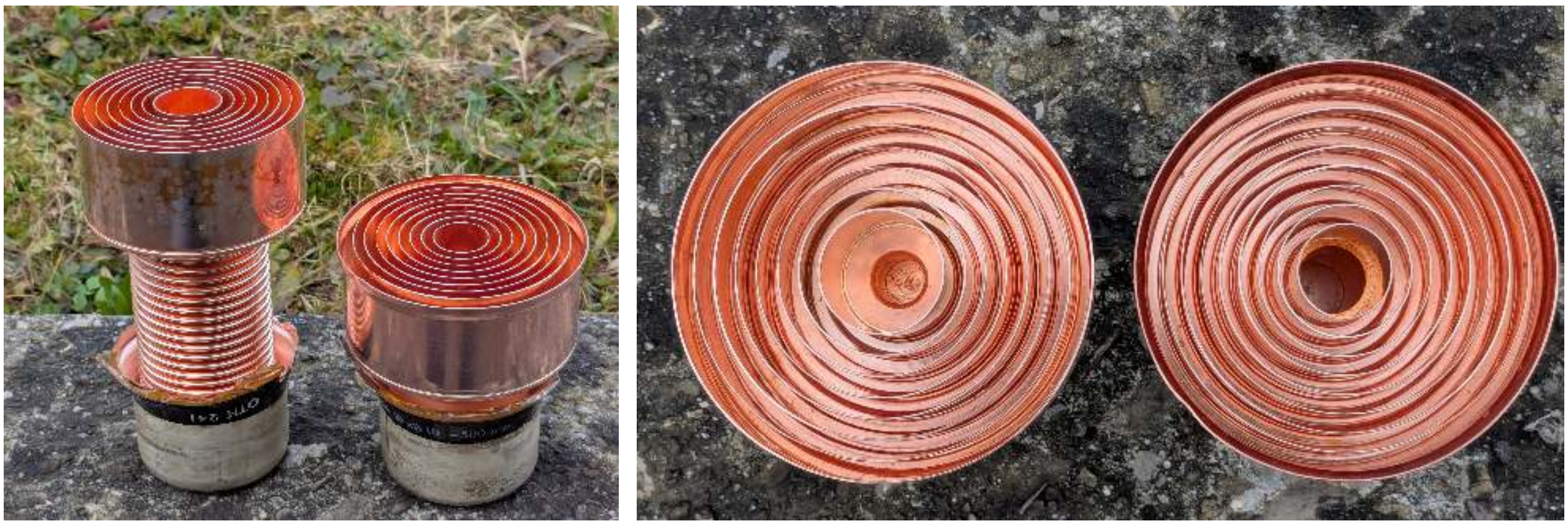

Figure 39: View of the concentric interleaving vanes.

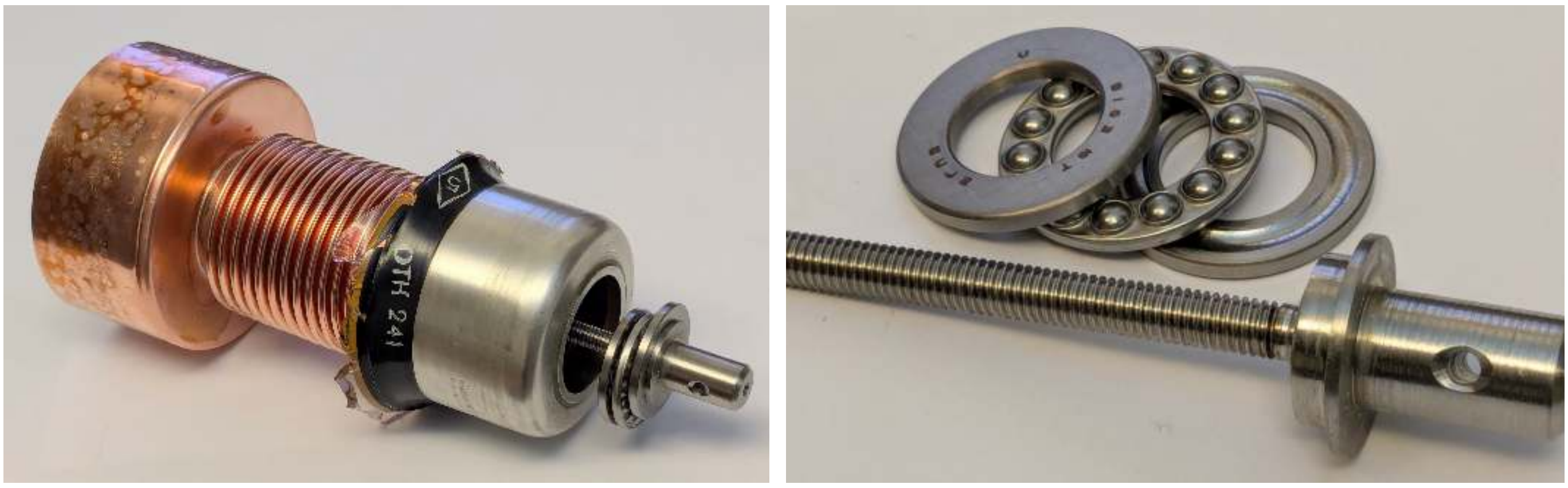

Figure 40: Turning the shaft clockwise retracts one set of vanes, reducing the capacitance. An internal spring (not visible in the photo) pushes the copper bellows back to full extension when the capacitance is increased.

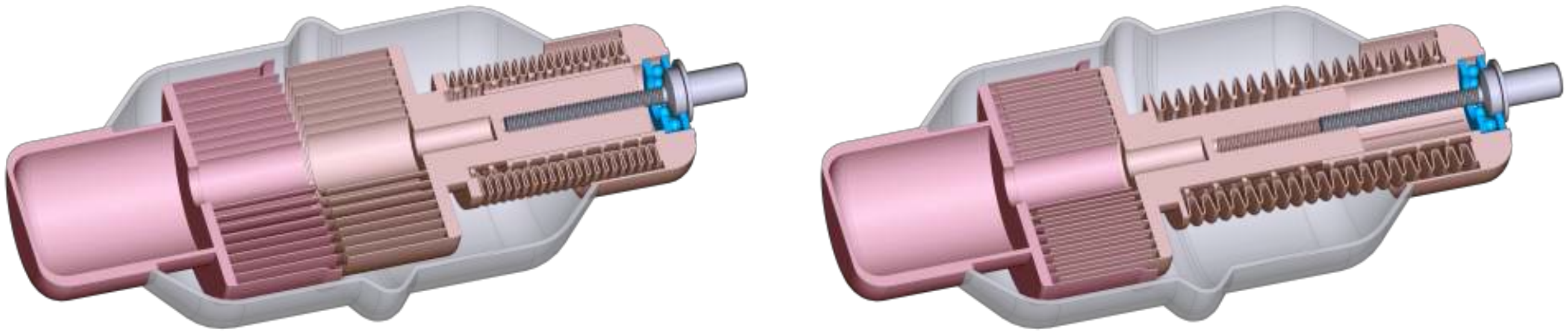

Figure 41: Left: vanes fully separated, 10 pF. Right: vanes fully interleaved, 500 pF. The spring that extends the bellows is visible.

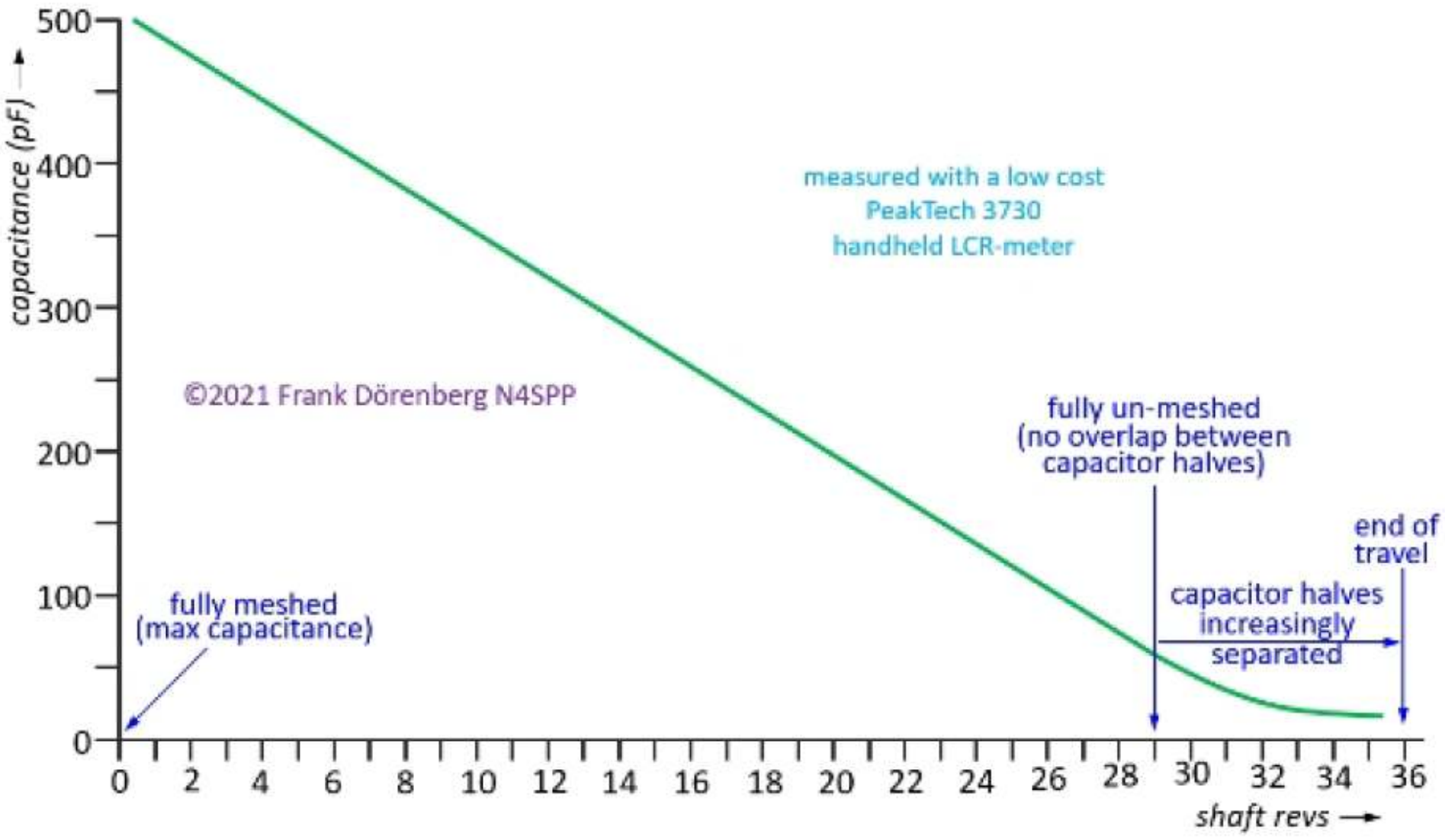

Figure 42: Capacitance characteristics versus turns of the axis. Reproduced with kind permission of Frank Dörenberg [3].

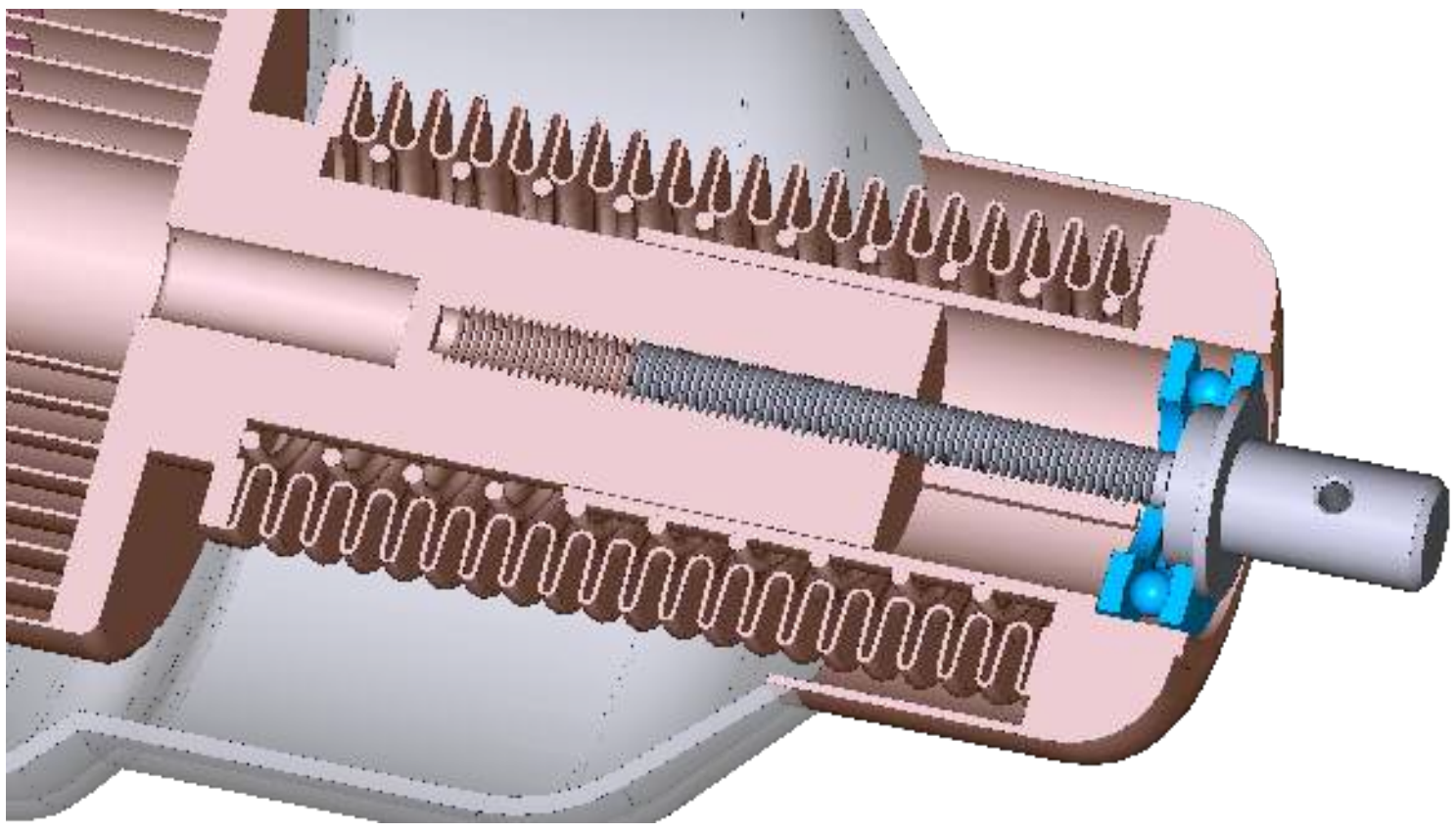

Figure 43: Cross-sectional detail showing the movable copper bellows, which not only provides the vacuum seal but also carries the RF current to the vanes. The thrust bearing absorbs the axial force of the shaft.

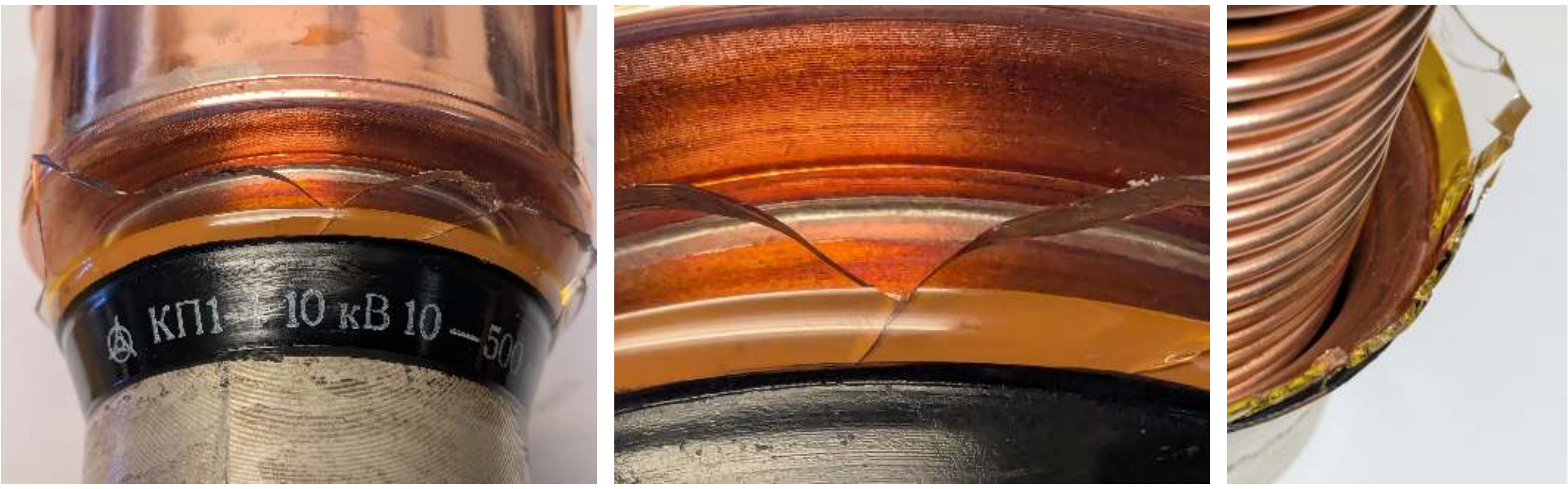

Figure 44: The transition from glass to copper was made by fusing glass and copper together.

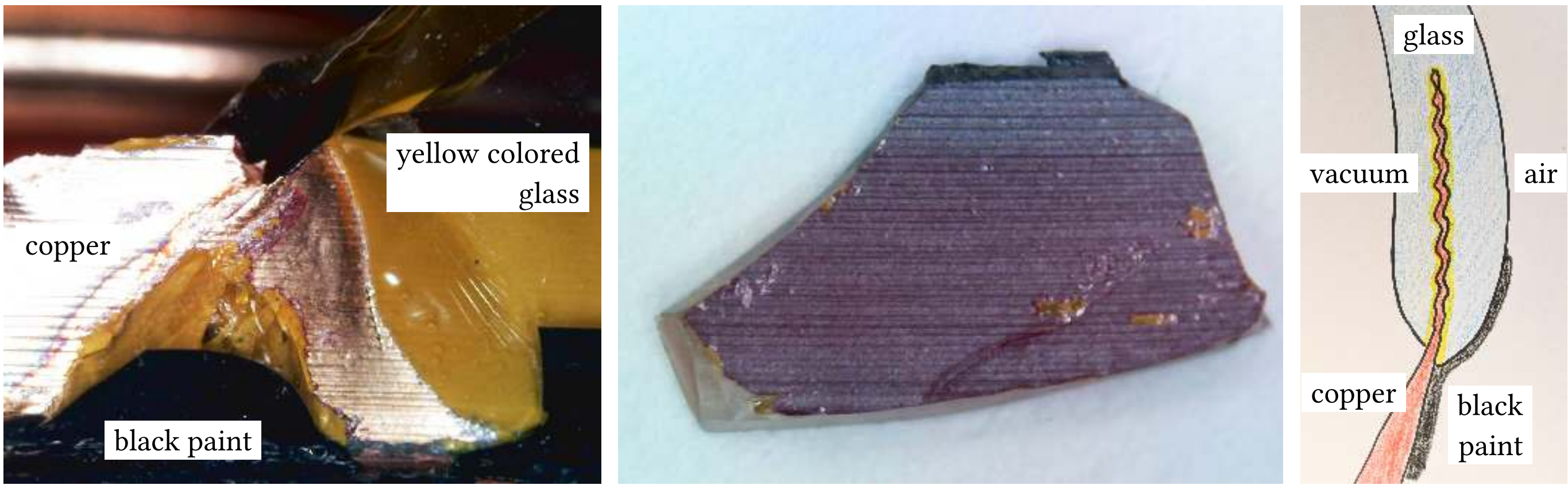

Figure 45: Left: view of a fracture point; when looking through the glass from outside, a yellow coloring is visible in the copper region, presumably caused by the fusion of glass and copper oxide. Center: a glass shard from the copper side, showing violet coloring with visible rings. Right: cross-sectional sketch. The copper lip must be very thin so that the stresses caused by the different thermal expansion coefficients of glass and copper do not become too high. Remarkably, this seal has remained vacuum-tight for decades.

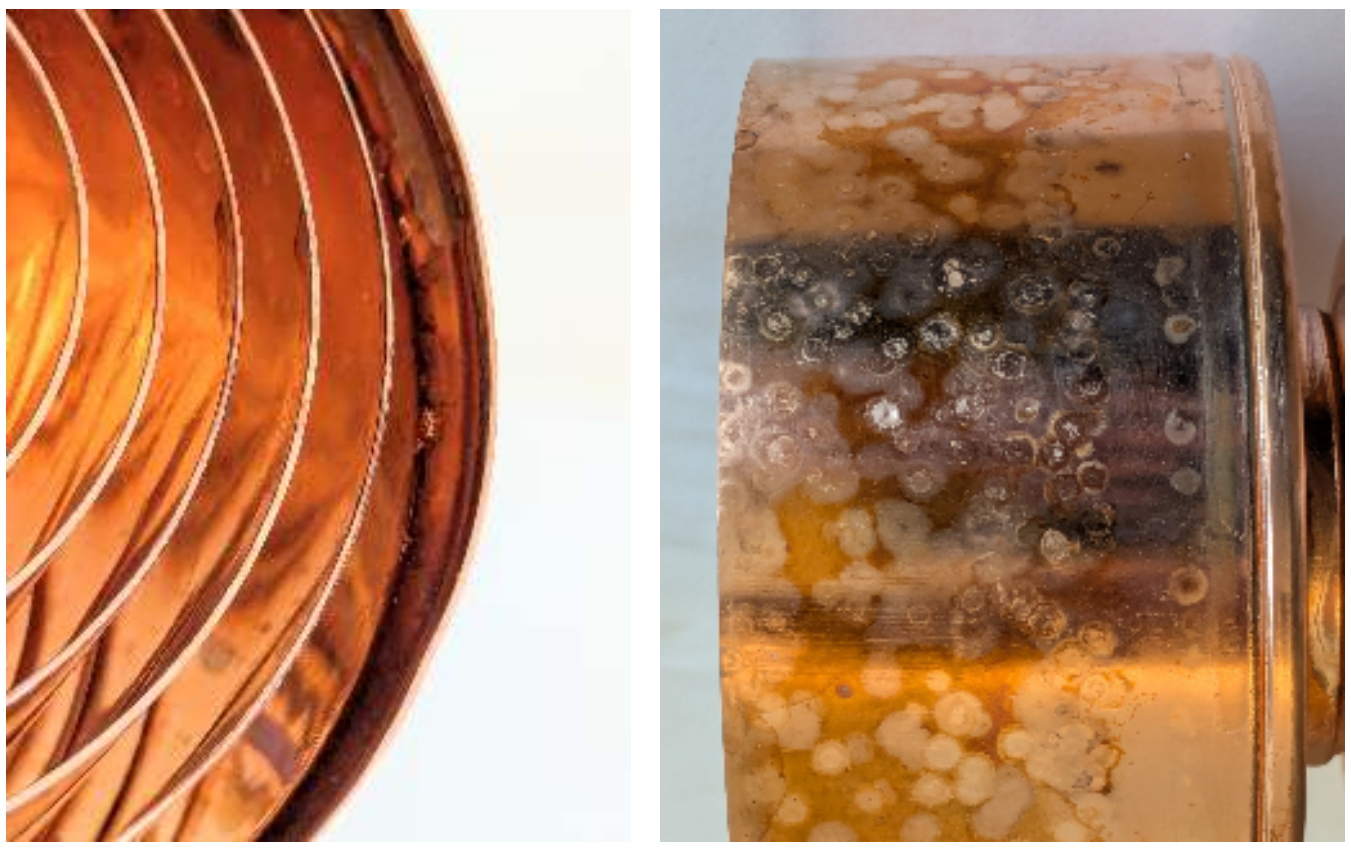

Figure 46: On this capacitor specimen, arcing marks are clearly visible.

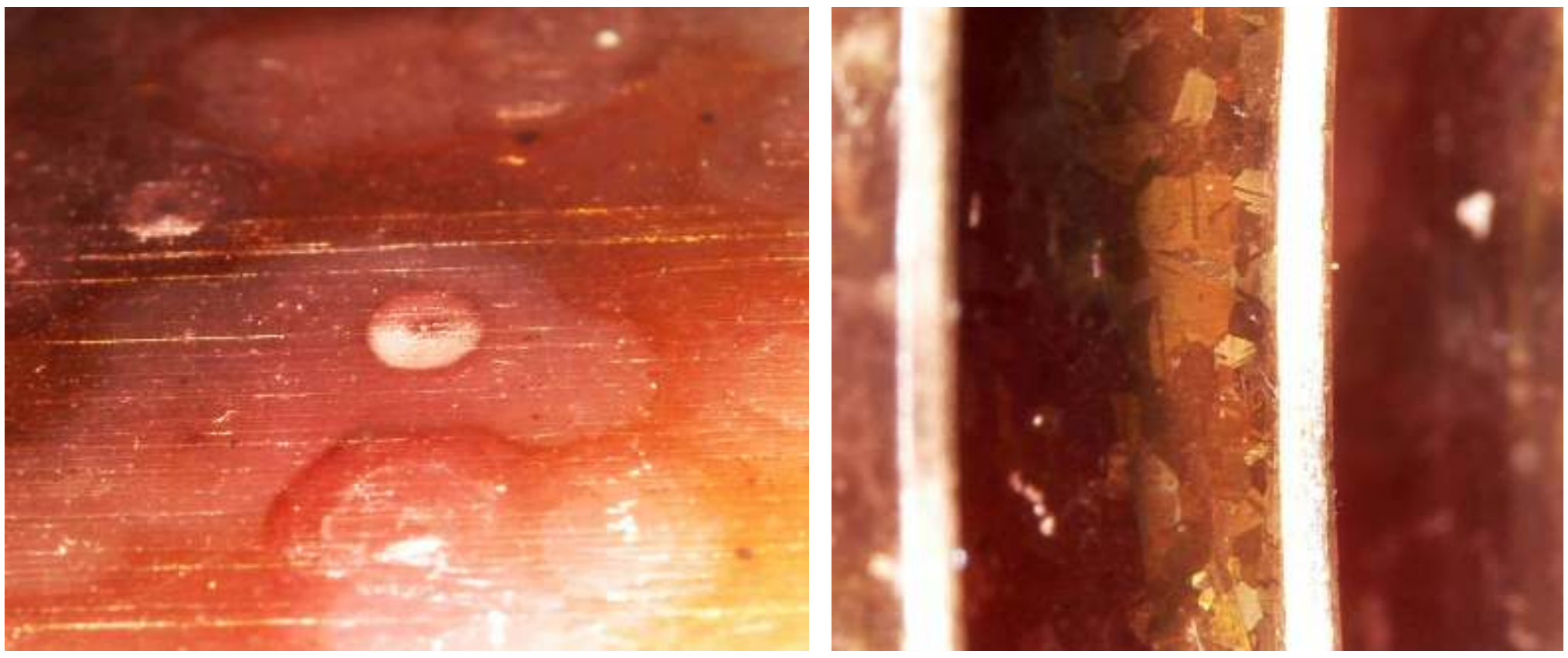

Figure 47: Left: close-up of an arcing pit, finely recessed and barely visible in the photograph. Right: view of a vane showing visible copper crystals — such large crystals indicate that very pure copper was annealed. Copper with impurities does not form crystals of this size. Pure copper conducts electricity well.

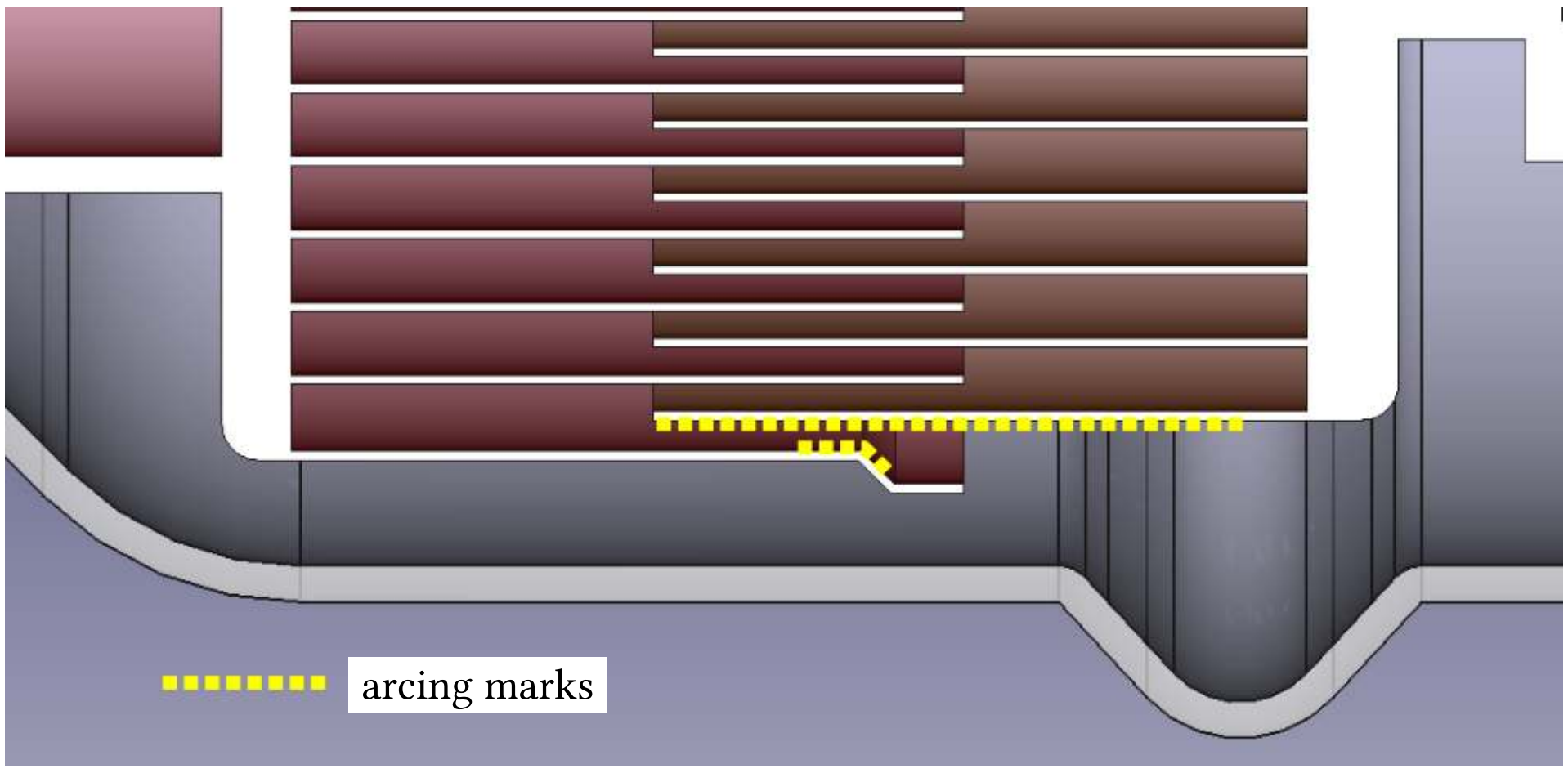

Figure 48: The arcing marks appear only at the outermost gap. The outermost vane is slightly bent outward — a feature presumably intended for electric-field grading. Despite this measure, flashovers still occur exclusively at the outermost vanes. It has been reported that vacuum capacitors can be conditioned by slowly increasing DC voltage through a series resistor, which smooths protruding surface irregularities.

| Detail | Observation |
| --- | --- |
| Type designation | КП1-4 (KP1-4); "КП" stands for "Конденсатор Переменный" (variable capacitor) |
| Glass | Wall thickness 2 mm |
| Copper | Very soft, presumably pure — as indicated by the large visible crystals — and fully annealed, then electropolished |
| Vanes | Plate thickness 0.5 mm; nominal free gap between plates 1.5 mm, reduced by mechanical tolerances |
| Solder joints | Likely silver brazing alloy, performed in vacuum without flux |
| Glass-to-copper transition | Glass and copper fused together, perfectly airtight. A thin copper lip is embedded in glass on both sides. |
| End caps | Silver-plated copper |
| Shaft | Stainless steel with M6 thread; 3.2 mm hole for pinning the drive coupling |
| Bearing | Axial thrust bearing; thread and bearing were completely packed with grease |
| Estimated origin | Presumably manufactured between approximately 1970 and 1990 in the USSR during the Cold War |

Table 1: Notable construction details of the KP1-4 vacuum capacitor.

I am impressed by this capacitor — a masterpiece of engineering.

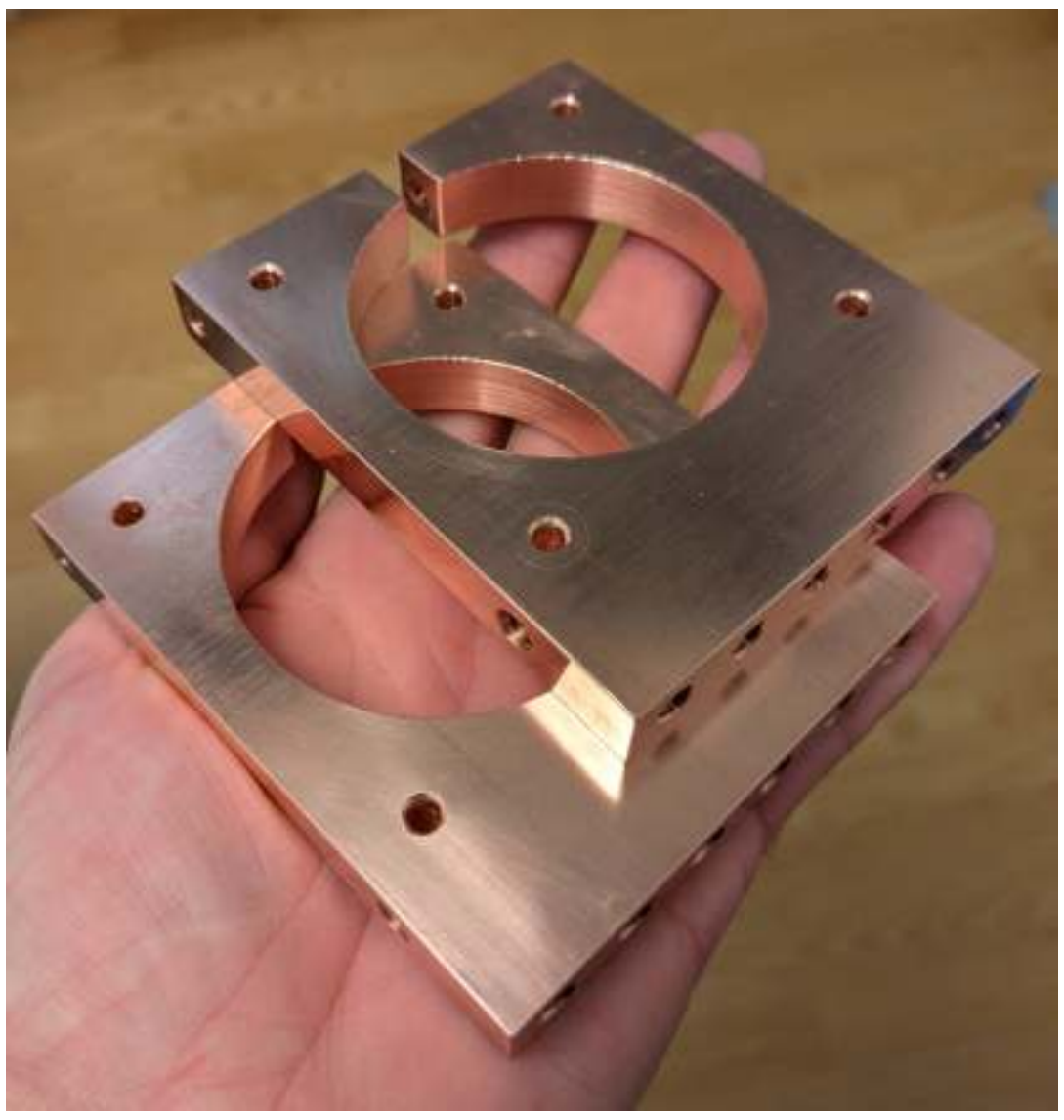
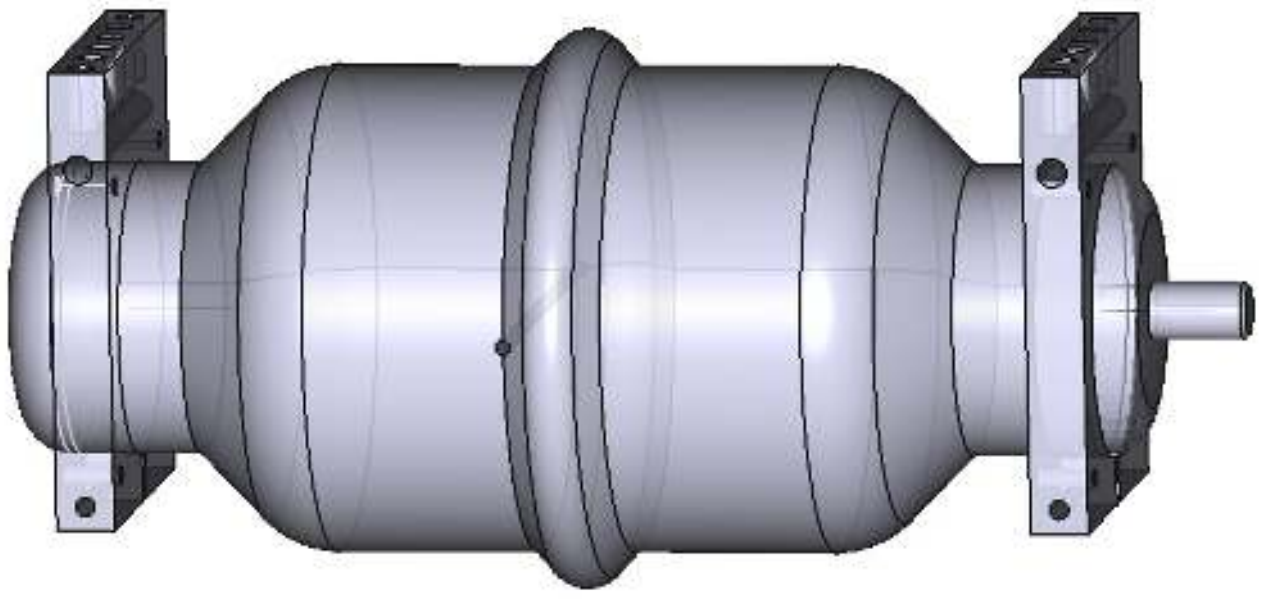

Figure 49: Clamps manufactured from 12 mm OF copper as a transition from the capacitor ends to the copper tubes.

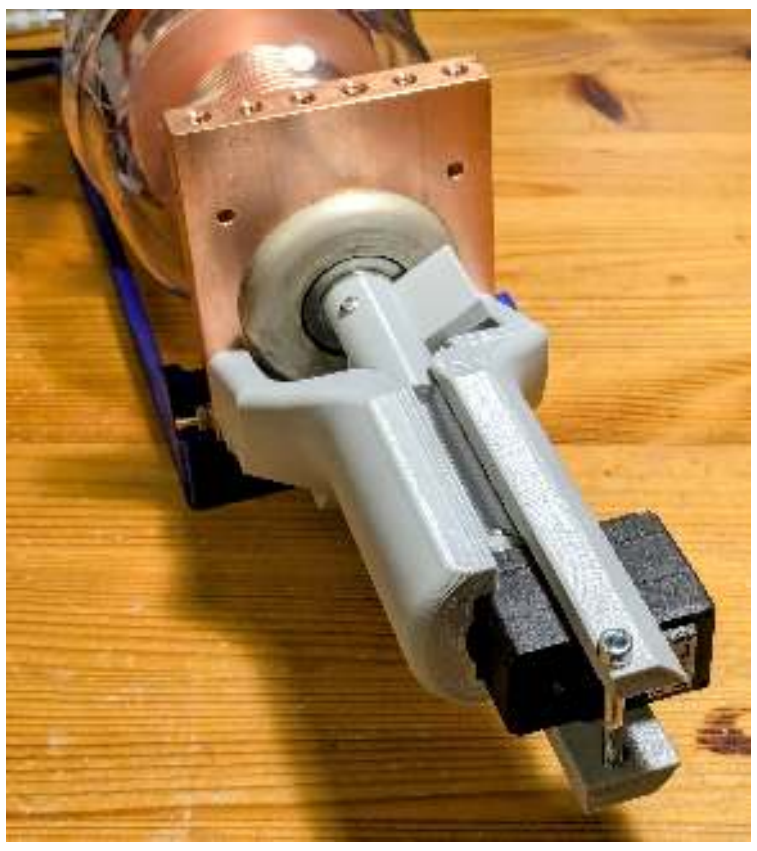
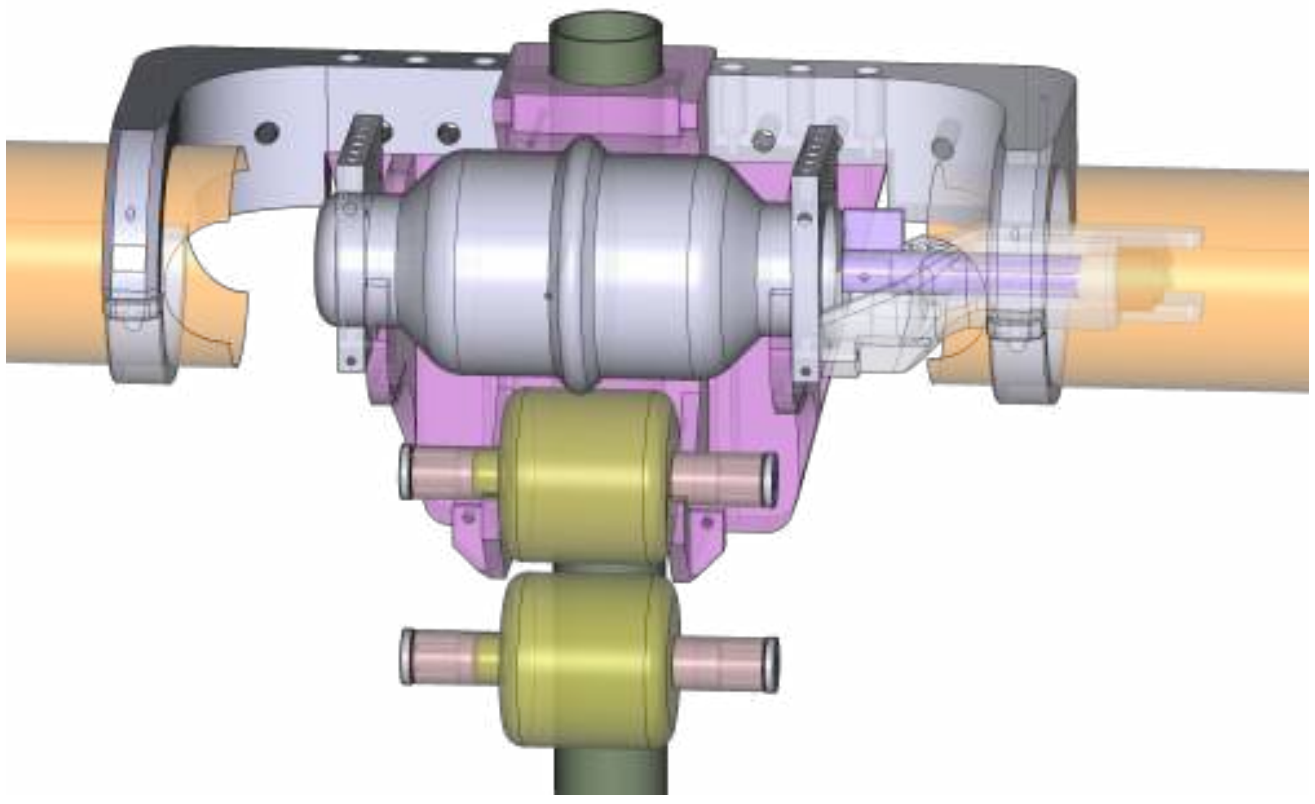

Figure 50: Capacitor assembly. The servo_f is located inside the copper tube and is therefore in a field-free region.

Servo end stop. Clockwise rotation: the capacitor shaft reaches its stop. The servo torque is artificially limited. Counterclockwise rotation: at the end of the range, the shaft is unscrewed from the capacitor. A pin on the shaft then reaches a mechanical stop so that the shaft cannot be unscrewed too far from the capacitor.

The servo has a built-in 12-bit magnetic encoder for angle measurement over 360°. The absolute position is tracked in a file on the PC so that the number of full rotations is retained even after a power interruption. Therefore, no homing is required when restarting the control software.

### 2.2.2 Fixed Capacitors Jennings 500 pF

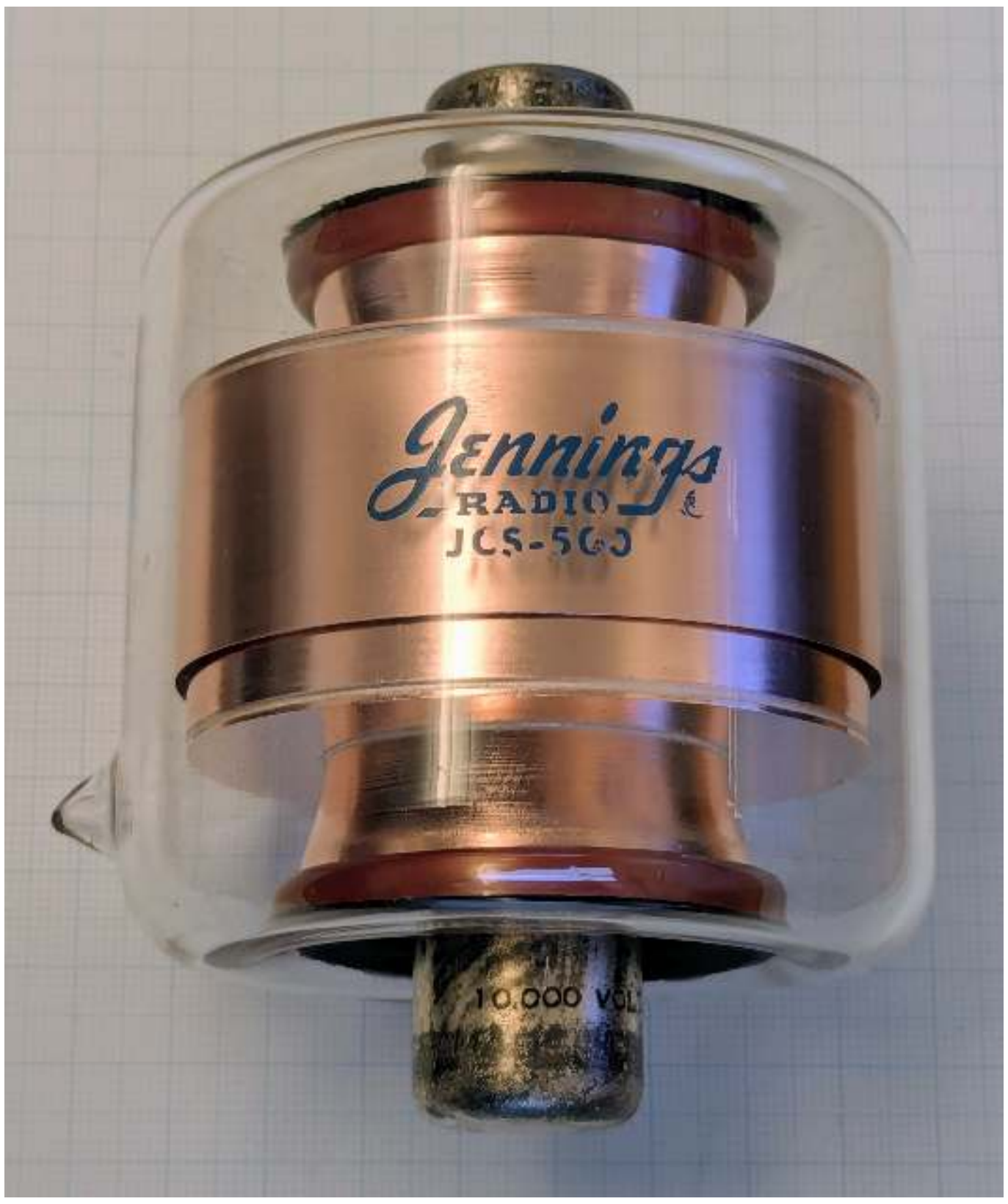

Figure 51: Jennings JCS-500-10S vacuum capacitor. Two of these are used.

Note the different colors of the metal-to-glass fusion. In the variable capacitor, the transition was yellow-colored — copper fused directly onto glass. The yellow coloring is presumably caused by copper oxide together with iron oxide in the glass.

In the Jennings capacitor, however, the fusion zone is dark red. Here, a Fe–Ni–Co alloy (Kovar) was likely used. This alloy is thermally better matched to glass. It colors dark red and provides a mechanically more robust bond than the direct fusion of copper with glass.

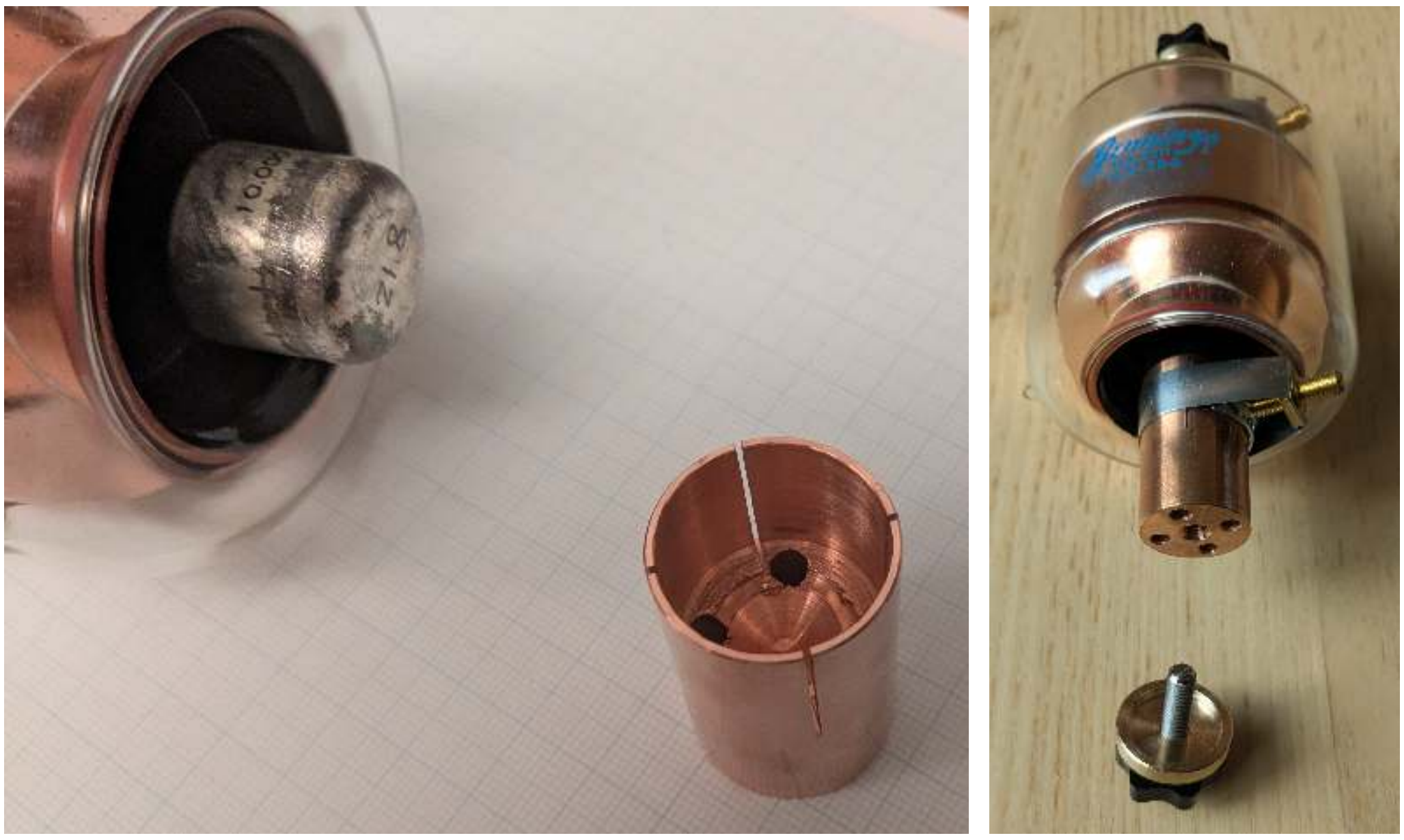

Figure 52: Turned sleeves made of OF copper serving as transition from vacuum capacitor to copper tubes. The sleeves are clamped onto the capacitor caps with hose clamps.

### 2.2.3 Fixed Capacitor 4700 pF PTFE, Unsuccessful

I conducted a test with this 4700 pF capacitor on the 160 m band. The PTFE dielectric is known for its low losses.

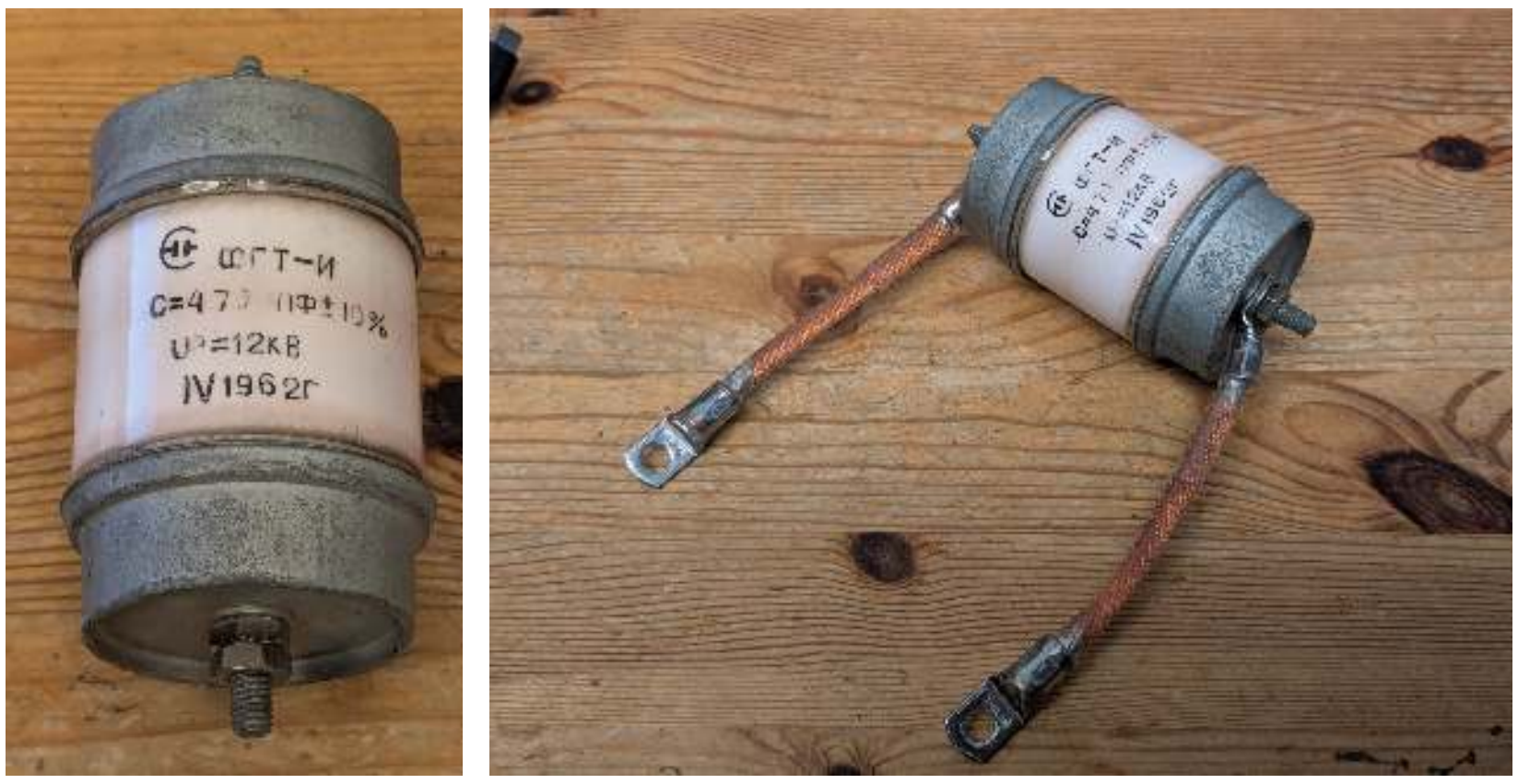

Figure 53: Markings: type designation ФГТ-И (FGT-I series, PTFE dielectric), 4700 пФ ±10% (4700 pF ±10%), Up = 12 кВ (peak voltage 12 kV), IV 1962 г (4th quarter 1962). Translation without guarantee.

The result was sobering — this capacitor is unsuitable for use in a magnetic loop antenna. The losses were so large that even the maximum coupling of the gamma match was insufficient to match the impedance to 50 ohms. Still, it is a beautiful sight — such lovingly crafted vintage components.

### 2.2.4 Fixed Capacitor 4000 pF PCB, Unsuccessful

A capacitor was fabricated from PCB material for the 160 m band.

30 circuit boards, each 1.2 mm thick and clad on both sides with 35 µm copper, active area 135 mm × 120 mm. Nominal gap between the plates: 1 mm. Side plates with 70 µm copper. Soldered with Stannol Kristall 611, Sn96.5Ag3Cu0.5.

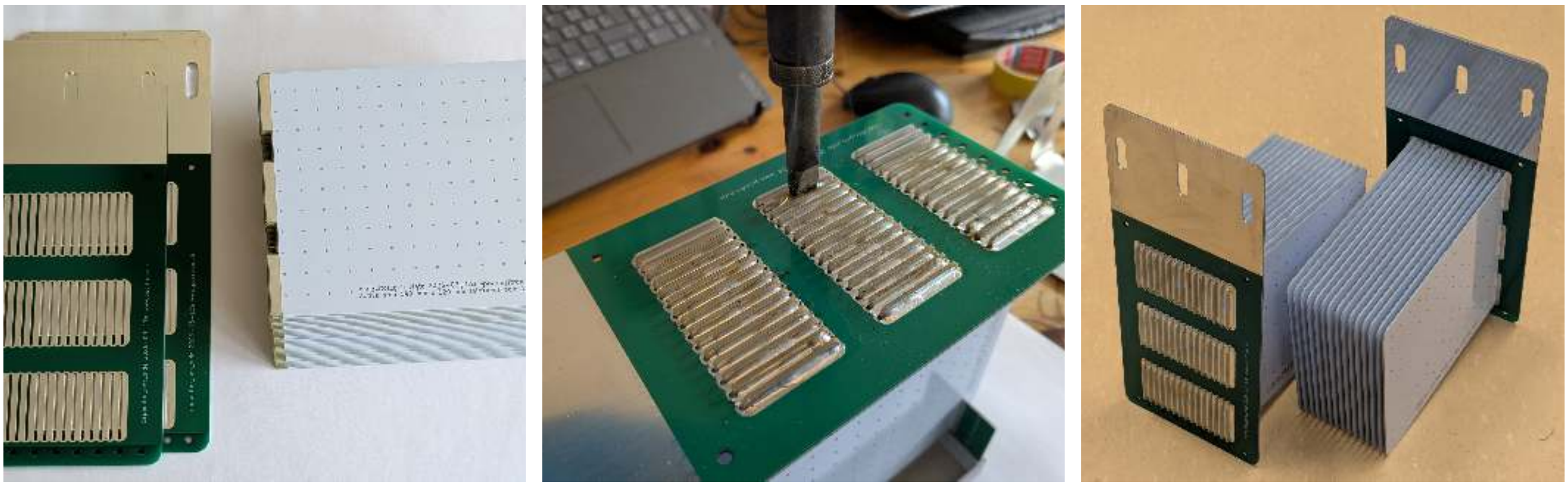

Figure 54: Left: circuit boards as delivered by the manufacturer. Center: soldering the plates with a powerful soldering iron — the solder flows very nicely. Right: finished plate stack. The capacitance can be adjusted in steps by mounting the plate stacks vertically with a slight offset.

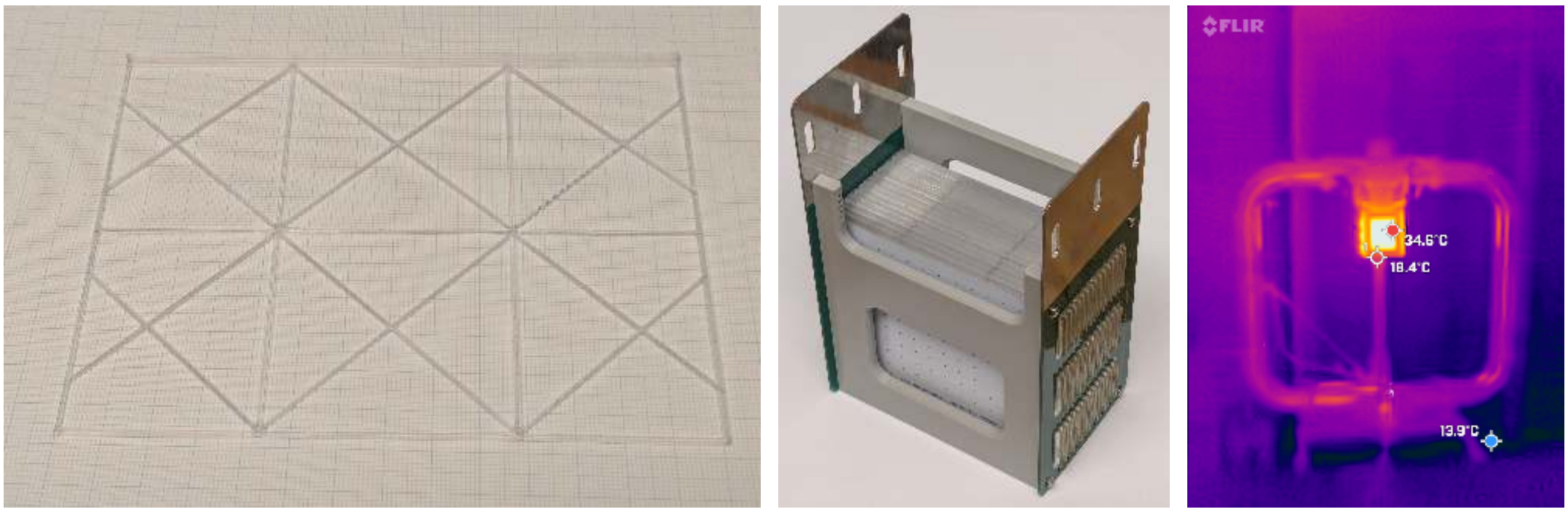

Figure 55: Left: spacers 3D-printed from PETG; the spacers prevent the plates from touching. Center: fully assembled capacitor. Right: significant heating after brief transmission.

Initial tests showed large losses — the capacitor heated up very quickly. The contact points and solder joints remained relatively cool. The losses evidently originate at the plates, presumably in the spacers.

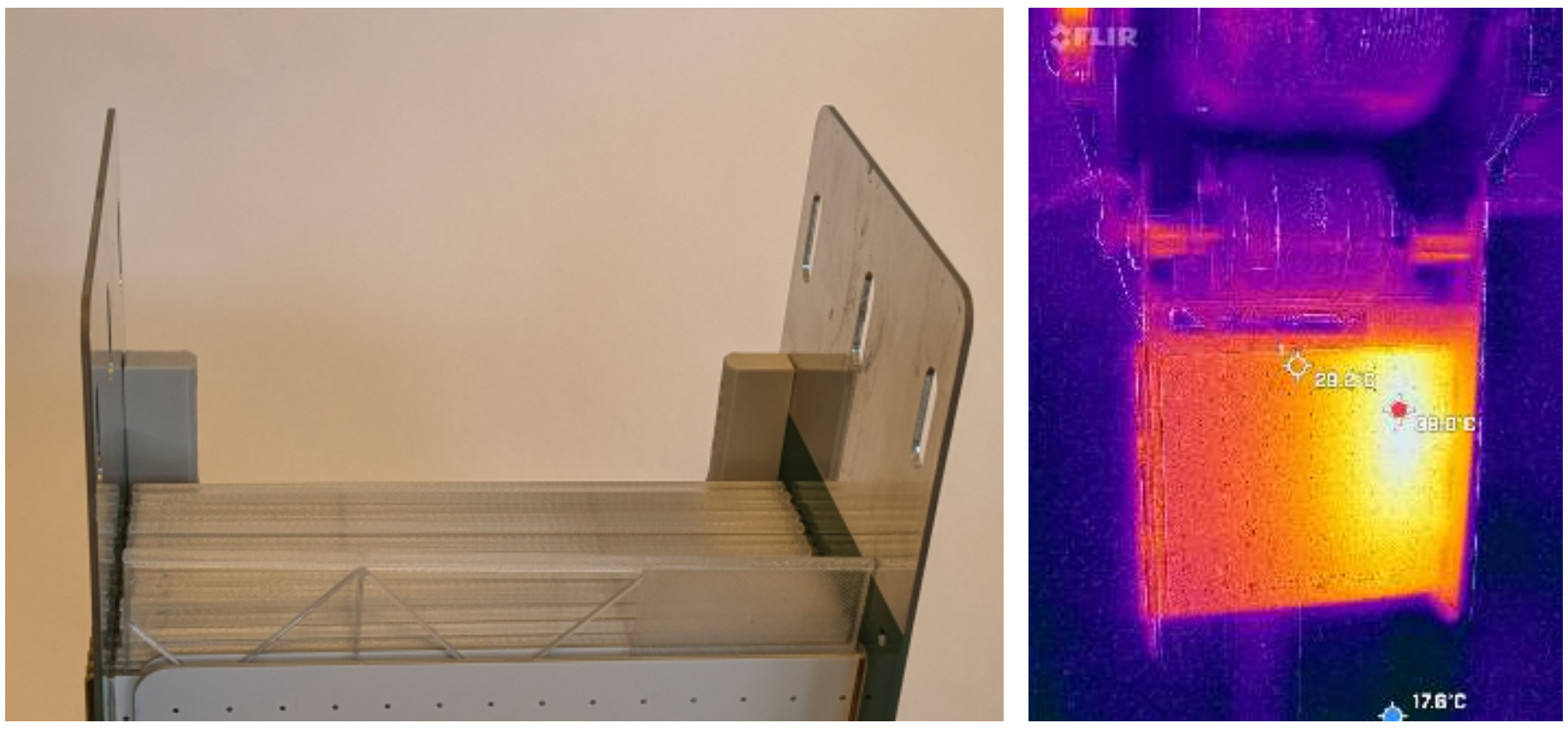

Figure 56: To corroborate the suspicion of losses in the spacers, two special spacers were fabricated with solid plastic filling the right quarter of the area. Left: the first spacer is only partially inserted into the capacitor; the solid-filled area is visible on the right. Right: after brief transmission, strong heating appeared precisely at the locations with solid plastic. The spacer losses are dominant.

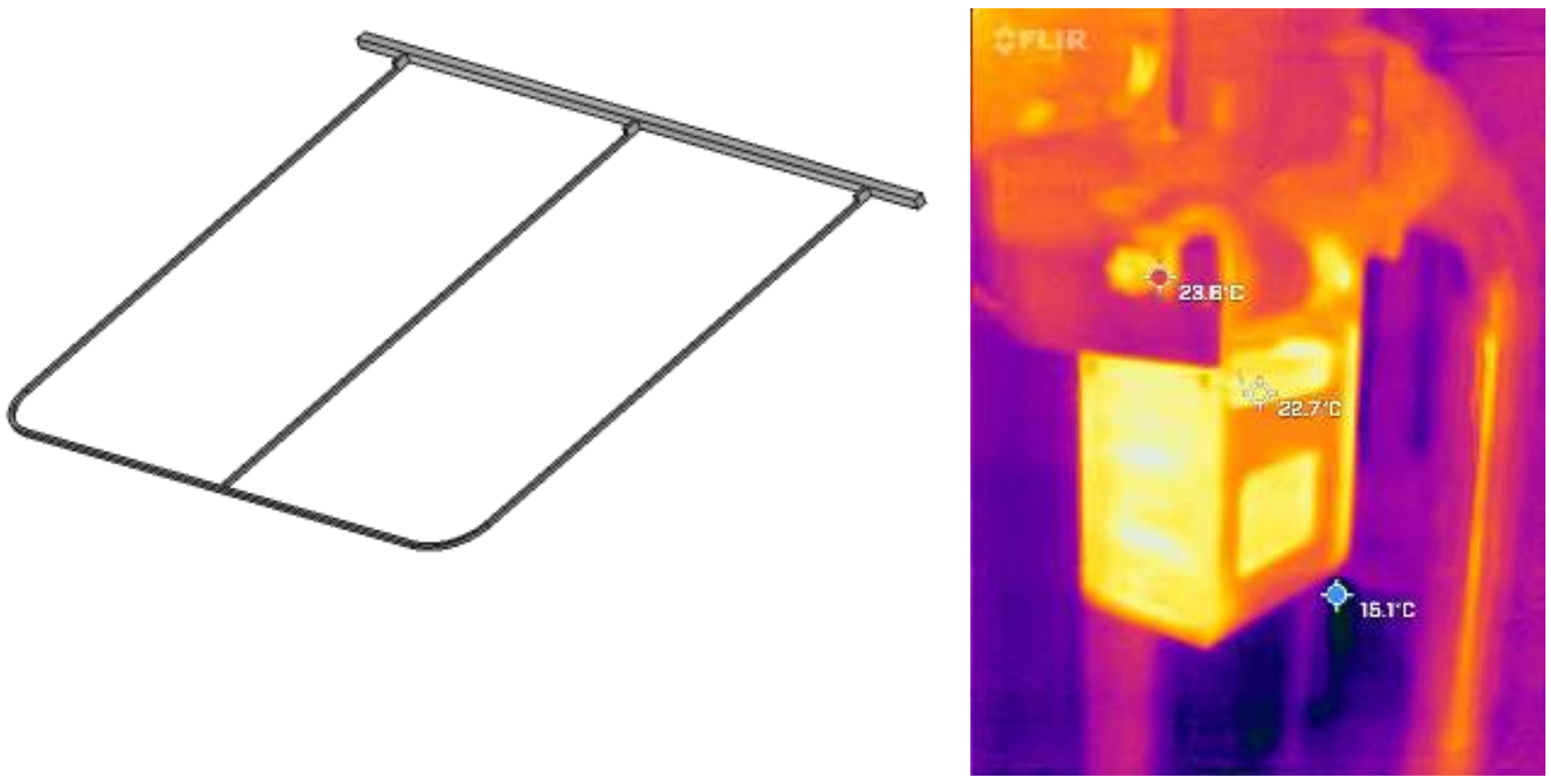

Figure 57: Left: new spacers with less material, made from COC (Cyclo Olefin Copolymer). COC should exhibit significantly lower dielectric losses than PETG. Right: the capacitor now heats up by only a few °C after 20 minutes of continuous transmission.

Above 50 W transmit power, the SWR deteriorates. There is apparently a non-linearity. The voltage across the capacitor is less than 900 $V_{rms}$, which should still be acceptable at a 1 mm plate gap. Heating and detuning due to losses can be ruled out because the SWR was already poor immediately at the start of transmission. Detuning caused by electrostatic forces changing the plate spacing? Could it be corona discharge? Nothing was visible or audible from a distance, and no ozone smell was detected.

The capacitor was tested for DC voltage withstand: 2500V DC showed no measurable leakage, at 5000V DC a breakdown occurred.

The experiment with the PCB capacitor was discontinued at this point due to heating and the resulting detuning, and due to the deteriorating SWR above 50 W transmit power.

### 2.2.5 Fixed Capacitor Vacuum 4200 pF

A 4200 pF capacitor is formed by combining multiple capacitors.

3 units Comet, CFMN-2800BAC/8-DE-G, 2800 pF, 8/4.8 kV, 75 mm diameter, 52 mm length, weighing 819 g each, M6 threads on both sides. For indoor use, these capacitors are massively oversized, but the optimal values for my application were not available on the second-hand market.

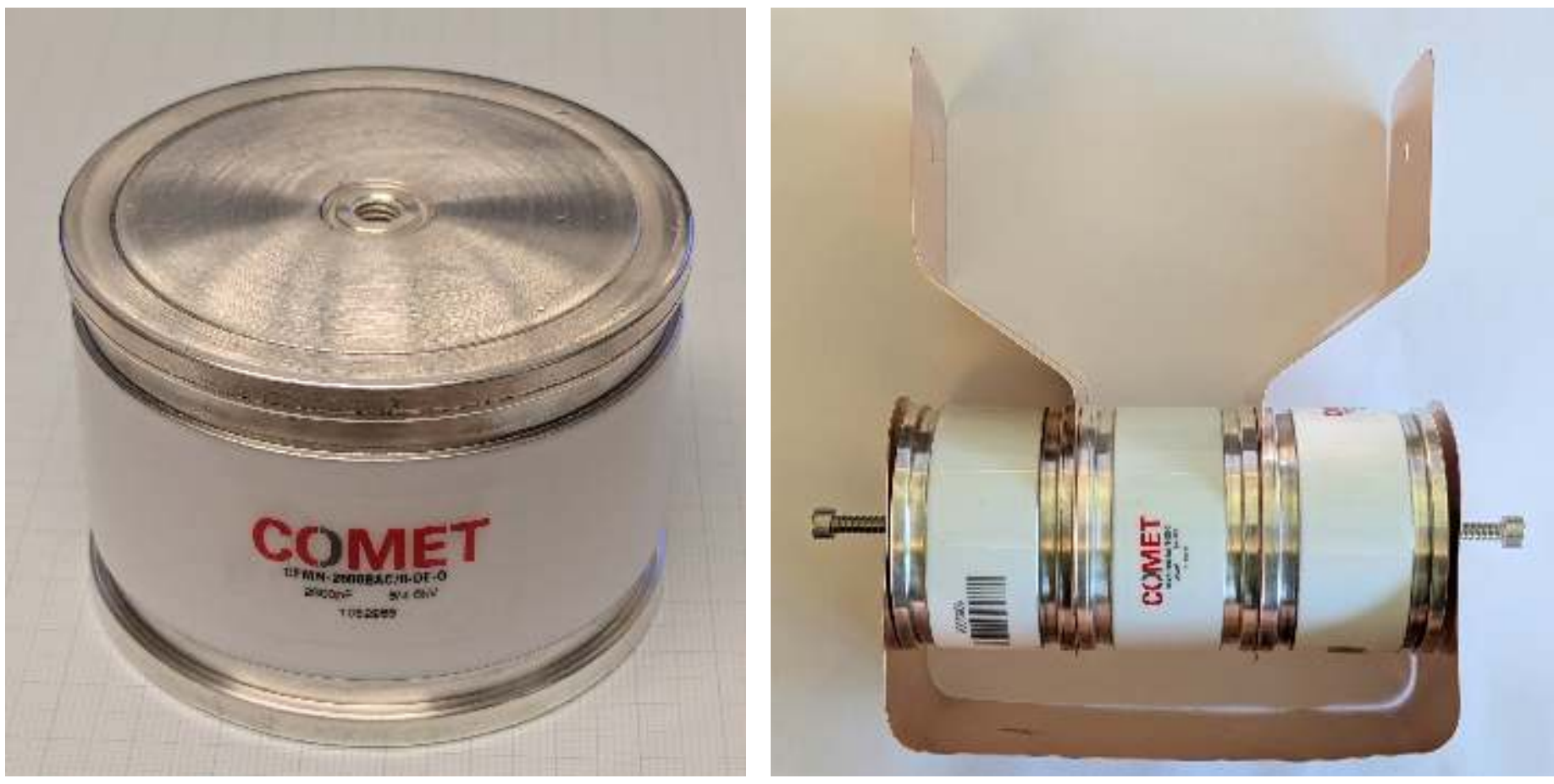

Figure 58: Left: a single capacitor with 2800 pF. Right: capacitors loosely connected with metal sheets to illustrate the wiring.

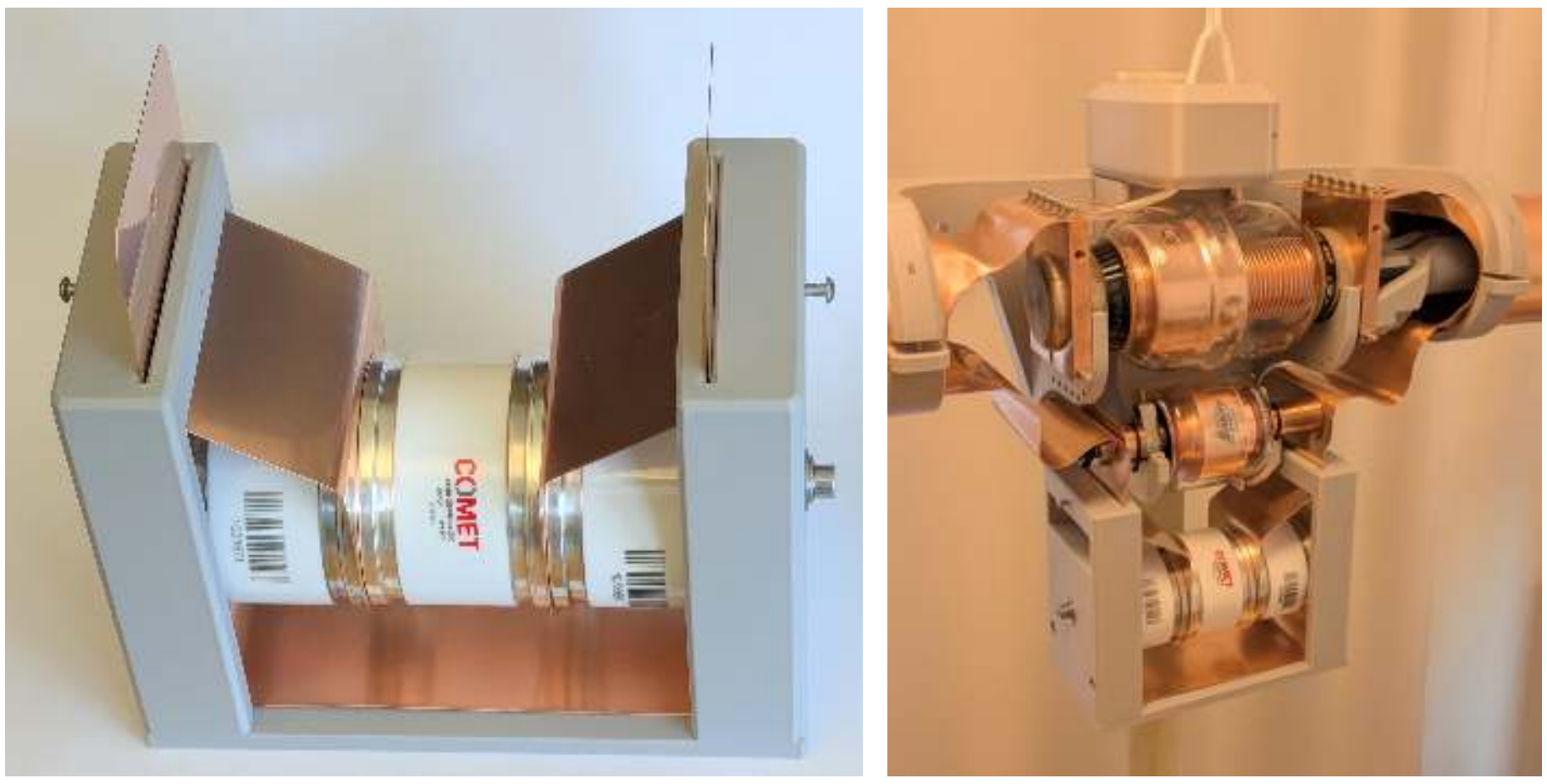

Figure 59: Left: assembled 4200 pF capacitor, weighing 2.3 kg. The capacitors are firmly bolted together; the M6 threads provide solid mechanical strength. Right: capacitor mounted on the antenna.

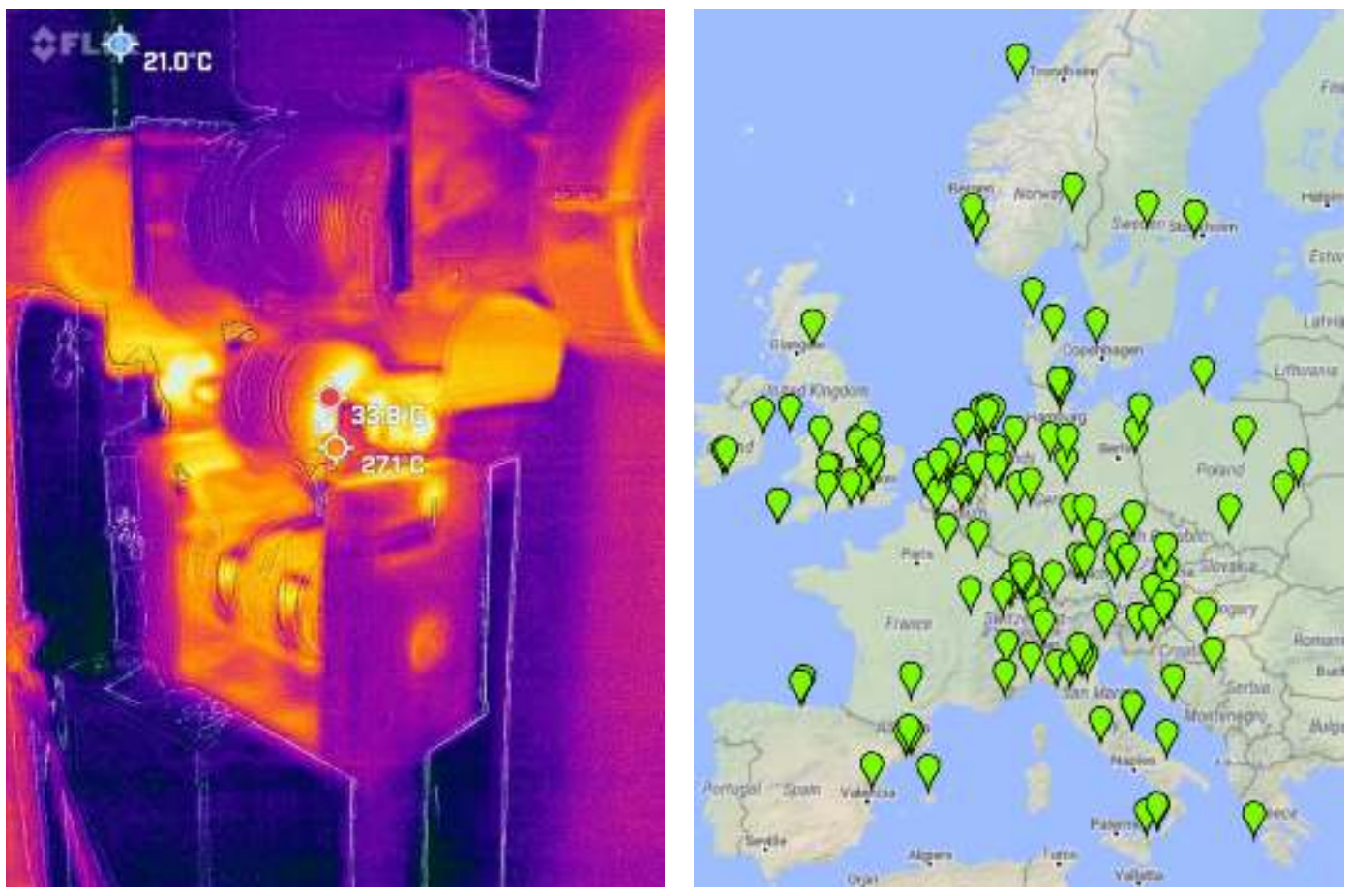

Figure 60: Left: after 10 minutes of transmitting at 100 W, the Comet capacitors heated up by 3.7 °C, the Jennings capacitor in the center by 0.3 °C, and the metal sheet connecting to the Jennings capacitors by 8 °C. Right: during a test on April 10, 2026, FT8 signals were decoded within a radius of approximately 1500 km. This is quite remarkable considering that the radiated power is at most 70 mW. Source: pskreporter.info.

The weak point is the junction between the main loop sheet and the Comet capacitor sheet, which is connected by only a single M6 screw. This contact point produces the most heat and indirectly warms the Jennings capacitor. Most of the current flows through the Comet capacitors, yet they heat up only slightly. The largest share of the losses occurs in the building, not in the antenna. A better junction would not make a significant difference. The resonant frequency of the antenna shifts downward by 225 Hz due to the heating — negligible for operation.

## 2.3 Gamma Match

This section shows a gamma match implementation corresponding to the concept in Figure 14.

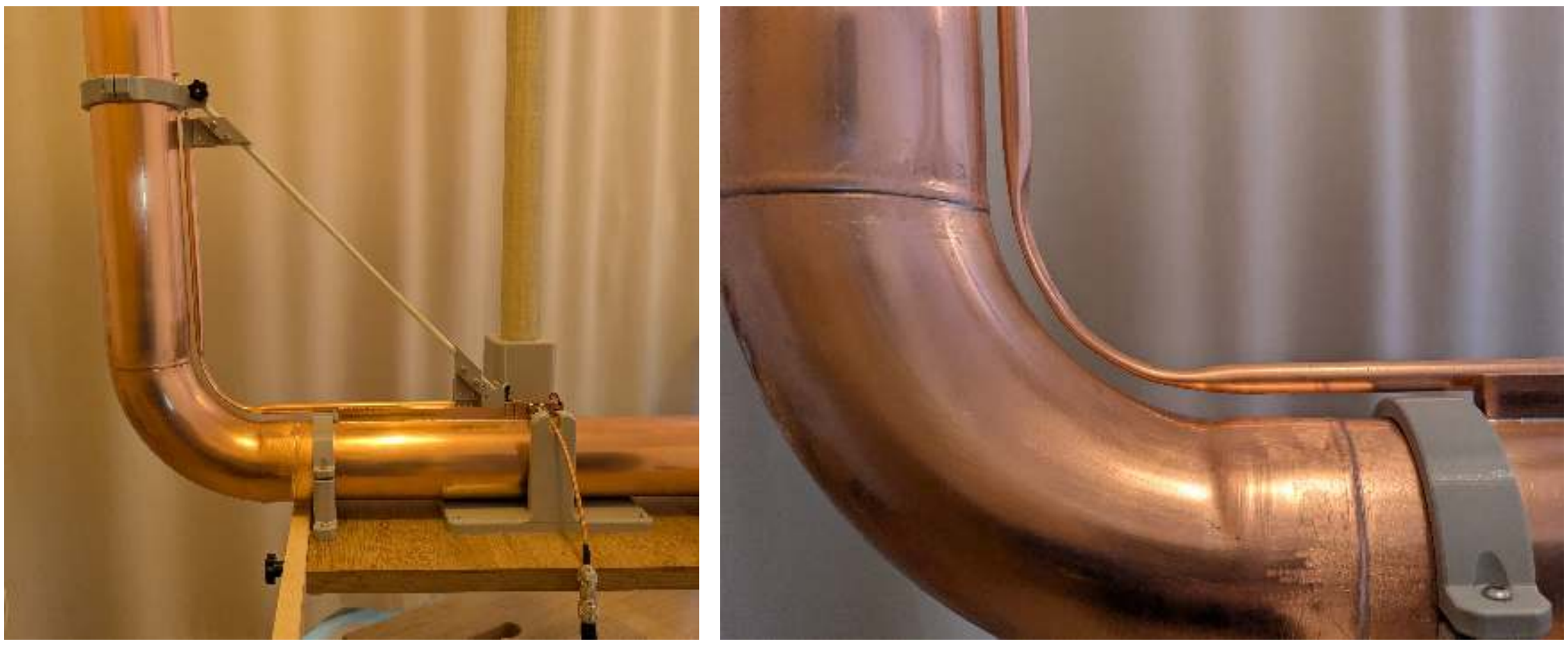

Figure 61: Gamma match constructed from 12 mm diameter copper tubing with a total length of 0.7 m.

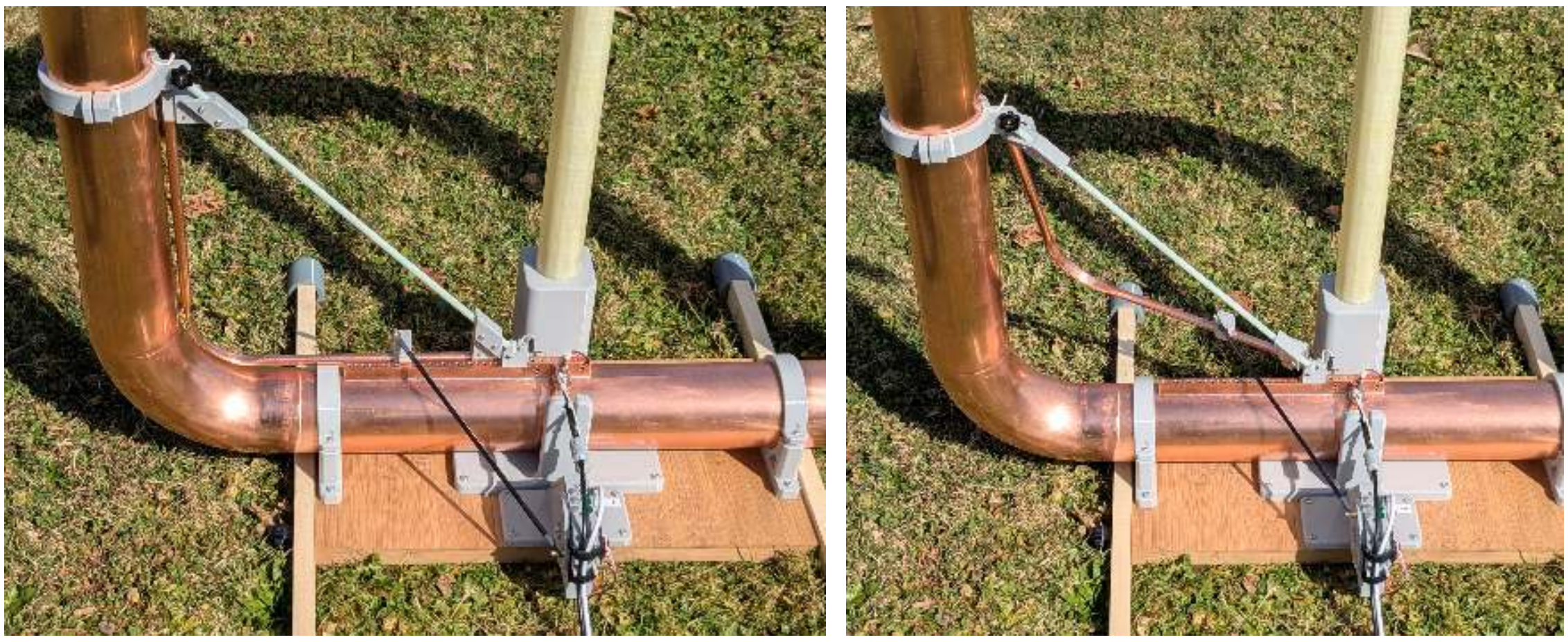

Figure 62: Left: small coupling; right: large coupling. The servo_z automatically adjusts the coupling and is connected via a black fiberglass rod (not carbon fiber, which would be electrically conductive).

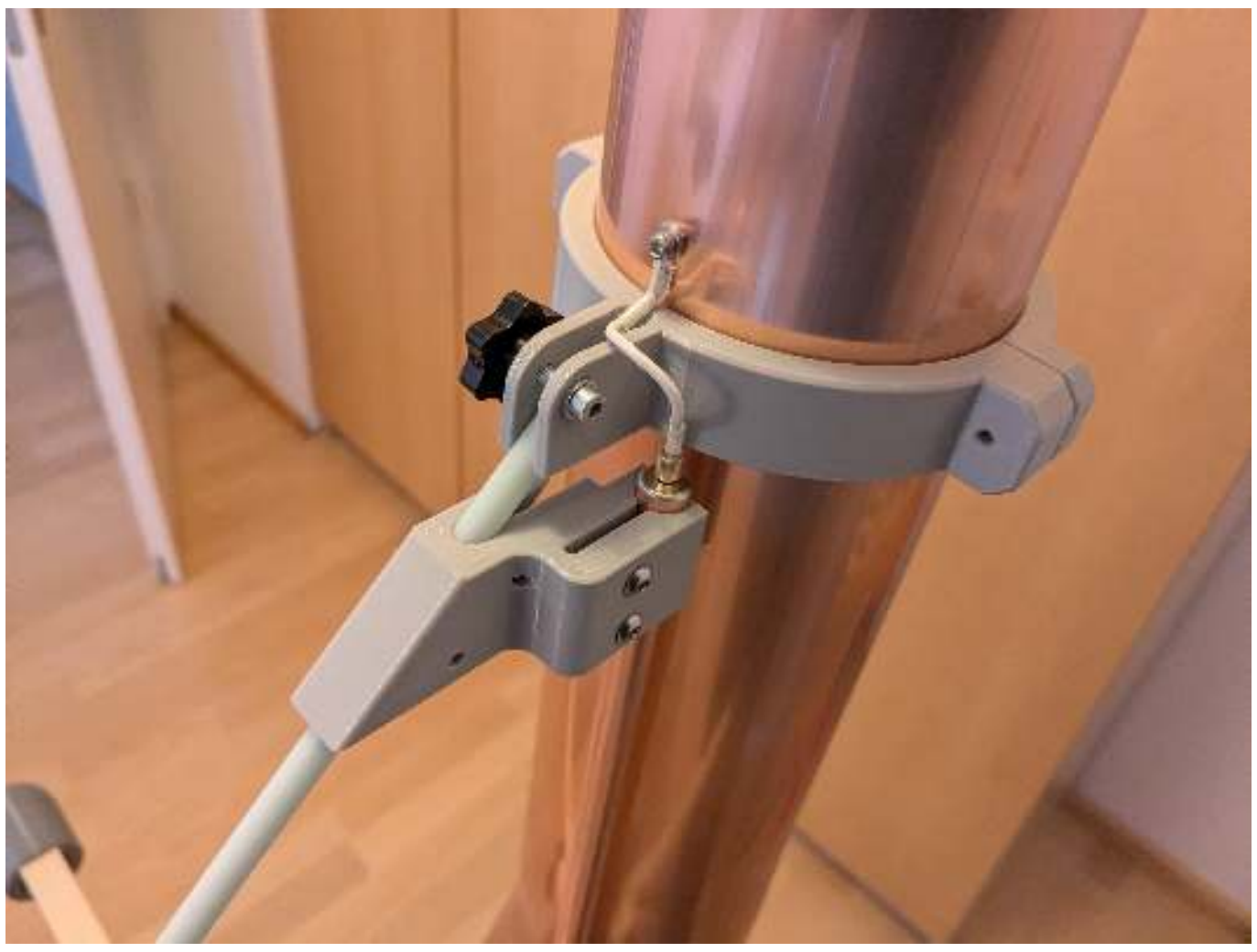

Figure 63: Pivot joint in the center of the loop with galvanic connection to the main loop.

The gamma match is quite large. When set to maximum coupling, it encloses 1/4 of the magnetic flux lines of the main loop; such a high coupling is never needed. However, an oversized gamma match does not impose any disadvantages.

In some measurements, far from damping materials, the coupling could not be reduced sufficiently, so I temporarily installed a smaller coupling element.

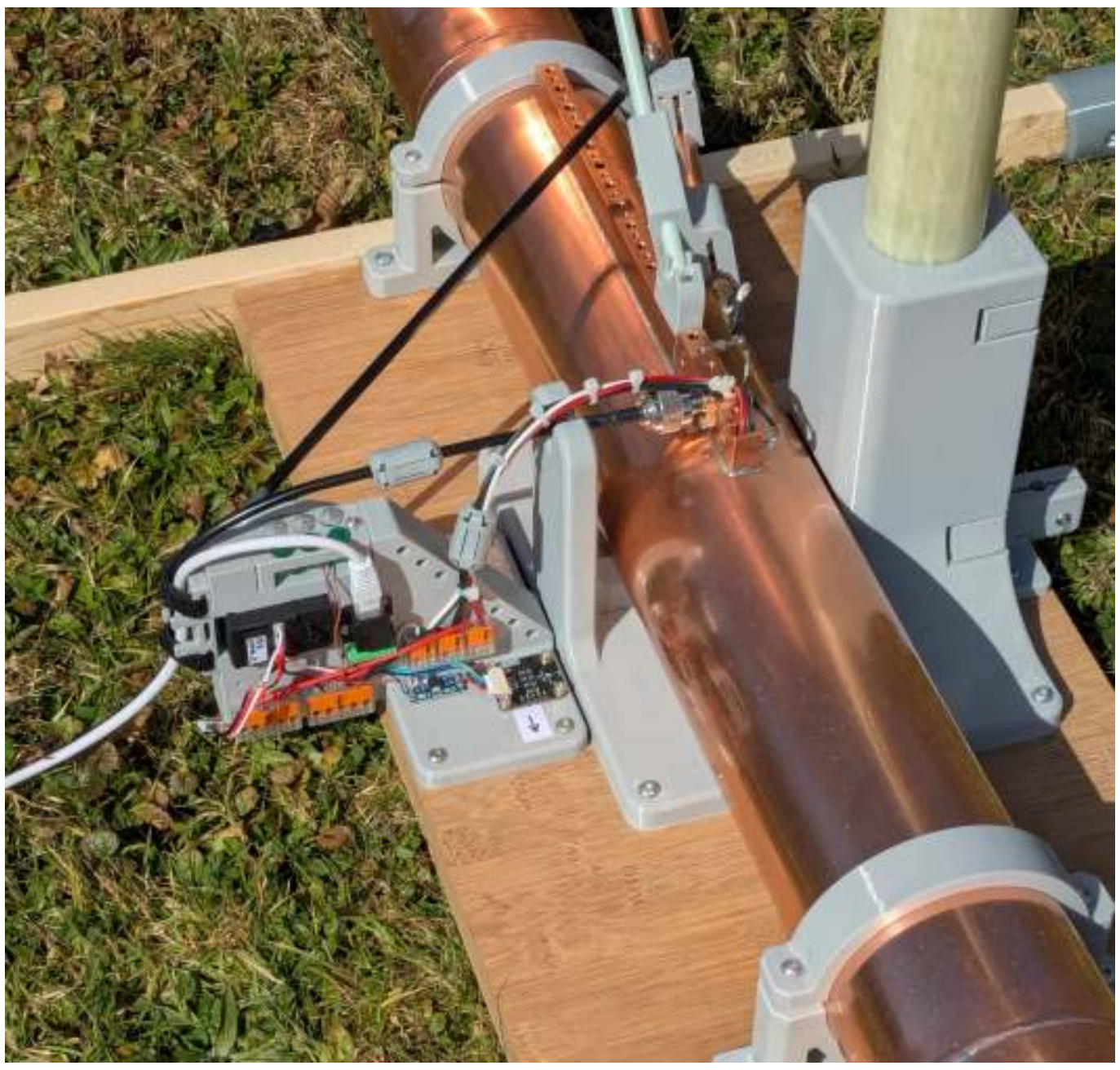

Figure 64: Cable routing and placement of the electronic components are exactly in the symmetry plane of the antenna. Left, black: servo_z for coupling adjustment; below: magnetometer for azimuth control. Not shown in the image: servo_h for azimuth control.

## 2.4 Azimuth Control

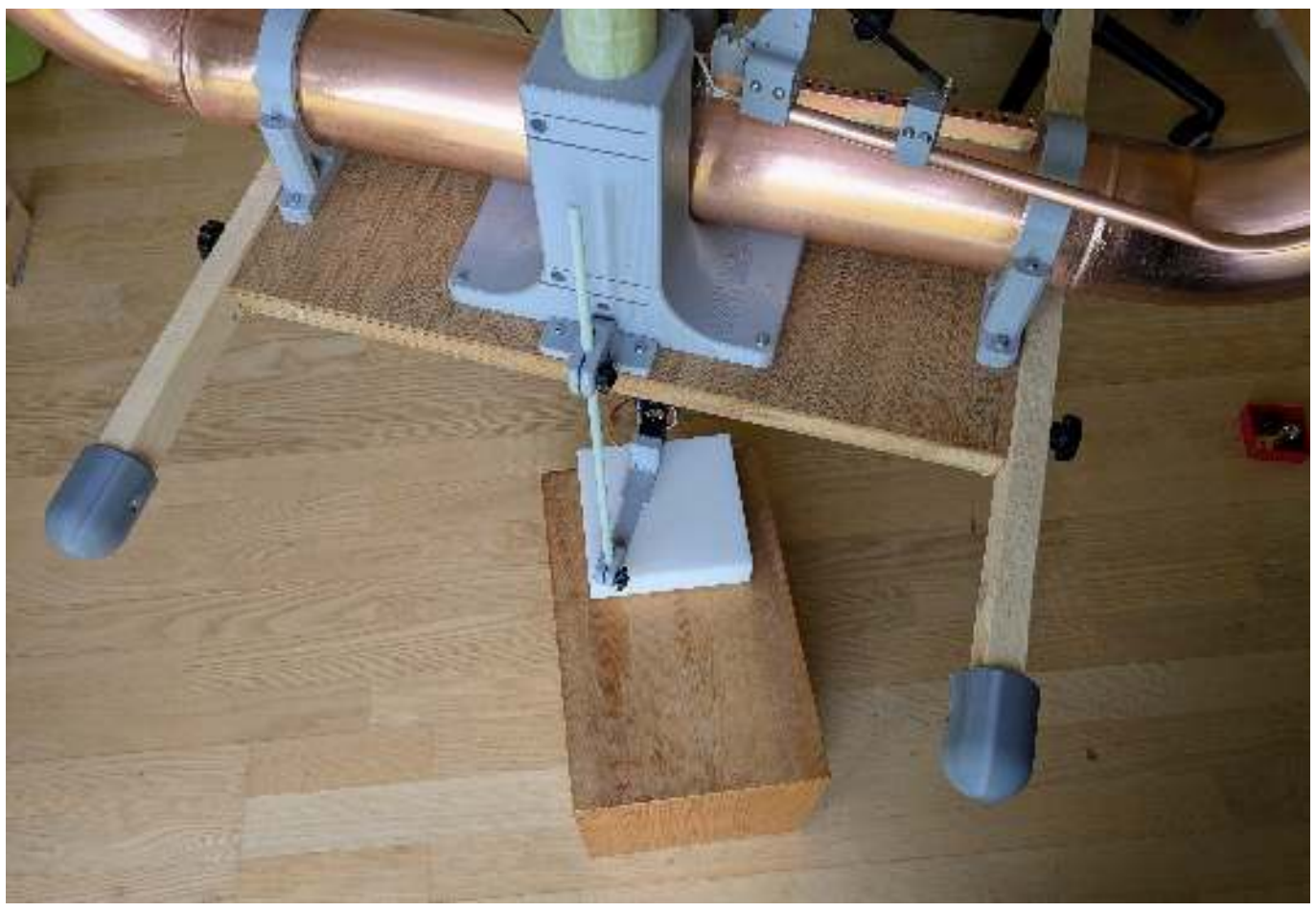

Figure 65: Below the antenna, the servo_h controls the azimuth. The servo axis is centered directly below the rope from which the antenna is suspended from the ceiling. The servo_h is connected to the antenna via a lever and a fiberglass rod. This allows the antenna to be rotated over a range of 220°. Because the antenna radiation pattern is symmetric, a sweep of 180° is sufficient.

## 2.5 Schematic

### 2.5.1 Overview

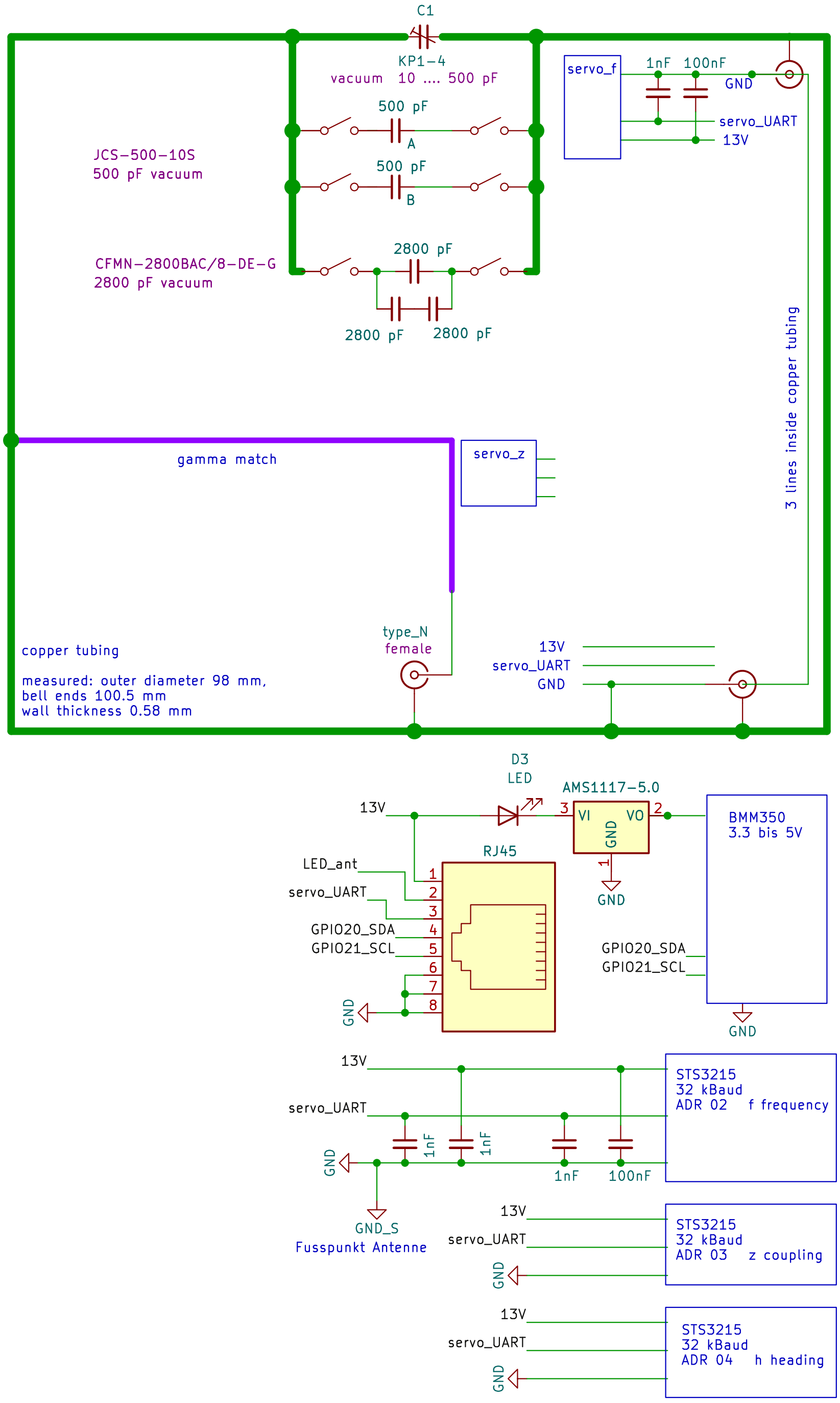

Figure 66: Schematic of the antenna system including tuning servos and magnetometer.

The upper part of the schematic mirrors the physical layout of the magnetic loop; the components are arranged in positions corresponding to their actual placement on the antenna.

For the data line, a readily available unshielded AWG24 Ethernet patch cable is used. In an additional switching box (not shown in the schematic), relays connect the antenna either to the VNA or to the transmitter output. The transmitter is also inhibited during the tuning process. LEDs mounted on the antenna (not shown in the schematic) indicate the operating state: red signals that transmission is possible, green signals that the transmitter is inhibited and the antenna can be safely approached. These LEDs are always illuminated. The servo and magnetometer supply voltage (“13 V”) is only applied during tuning, because the servo electronics could otherwise interfere with the received signal.

### 2.5.2 Cable Routing to Frequency Tuning Servo

A common challenge in magnetic loop antenna design is the electrical insulation between the motor and the capacitor.

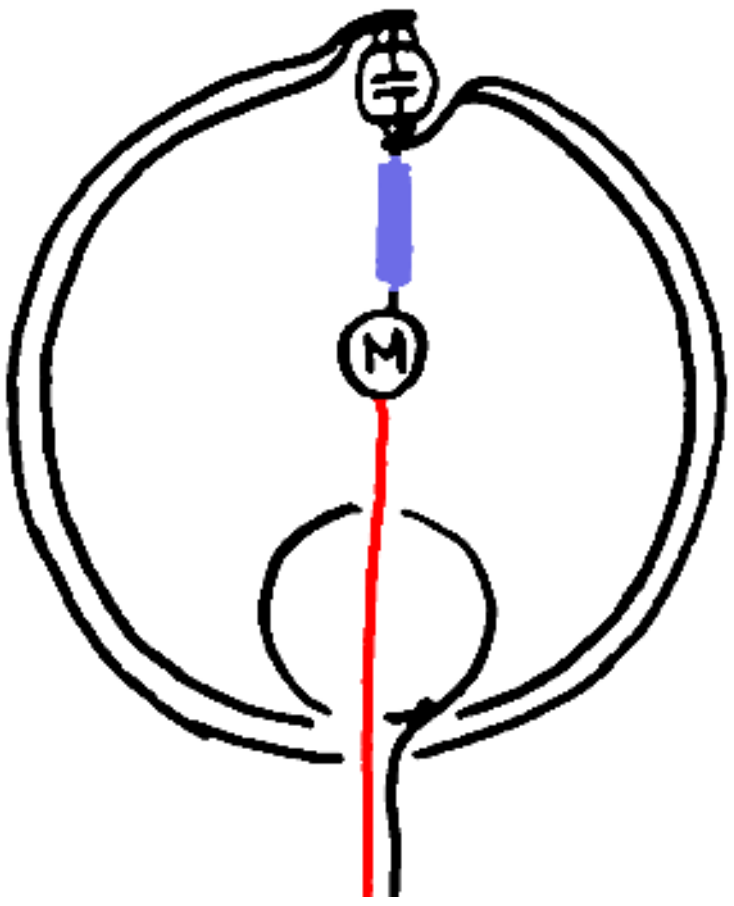

Figure 67: Example of a common design. The tuning capacitor is mounted vertically at the top of the loop. The motor shaft is coupled to the capacitor via an insulating shaft (blue). The motor cable (red) runs downward along the antenna's symmetry axis. During transmission, the full loop voltage $U_{\text{loop}}$ appears across the capacitor. Because the motor and its cable run along the symmetry axis of the loop, a voltage of $U_{\text{loop}}/2$ is present between the motor and the nearer capacitor terminal. The insulating shaft must withstand this voltage. Furthermore, the capacitive coupling between the motor body and the capacitor introduces an asymmetry in the antenna.

In the magnetic loop antenna presented in this paper, the control cable is routed through the right vertical section of the loop conductor (see center of Figure 68). Because the tube and cable follow the same path, no voltage is induced between them. No insulation between the motor and capacitor is required.

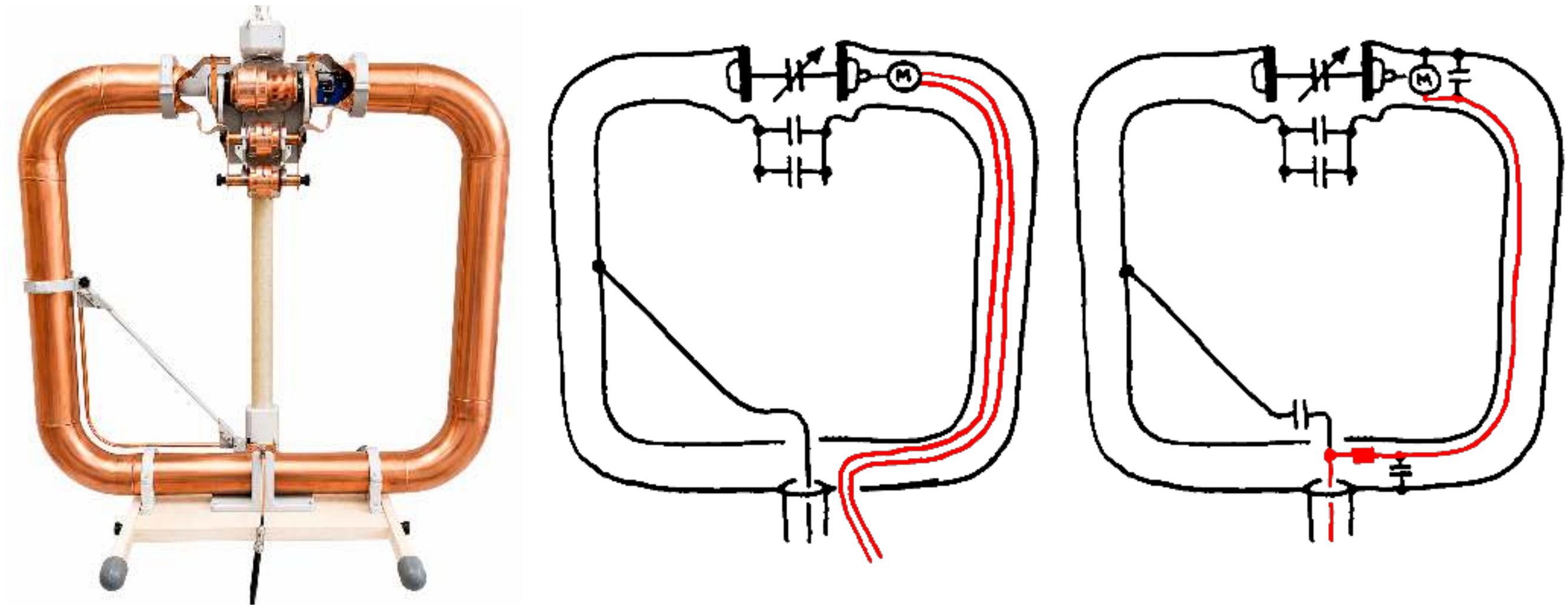

Figure 68: Left: the antenna presented in this paper. Center: the control cables (red) run through the right vertical section of the loop conductor; the drive motor can be mounted directly on the capacitor without any insulation. Right: an optional arrangement where the tuning motor is powered via the feed coaxial cable; the RF component is routed to the gamma match while the DC component is carried by the red cable to the motor, with a series inductor and blocking capacitors keeping the RF away from the motor. The loop conductor itself serves as the return path for the motor current.

Nevertheless, a fairly long shaft is used between the servo and the capacitor so that the servo is located inside the tube, i.e., in a field-free region, and therefore does not cause additional losses (see also Figure 50).

## 2.6 Key Parameters

| **Parameter** | **Value** |
|---|---|
| name | cu_1m_100mm_a |
| Antenna type | small magnetic loop |
| Shape | rectangular |
| Width (measured between tube centers) | 0.95 m |
| Height (measured between tube centers) | 0.85 m |
| Equivalent diameter | 1.006 m |
| Tubing diameter | 100 mm |
| Area | 0.795 $m^2$ |
| $L$ estimated from geometry and validated by measurements | 1.51 µH |
| Tubing wall thickness (thicker than skin depth) | 0.6 mm |
| Variable capacitor | KP1-4, 10-500 pF, 7 kVrms, 50 Arms |
| Fixed capacitor A, 500 pF, switchable | Jennings JCS-500-10S 500 pF, 4.2 kVrms, 80 Arms |
| Fixed capacitor B, 500 pF, switchable | Jennings JCS-500-10S 500 pF, 4.2 kVrms, 80 Arms |
| Fixed capacitor C, 4200 pF, switchable | 3× Comet CFMN-2800BAC/8-DE-G, 2800 pF 3.4 kVrms, 100 Arms |
| Tuning drives | Frequency servo_f: Feetech STS3215, 12 V<br>Coupling servo_z: Feetech STS3215, 12 V<br>Azimuth servo_h: Feetech STS3215, 12 V |
| Azimuth control | Bosch BMM350, 3-axis magnetometer |

| Tuning ranges | | Variable Capacitor only | | Variable Capacitor plus 500 pF A | | Variable Cap. plus plus 500 pF A plus 500 pF B | | Variable Cap. plus plus 4200 pF | | Variable Cap. plus plus 500 pF A plus 4200 pF | |
|---|---|---|---|---|---|---|---|---|---|---|---|
| Capacity | | min | max | min | max | min | max | min | max | min | max |
| KP1-4_10_500 | F | 10E-12 | 500E-12 | 10E-12 | 500E-12 | 10E-12 | 500E-12 | 10E-12 | 500E-12 | 10E-12 | 500E-12 |
| JCS-500-10S A | F | | | 500E-12 | 500E-12 | 500E-12 | 500E-12 | | | 500E-12 | 500E-12 |
| JCS-500-10S B | F | | | | | 500E-12 | 500E-12 | | | | |
| Comet 4200 pF | F | | | | | | | 4.2E-09 | 4.2E-09 | 4.2E-09 | 4.2E-09 |
| Parasitic | F | 7E-12 | 7E-12 | 7E-12 | 7E-12 | 7E-12 | 7E-12 | 7E-12 | 7E-12 | 7E-12 | 7E-12 |
| Total Capacity | F | 17E-12 | 507E-12 | 517E-12 | 1.007E-09 | 1.017E-09 | 1.507E-09 | 4.217E-09 | 4.707E-09 | 4.717E-09 | 5.207E-09 |
| Frequency | Hz | 31.0E6 | 5.7E6 | 5.6E6 | 4.0E6 | 4.0E6 | 3.3E6 | 1.97E6 | 1.86E6 | 1.86E6 | 1.77E6 |

| Band | | Variable Capacitor only | Variable Capacitor plus 500 pF A | Variable Cap. plus plus 500 pF A plus 500 pF B | Variable Cap. plus 4200 pF / Variable Cap. plus plus 500 pF A plus 4200 pF |
|---|---|---|---|---|---|
| 160 m | Hz | | | | 1.84E6 |
| 80 m | Hz | | | 3.7E6 | |
| 60 m | Hz | | 5.4E6 | | |
| 40 m | Hz | 7.1E6 | | | |
| 30 m | Hz | 10.1E6 | | | |
| 20 m | Hz | 14.2E6 | | | |
| 17 m | Hz | 18.1E6 | | | |
| 15 m | Hz | 21.2E6 | | | |
| 12 m | Hz | 24.9E6 | | | |
| 10 m | Hz | 29.0E6 | | | |

Figure 69: Calculated frequency ranges for different capacitor configurations.

# 3 Measurements and Findings

## 3.1 Outdoor Setup

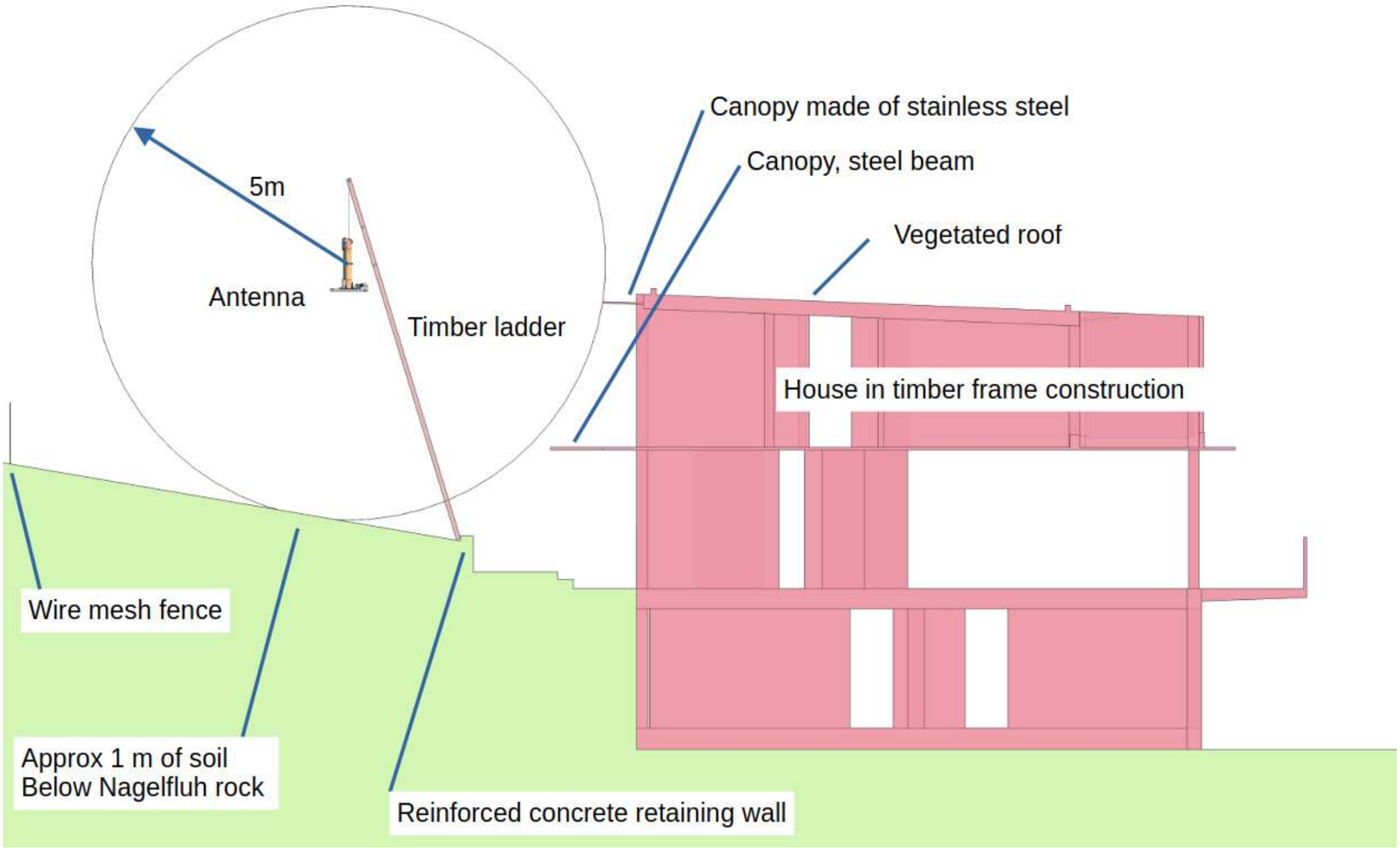

Figure 70: Cross-sectional house model used for the measurement setup.

A wooden ladder was secured to the house with ropes. At the top of the ladder, a pulley redirects the rope used to raise the antenna. At maximum elevation, the antenna is 5 m above ground and maintains at least 5 m distance from other metallic objects. The ladder itself only has small metal parts and screws.

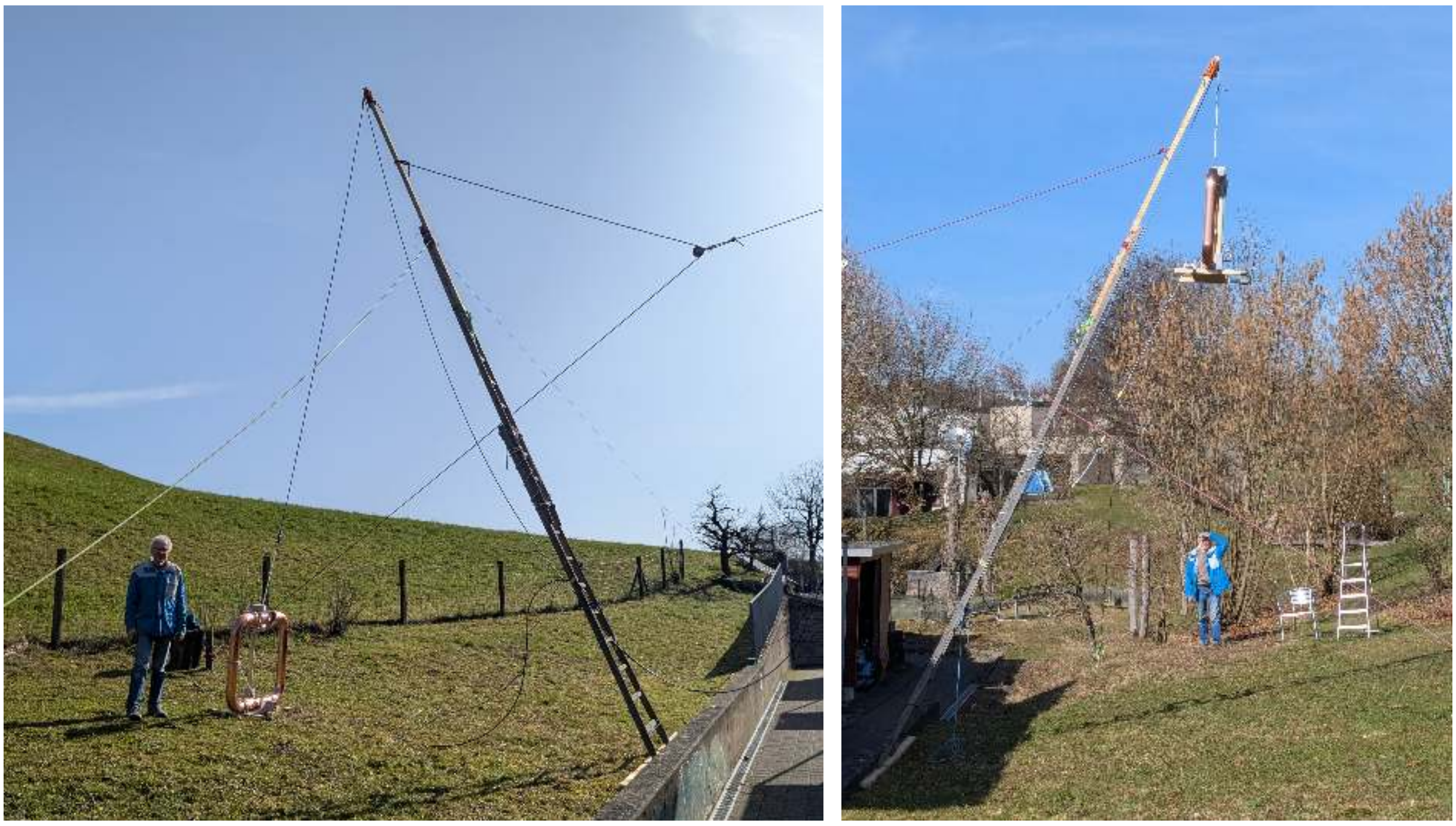

Figure 71: Antenna on the ground and at maximum height.

## 3.2 Indoor Setup

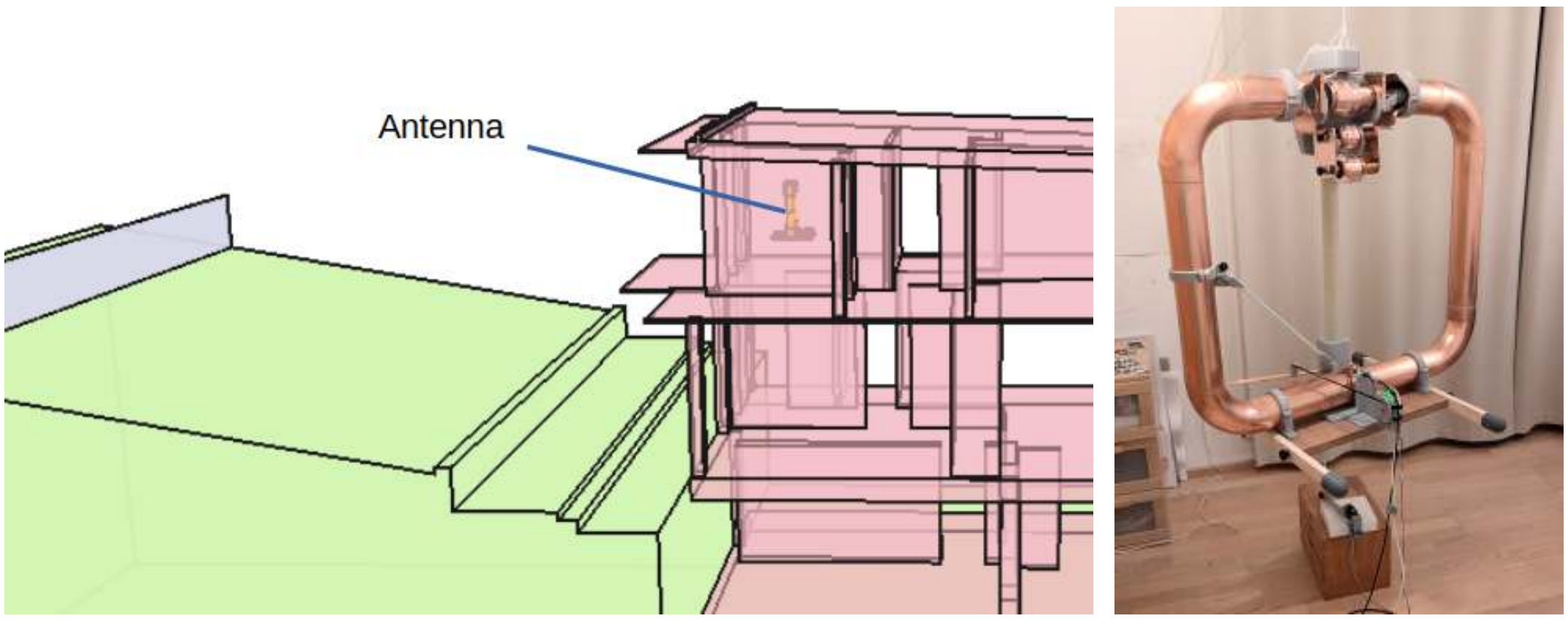

Figure 72: Indoor antenna setup: left, indoor measurement geometry; right, installation photo.

The antenna was suspended from the ceiling on the upper floor of the house, 0.8 m above the floor. The antenna was tested at various positions inside the house; at the position shown, the antenna damping is relatively low. Above the antenna, the roof has a green layer with moist soil. In the floor below the antenna, there is underfloor heating with plastic pipes coated with aluminum. In addition, a photovoltaic system is installed nearby on the roof, and the canopies are covered with stainless steel sheet metal. The window panes have aluminum frames. Although the house has timber-frame construction, these are still far from optimal indoor conditions.

## 3.3 Measured Bandwidths

| | | | Radiation Resistance King's Model | Bandwith @ SWR 2.62 @ Antenna Feedpoint | | Damping Resistance | | Loss Resistance | | Antenna Efficiency | |
|---|---|---|---|---|---|---|---|---|---|---|---|
| | | | | | | XL * df / f | | $R_T$-$R_R$ | | $R_R$ / $R_T$ | |
| Band | f | XL | $R_R$ | df | | $R_T$ | | $R_{loss}$ | | η | |
| | kHz | V/A | Ohm | kHz | | Ohm | | Ohm | | | |
| | | | | indoor | outdoor | indoor | outdoor | indoor | outdoor | indoor | outdoor |
| 160m | 1840 | 17.5 | 0.000028 | 3.7 | | 0.035 | | 0.035 | | 0.08% | |
| 80m | 3573 | 33.9 | 0.000399 | 5.7 | 2.4 | 0.054 | 0.022 | 0.053 | 0.022 | 0.7% | 1.8% |
| 60m | 5357 | 50.8 | 0.002016 | 11.9 | 5.5 | 0.113 | 0.052 | 0.111 | 0.050 | 1.8% | 3.8% |
| 40m | 7074 | 67.1 | 0.006136 | 17 | 6.7 | 0.159 | 0.063 | 0.152 | 0.057 | 3.9% | 9.7% |
| 30m | 10136 | 96.2 | 0.025940 | 40 | 9.3 | 0.377 | 0.088 | 0.351 | 0.062 | 6.9% | 29.4% |
| 20m | 14074 | 133.5 | 0.096930 | 87 | 12.1 | 0.829 | 0.115 | 0.732 | 0.018 | 11.7% | 84.3% |
| 15m | 21074 | 199.9 | 0.493869 | 231 | 66 | 2.189 | 0.623 | 1.695 | 0.130 | 22.6% | 79.2% |
| 12m | 24915 | 236.4 | 0.974116 | 359 | 201 | 3.407 | 1.904 | 2.433 | 0.930 | 28.6% | 51.2% |
| 10m | 28074 | 266.4 | 1.584435 | 645 | 350 | 6.122 | 3.317 | 4.537 | 1.733 | 25.9% | 47.8% |

Figure 73: Measured bandwidths. Derived resistances and estimated antenna efficiency.

Findings:

- An antenna efficiency of 84.3% outdoors at 20 m is a good value. With even greater distance from surrounding objects, it would likely increase further. This is where the elaborate antenna construction pays off.
- Outdoor efficiency: one would expect efficiency to continue increasing with frequency, but above 14 MHz it decreases again. This is very surprising.

## 3.4 Dependence on Height Above Ground

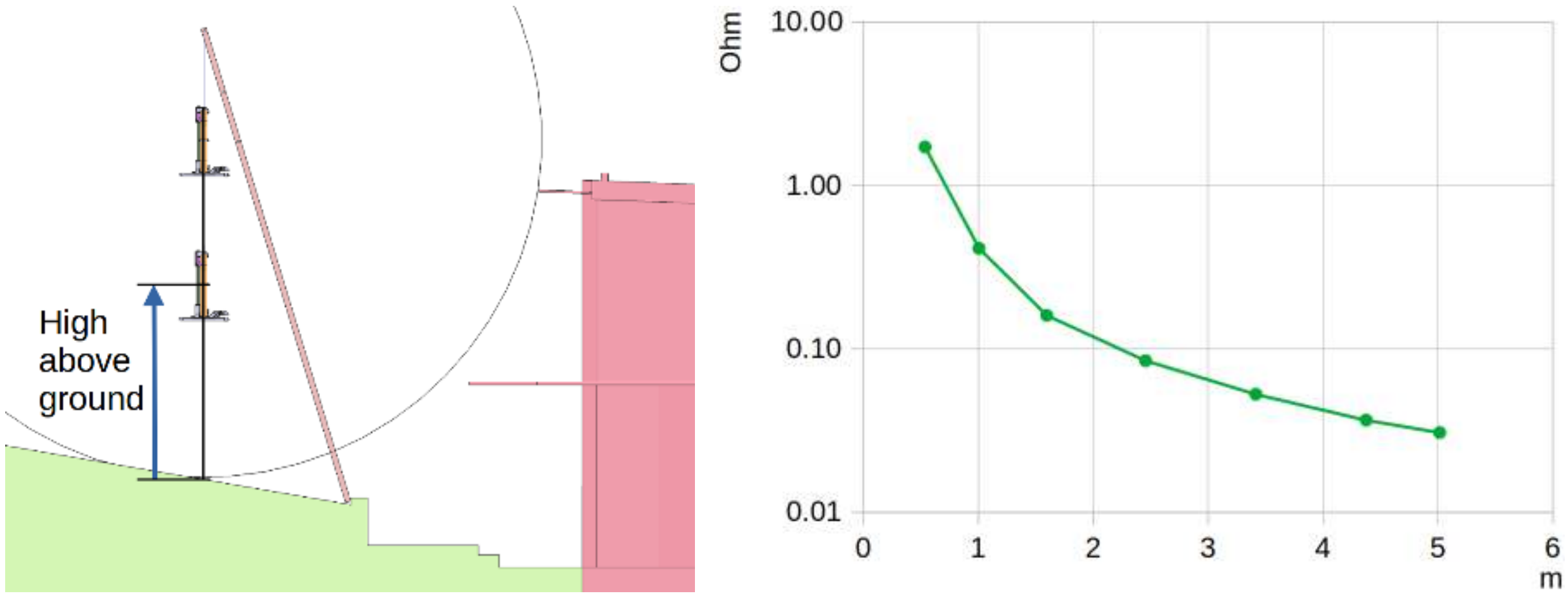

Figure 74: Estimated loss resistance $R_{\text{loss}}$ as a function of height above ground for two frequencies for the 20 m band (14 MHz).

Findings:

- The loss resistance decreases continuously with increasing height as expected. With even greater distance from surrounding objects, the loss resistance would likely decrease further.
- Measurements were also performed in other bands. There, the dependence was far less regular. It is likely that large metallic structures in the surroundings, such as the long fence or the stainless-steel-covered canopies spanning four row houses, can come into resonance at these frequencies and strongly influence the antenna behavior, even at distances beyond 5 m. A 5 m separation is likely too small considering the wavelengths involved.

## 3.5 Influence of Roof Moisture

As described in Figure 72, the antenna is installed on the upper floor directly below a green roof with moist soil. Measurement series were performed under contrasting soil-moisture conditions: one after an extended dry period (soil presumably dry) and one shortly after approximately 10 mm of rainfall. The measured bandwidths differed noticeably, with the direction of change varying by band — some bands showed a wider bandwidth, others a narrower one. The largest difference was observed in the 30 m band, where the bandwidth was approximately 25% wider under wet conditions than under dry conditions.

This suggests that the moist soil above the antenna acts as a lossy dielectric coupled into the near field of the loop, increasing $R_E$ and thereby reducing antenna efficiency.

### 3.6 Influence of Ferrite Cores on Underfloor Heating Pipes

As described in Figure 72, the floor below the antenna contains underfloor heating pipes with aluminium-coated plastic tubing. These conductive pipes lie within the near field of the antenna and are a potential source of environmental losses ($R_E$).

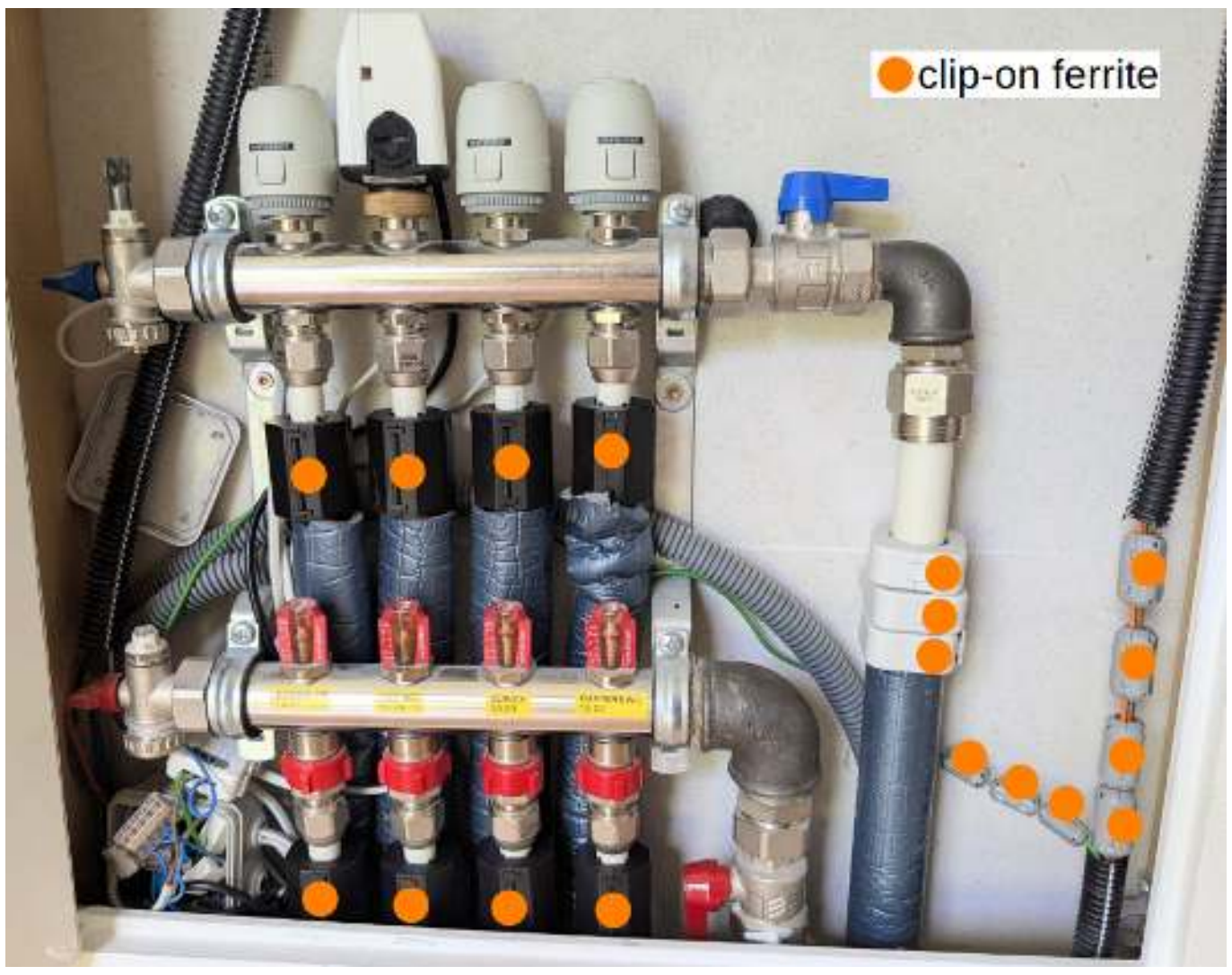

Figure 75: Clip-on ferrite cores attached to the exposed underfloor heating pipes on the upper floor. A few cores were also placed on the control wiring of the heating system.

In an attempt to reduce induced currents in these pipes, clip-on ferrite cores were attached to several exposed pipe sections and to the heating control wiring. Attaching the cores changed the measured bandwidths: some bands became narrower, others became wider. The largest observed change was an 11% bandwidth increase in the 20 m band.

The overall influence on antenna performance was small — and in most bands slightly detrimental — so the cores were subsequently removed.

## 3.7 Measurement of the H Field

Section 1.5 estimates the H-field under free-space conditions. In practice, however, the building contains numerous conductive objects that distort the near field. To quantify the extent of this distortion, the H-field was measured at several locations and compared with the theoretical predictions.

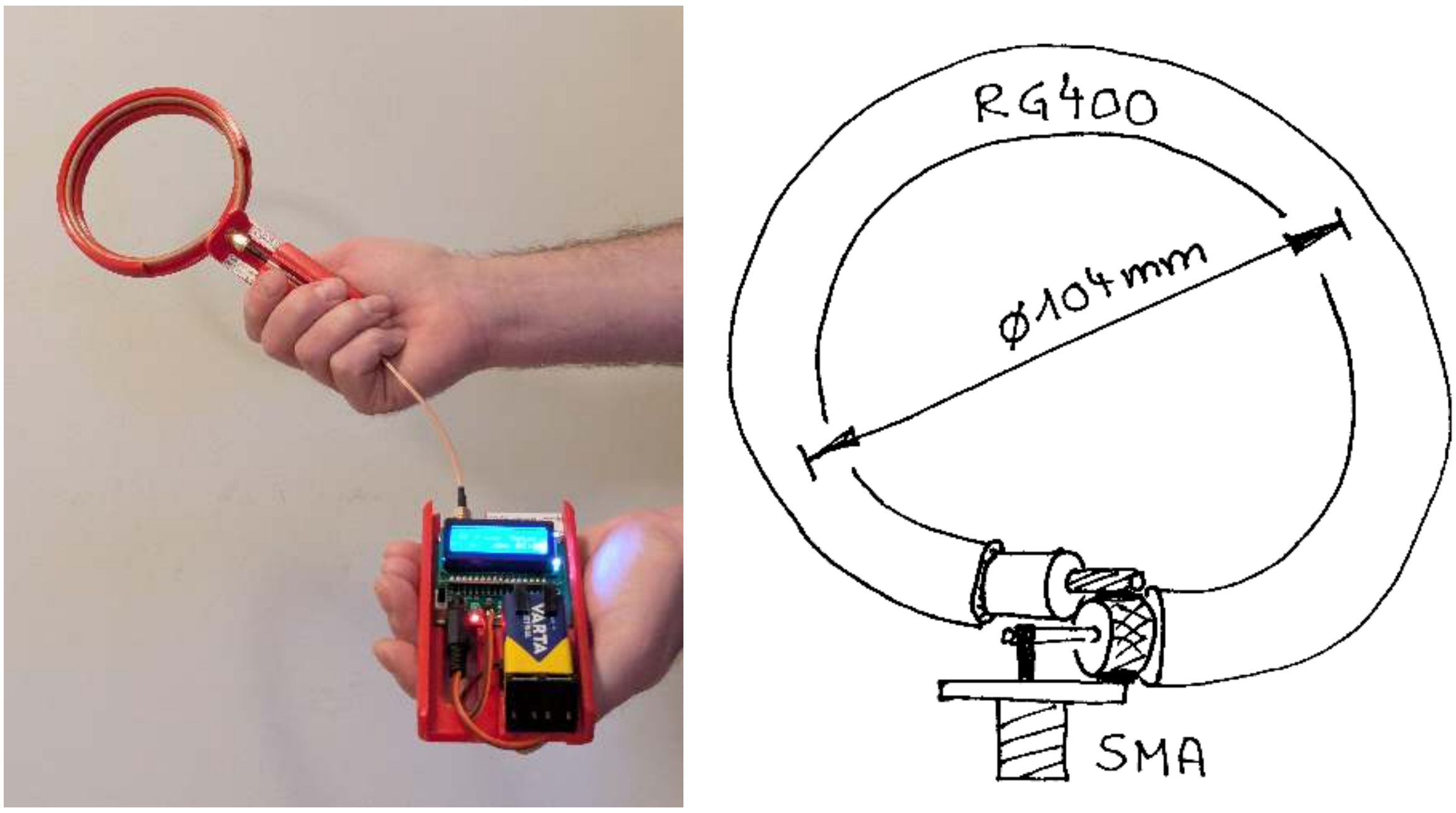

Figure 76: Left: simple self-made H-field probe. The induced voltage is measured with an inexpensive power meter containing an AD8307 logarithmic amplifier. The power meter was verified: despite its low cost, it is remarkably accurate. The combined measurement uncertainty of the probe and power meter is estimated at $\pm$ 0.5 dB, which is negligible compared to the field distortions caused by the building. Right: schematic of the sensing loop. The inner conductor forms the loop; the outer conductor provides shielding.

To determine the H-field magnitude, the H-field probe was held at arm's length and rotated until the maximum reading was obtained.

The transceiver displays 100 W at $f_0 = 14.1$ MHz. After cable losses, an estimated $P = 94$ W arrives at the antenna. The measured bandwidth $B_{\text{SWR 2.62}} = 86.6$ kHz yields $R_T = 0.842\ \Omega$ and $I_{\text{main_loop}} = 10.5\ \text{A}_{\text{rms}}$. From these values, the expected magnetic field strength is calculated using the full retarded magnetic dipole solution, applied to the equivalent-diameter circular approximation of the rectangular loop.

| | X | Y | Z | expected | measured | |
|---|---|---|---|---|---|---|
| | m | m | m | A/m | A/m | factor |
| A | 2 | 0 | 0 | 0.1909 | 0.2154 | 1.1 |
| B | 3 | 1 | 0 | 0.0544 | 0.0482 | 0.9 |
| C | 3 | 0 | 0 | 0.0652 | 0.0705 | 1.1 |
| D | 3 | -1 | 0 | 0.0544 | 0.0621 | 1.1 |
| E | 3 | -2 | 0 | 0.0352 | 0.0435 | 1.2 |
| F | 3 | -3 | 0 | 0.0213 | 0.0430 | 2.0 |
| G | 5 | 1 | 0 | 0.0177 | 0.0134 | 0.8 |
| H | 6 | 1 | 0 | 0.0120 | 0.0095 | 0.8 |
| I | 7 | 1 | 0 | 0.0086 | 0.0085 | 1.0 |
| K | -11 | 0 | 0 | 0.0034 | 0.0057 | 1.7 |
| L | -3 | -38 | 0 | 0.0015 | 0.0023 | 1.5 |
| M | 0 | 0 | 3 | 0.0223 | 0.0179 | 0.8 |
| N | 0 | 0 | -5.7 | 0.0088 | 0.0063 | 0.7 |
| O | 5.5 | 0 | -5.7 | 0.0067 | 0.0046 | 0.7 |

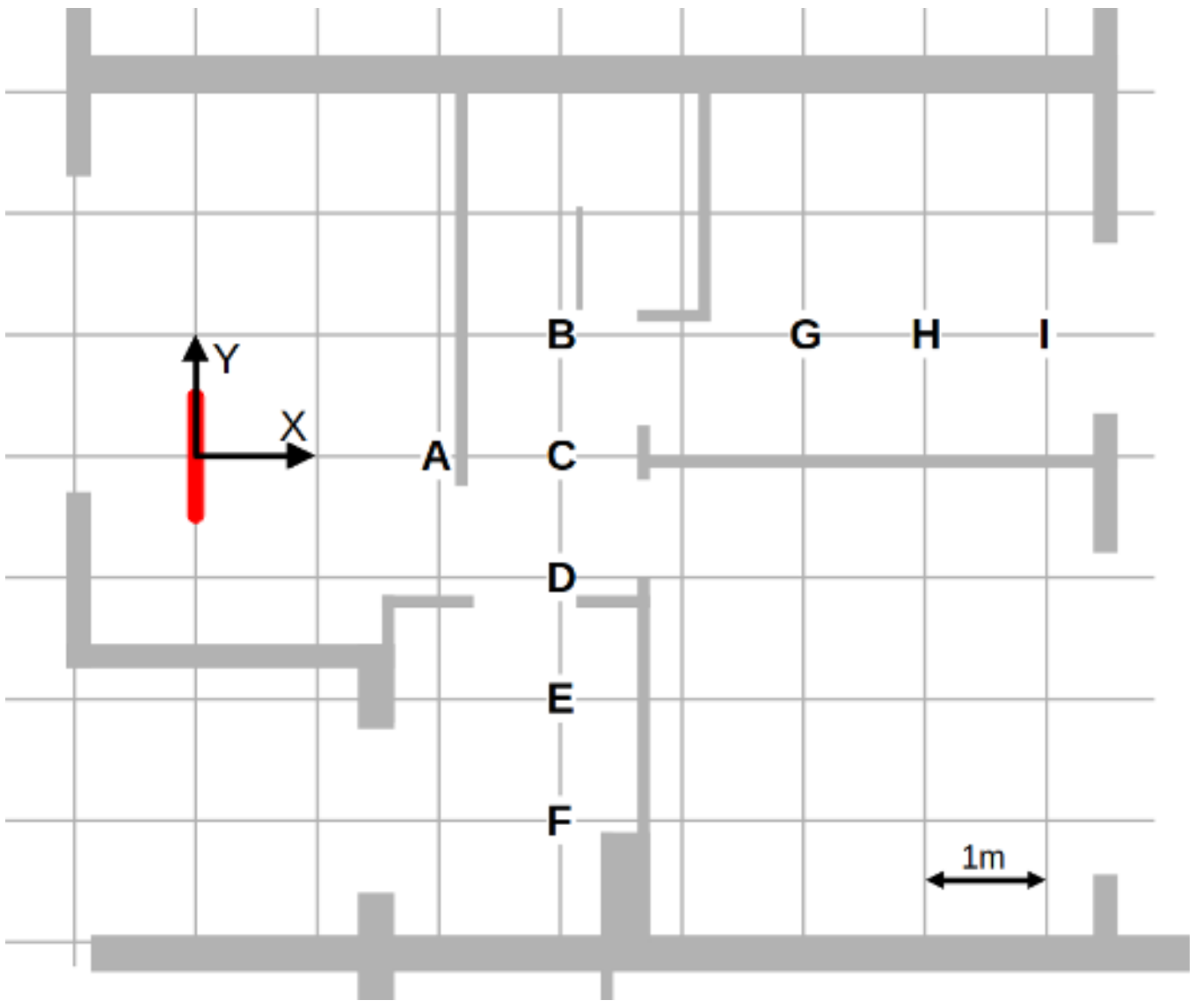

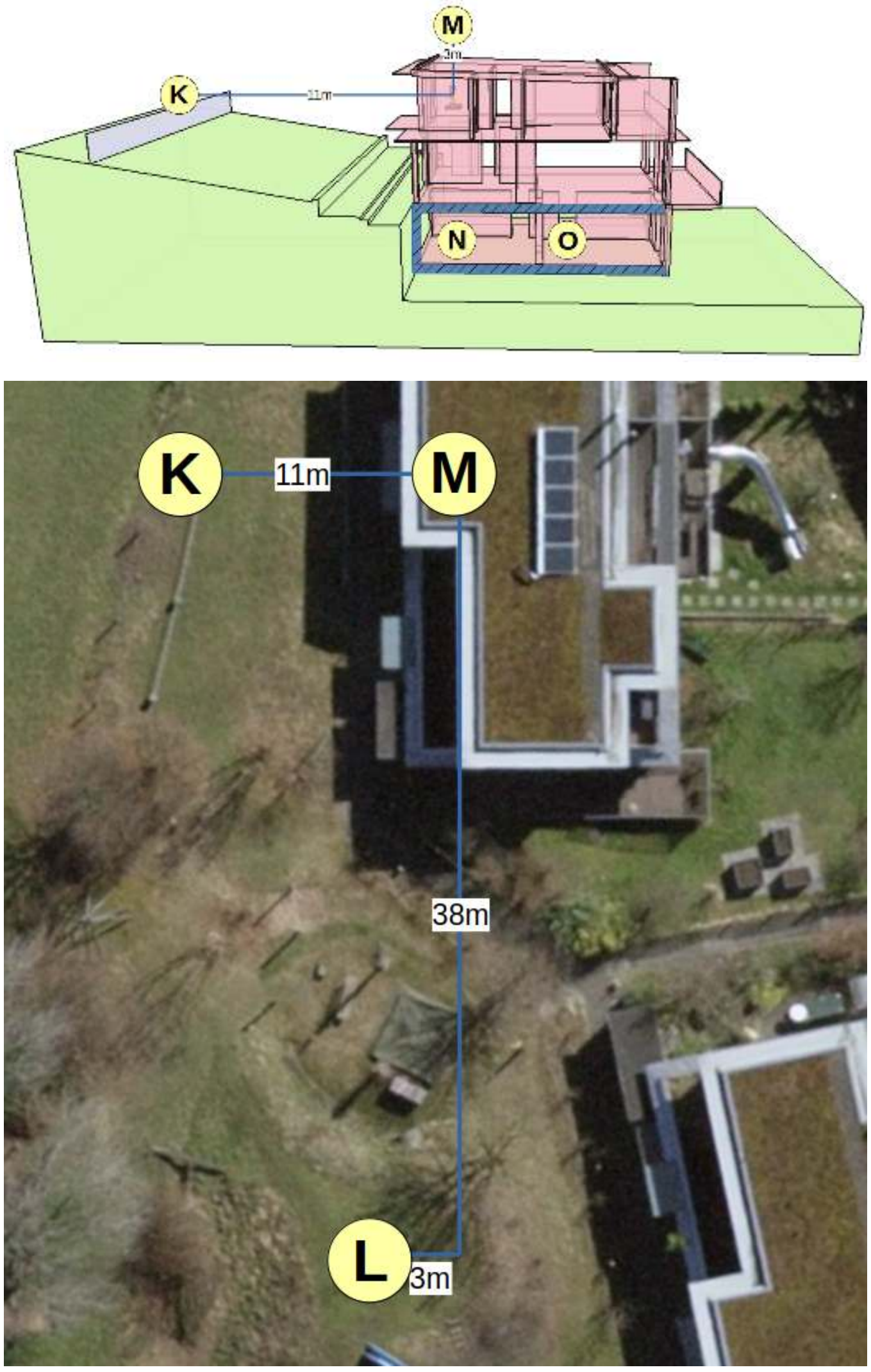

Figure 77: Left column, top: measurement points with calculated and measured H-field; the factor shows the ratio of measured to calculated field strength. Left column, bottom: floor plan of the upper story, walls shown in gray; the antenna is shown in red; at the height of the antenna ($Z = 0$, coordinate system as defined in Figure 15), several points are measured. Right column, top: measurement points K, M, N, and O in the basement; reinforced concrete is shown with blue hatching. Right column, bottom: measurement points K, L and M.

In summary, the measured field strength agrees surprisingly well with the predicted values; larger deviations are more typical in practice.

Near point F, the building's riser pipes are located — thick copper pipes supplying the thermal solar system on the roof. The factor of 2.0 may be explained by this.

When the H-field probe is moved directly along the floor, a strong influence of the underfloor heating — which uses aluminum-coated plastic pipes — is observed.

The measured values at points K, L, and M — located outside the building — are notably large. The building does not appear to attenuate the field significantly.

While the building significantly damps the magnetic loop, it provides only limited shielding of the magnetic field. Thus, damping of the loop and attenuation of the magnetic field are different phenomena.

In contrast to the two upper timber-frame stories, the basement is made of reinforced concrete.

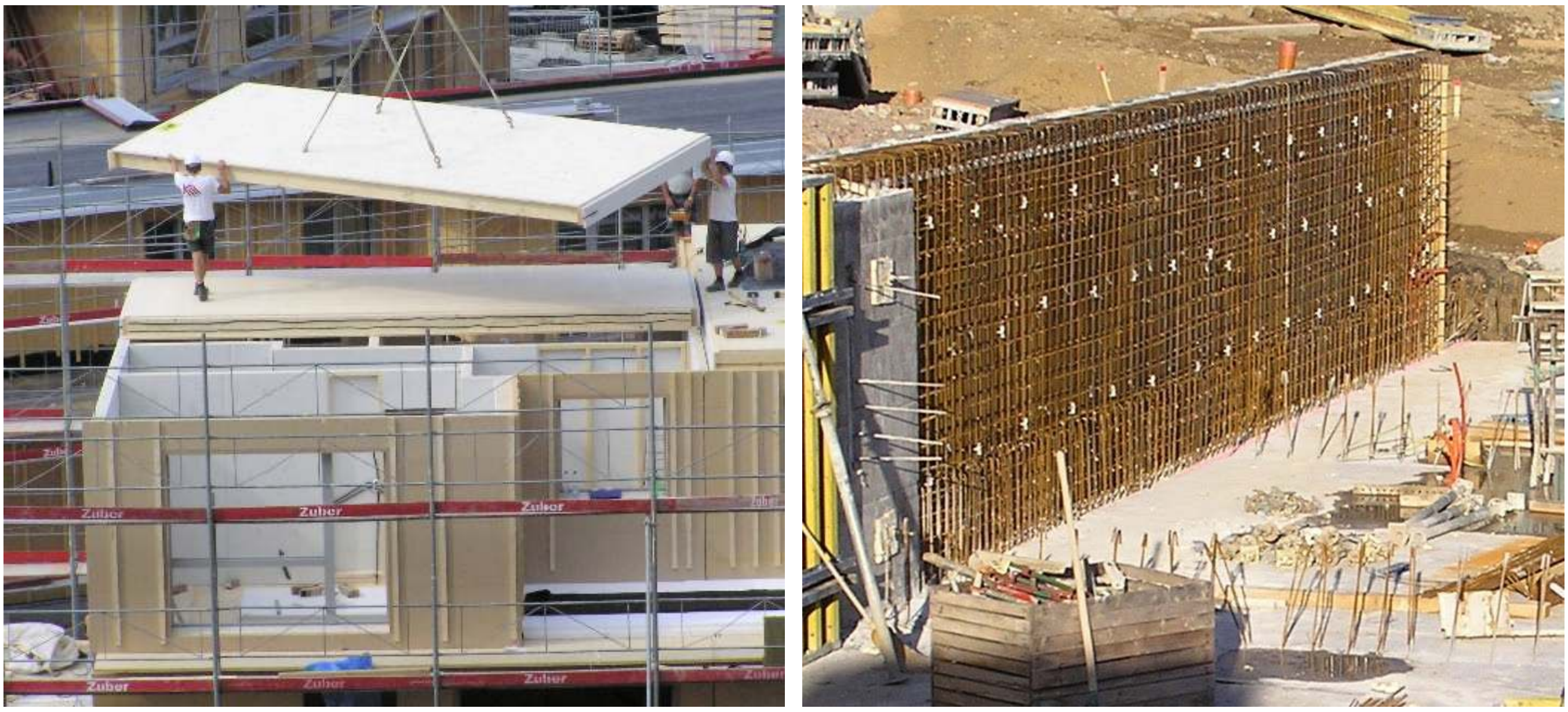

Figure 78: Left: impression of the top timber-frame story; at this stage the room was still open, and the antenna was later installed there. The wooden panel still suspended by ropes later became part of the roof. Right: impression of the reinforced-concrete basement level, showing the dense steel reinforcement of one side wall before the concrete was poured.

Points N and O are located in the basement, where the measured factor reaches its minimum value of 0.7. The H-field is likely attenuated by the steel reinforcement, and these points are also closest to the surrounding soil.

Given the large amount of steel reinforcement, a substantially stronger attenuation might have been expected. However, the measured factor of 0.7 indicates that most of the field still penetrates the reinforced structure.

## 3.8 Was the elaborately constructed antenna justified?

I use the antenna only indoors because regulations do not allow me to install an antenna on the roof. The effort I invested in reducing the loss resistances is too high for indoor operation.

Here is a concrete example to illustrate this:

At 20 m, I measured an outdoor loss resistance of 0.024 Ω. With $R_{\text{loss}} = R_L + R_C + R_E$, and assuming $R_E = 0\Omega$, the maximum combined loss resistance of the inductor and capacitor is 0.024 Ω. When I use the antenna indoors, $R_{\text{loss}}$ increases to 0.757 Ω.

This means the loss resistance contributions from the inductor and capacitor become only a very small part of the total losses. The indoor antenna efficiency would therefore remain very similar even if $R_L$ and $R_C$ were higher.

If I were to build another magnetic loop, I would use thinner tubing and air capacitors, since the voltages encountered indoors are sufficiently low.

### 3.9 Is the loop diameter well chosen?

A loop diameter of 1 m was chosen as a practical compromise for indoor use: not so small that capacitor voltages become problematic, and not so large that it sits close to the walls — which, as discussed below, can increase environmental losses.

A smaller loop would be less expensive and easier to handle, though it might perform slightly worse for reception — on HF, especially at lower frequencies where atmospheric noise dominates, this difference is often negligible in practice. A further disadvantage of a smaller loop is the higher voltage that appears across the tuning capacitor for the same transmitted power.

Increasing the loop diameter does not reliably improve the radiated power of an indoor loop. While a larger diameter raises the radiation resistance, it also raises the environmental loss resistance by a similar factor. As the total resistance increases, the loop current decreases accordingly, and the radiated power remains approximately the same.

This may seem to contradict the common claim in the literature that a larger loop diameter improves efficiency — a claim that holds only when antenna-internal losses dominate.

The following thought experiment illustrates why, in the indoor case, antenna efficiency depends only weakly on loop diameter.

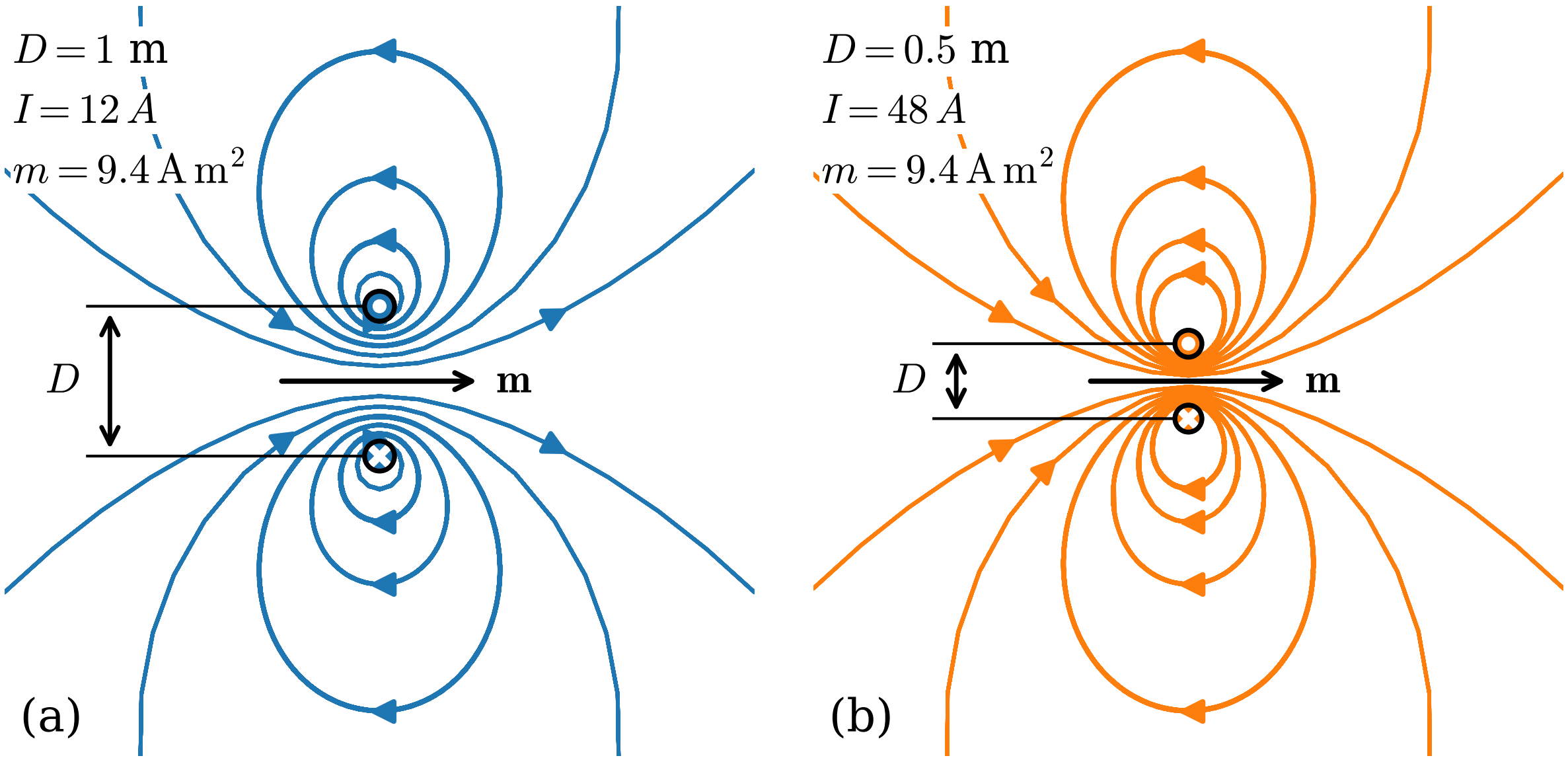

Figure 79: Comparison of two magnetic loops with diameters of 1.0 m (a) and 0.5 m (b). Both loops produce the same magnetic dipole moment, requiring the smaller loop to carry four times the current. The magnetic field lines differ primarily near the loops. The field distribution was calculated using the Biot–Savart law, which is a good approximation in the near-field region (r≪λ).

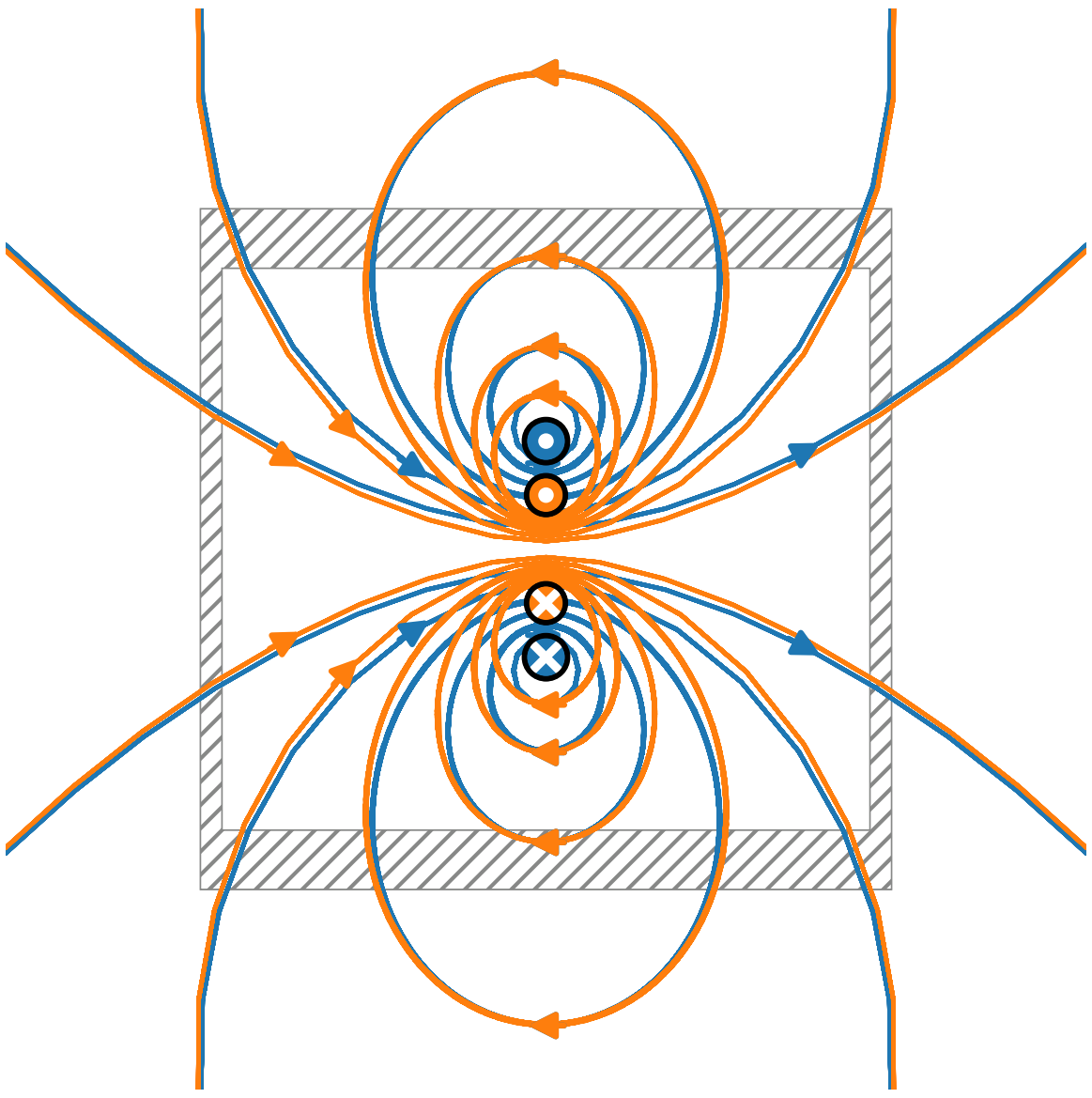

Figure 80: Overlay of the field lines of both loops from Figure 79. The gray hatched region indicates a typical room boundary. Near walls, ceiling, and floor, the field-line pattern is nearly identical for both loops.

Because the field-line pattern at the walls is nearly identical for both loops, wall absorption is also similar. In addition, far-field radiation is similar because radiated power is proportional to the square of the magnetic dipole moment.

The key point is this: for a given power delivered to the antenna (for example 100 W), a corresponding magnetic dipole moment is established. If antenna-internal losses are small ($R_L, R_C \ll R_T$), most of that power is shared between radiation ($R_R$) and environmental losses ($R_E$). A large and a small loop can therefore produce approximately the same dipole moment, and the loop diameter is then not a dominant factor for efficiency.

In some practical indoor cases, increasing the loop size is even expected to reduce efficiency: a larger loop placed closer to a wall will produce a stronger local field at that wall, increasing wall absorption and thus lowering efficiency.

## 3.10 Common-mode currents

A few words on antenna symmetry and common-mode currents: I made an effort to build the antenna as symmetrically as possible. However, the gamma match alone introduces some asymmetry. For indoor operation, I therefore use a self-built choke combination on both the coax cable and the servo control line: 17 turns on 2 stacked FT240-31 cores, 9 turns on 4 stacked FT240-43 cores, and 9 turns on 3 stacked FT240-52 cores. This combination provides high impedance across the entire frequency range.

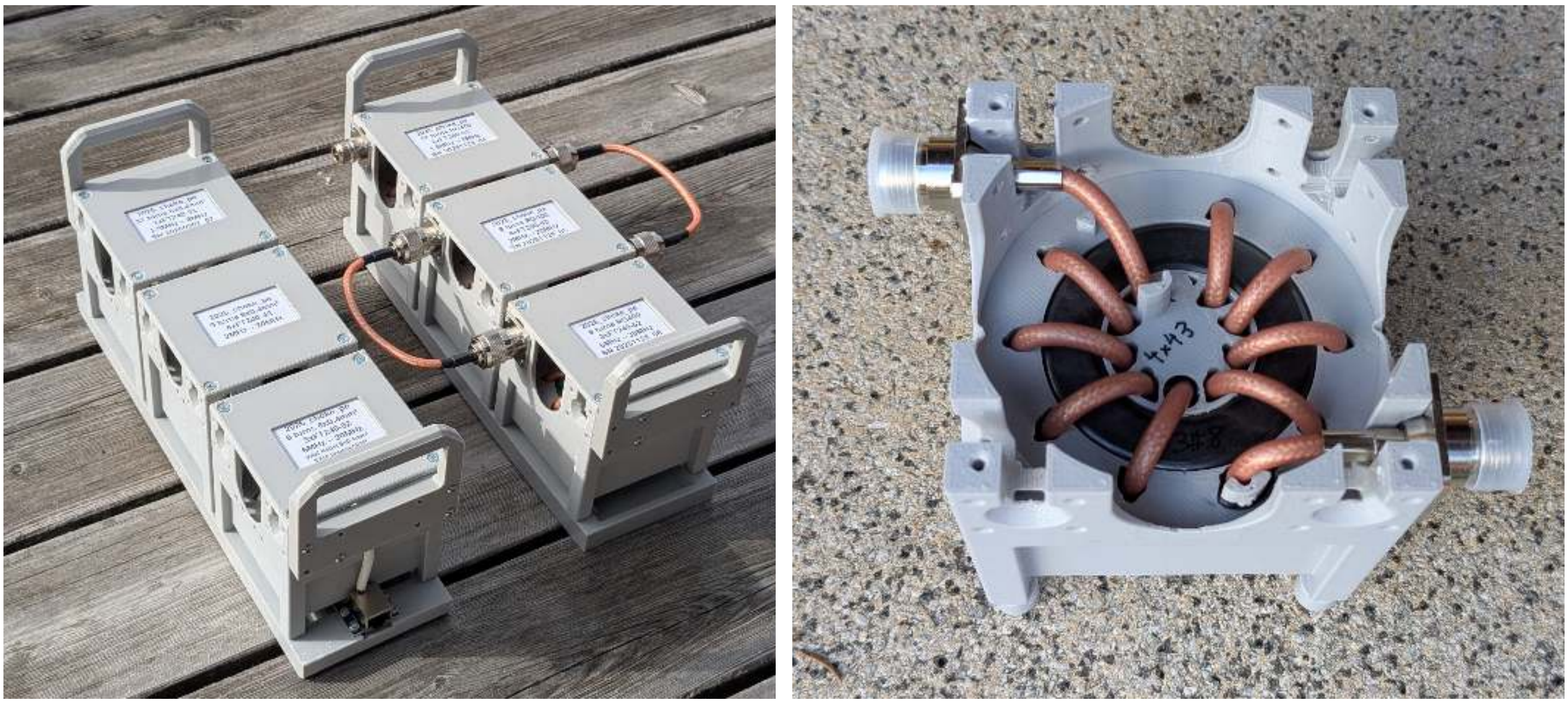

Figure 81: Common-mode choke assemblies used on the coax feed and servo control line. Right: example with RG400, 9 turns on 4 stacked FT240-43 cores. The 3D-printed PETG part keeps the cables in position and separated, thereby reducing parasitic capacitance. I later replaced the RG400 cables with Messi & Paoloni Hyperflex 5, which have lower losses and are more flexible, making them easier to install

For the outdoor measurement values, I did not use the chokes. I made sure that the feed cables were hanging as centrally as possible relative to the antenna. I installed and removed the chokes again for several test measurements and could not observe any significant influence on the measured bandwidth.

We also performed measurements in which a person held the feed cables by hand. This also showed no significant influence. I therefore conclude that common-mode currents did not have a significant influence on the measured bandwidth.

## 3.11 Measurement Setup, Bandwidth Correction

The measurements are performed with a NanoVNA V2 Plus4. The calibration was performed directly at the VNA, not at the antenna feed point. In retrospect, this was incorrect, and I do not recommend performing the measurement this way.

Unfortunately, I forgot to do this, and the outdoor setup has already been dismantled again.

Cables
- 10 m RG8U,
- 5.5 m RG400 (in the chokes, for indoor measurements)
- 2 m RG58.
- About 10 cable connections along the feed line

The cable losses in the following table were measured. These agree very well with the losses specified by the cable manufacturers. A bandwidth correction factor based on these losses was applied, because the measured SWR 2.62 bandwidth at the VNA is larger than at the antenna feed point.

| Band | f | Bandwith Measured @ SWR 2.62 @ VNA df (kHz) | | Cable Losses One way Measured (dB) | | Bandwith Correction Factor Cable Losses | | Bandwith @ SWR 2.62 @ Antenna Feedpoint df (kHz) | |
|---|---|---|---|---|---|---|---|---|---|
| | kHz | indoor | outdoor | indoor | outdoor | indoor | outdoor | indoor | outdoor |
| 80m | 3573 | 6.4 | 2.5 | 0.40 | 0.26 | 0.89 | 0.93 | 5.7 | 2.4 |
| 60m | 5357 | 14 | 6.0 | 0.42 | 0.30 | 0.88 | 0.92 | 11.9 | 5.5 |
| 40m | 7074 | 19 | 7.3 | 0.49 | 0.31 | 0.86 | 0.91 | 17 | 6.7 |
| 30m | 10136 | 48 | 10.2 | 0.63 | 0.33 | 0.83 | 0.91 | 40 | 9.3 |
| 20m | 14074 | 108 | 14 | 0.71 | 0.41 | 0.81 | 0.89 | 87 | 12.1 |
| 15m | 21074 | 301 | 76 | 0.87 | 0.47 | 0.77 | 0.87 | 231 | 66 |
| 12m | 24915 | 470 | 233 | 0.88 | 0.50 | 0.76 | 0.86 | 359 | 201 |
| 10m | 28074 | 877 | 407 | 0.99 | 0.51 | 0.74 | 0.86 | 645 | 350 |

Figure 82: Measured indoor and outdoor bandwidths at the VNA, measured cable losses and the resulting correction factors for the bandwidth, corrected bandwidth valid at the antenna feed point.

Details on the bandwidth correction factor can be found in Appendix Section 7.3.

I later replaced the cables with lower-loss and more flexible ones: Messi & Paoloni Hyperflex 10, Ultraflex 7, and Hyperflex 5.

## 3.12 Measurement Results: Heating

I would like to examine one indoor example in more detail.

- Transmit power: 100 W CW @ 28 MHz
- Power loss in cables and connectors: 1 dB, 20 W
- Power delivered to antenna: 80 W
- Damping resistance measured indoors $R_T$: 4.4 Ω
- Power in $R_T$ (damping), all the power fed to the antenna: 80 W
- Radiation resistance calculated $R_\mathrm{R}$: 1.54 Ω
- Power in $R_\mathrm{R}$ (radiation): 28 W
- Power in $R_\mathrm{loss}$, remaining power: 52 W

So I expect that 52 W of power is converted into heat in the antenna or in the surrounding environment.

I would like to estimate how much of this power remains in the antenna itself. For this purpose, I transmit for 10 minutes at 100 W and observe what heats up.

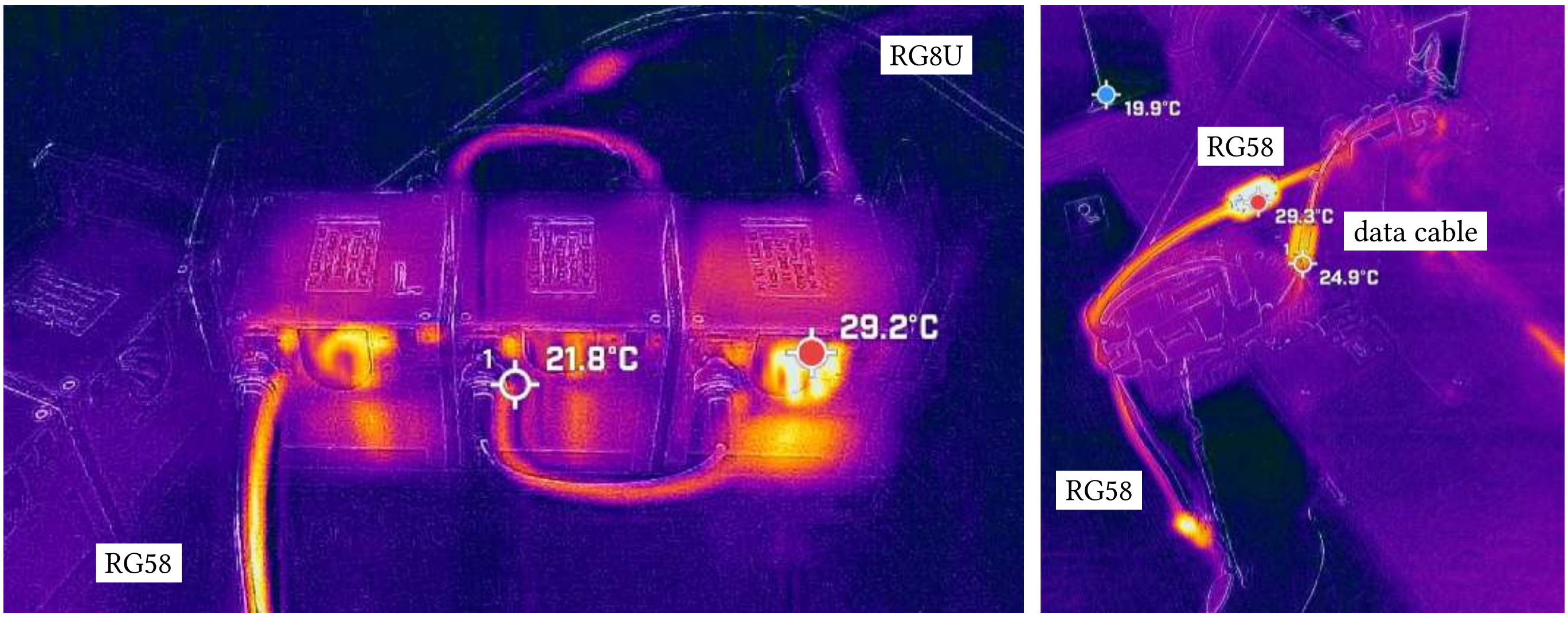

Figure 83: Left: the RG8U feed cable warmed by 1.1°C and the 17-turn choke by 8.2°C, consistent with about 20 W cable/connector loss. Right: three ZCAT1325-0530A clip-on chokes at the antenna warmed by about 8.1°C due to common-mode currents; a control test with a ZCAT choke near the main loop alone showed no noticeable heating.

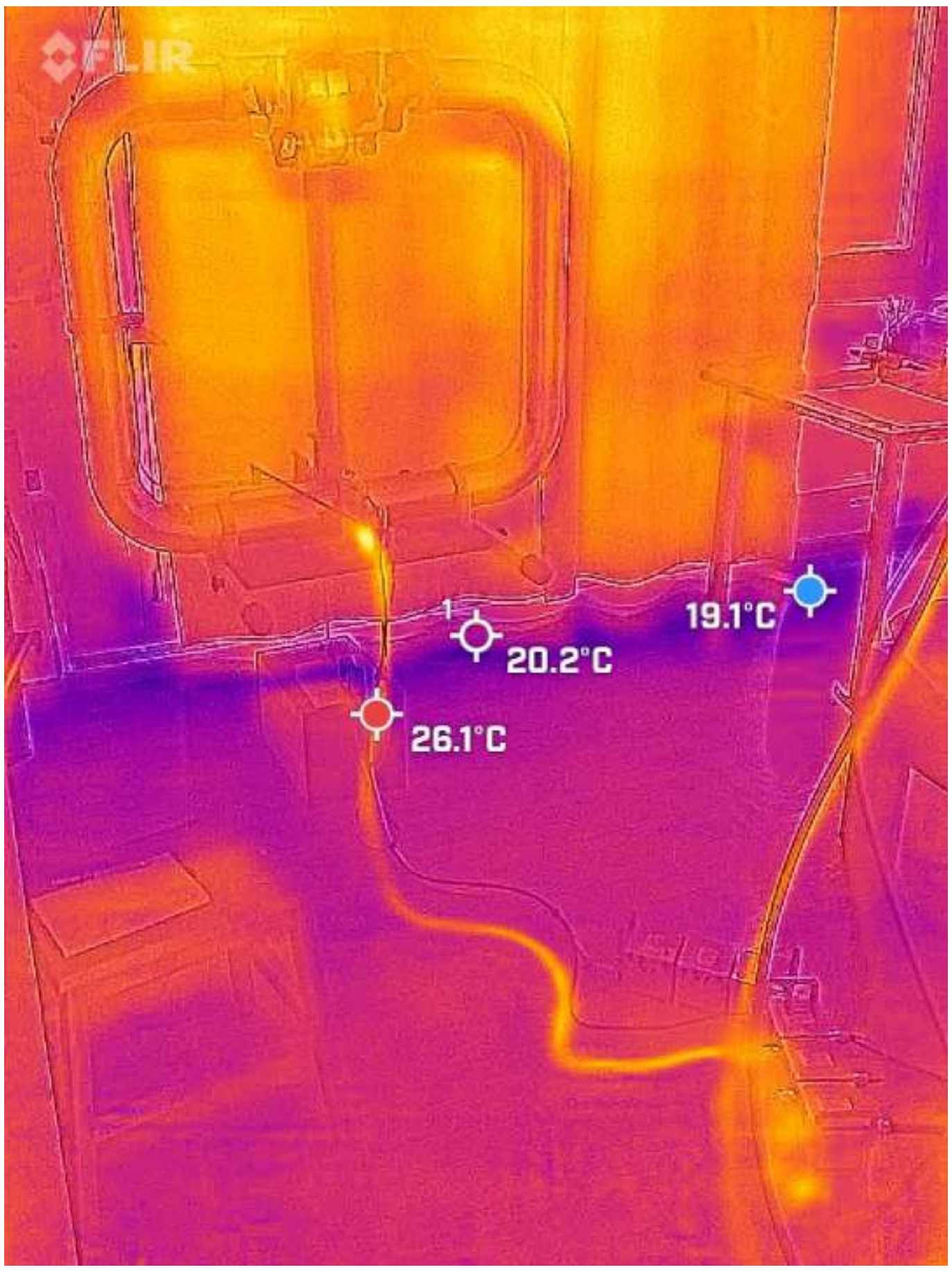

Figure 84: The antenna itself shows no visible heating (below about 0.3°C at taped measurement points), while the RG58 section from the chokes to the antenna is visibly warmer.

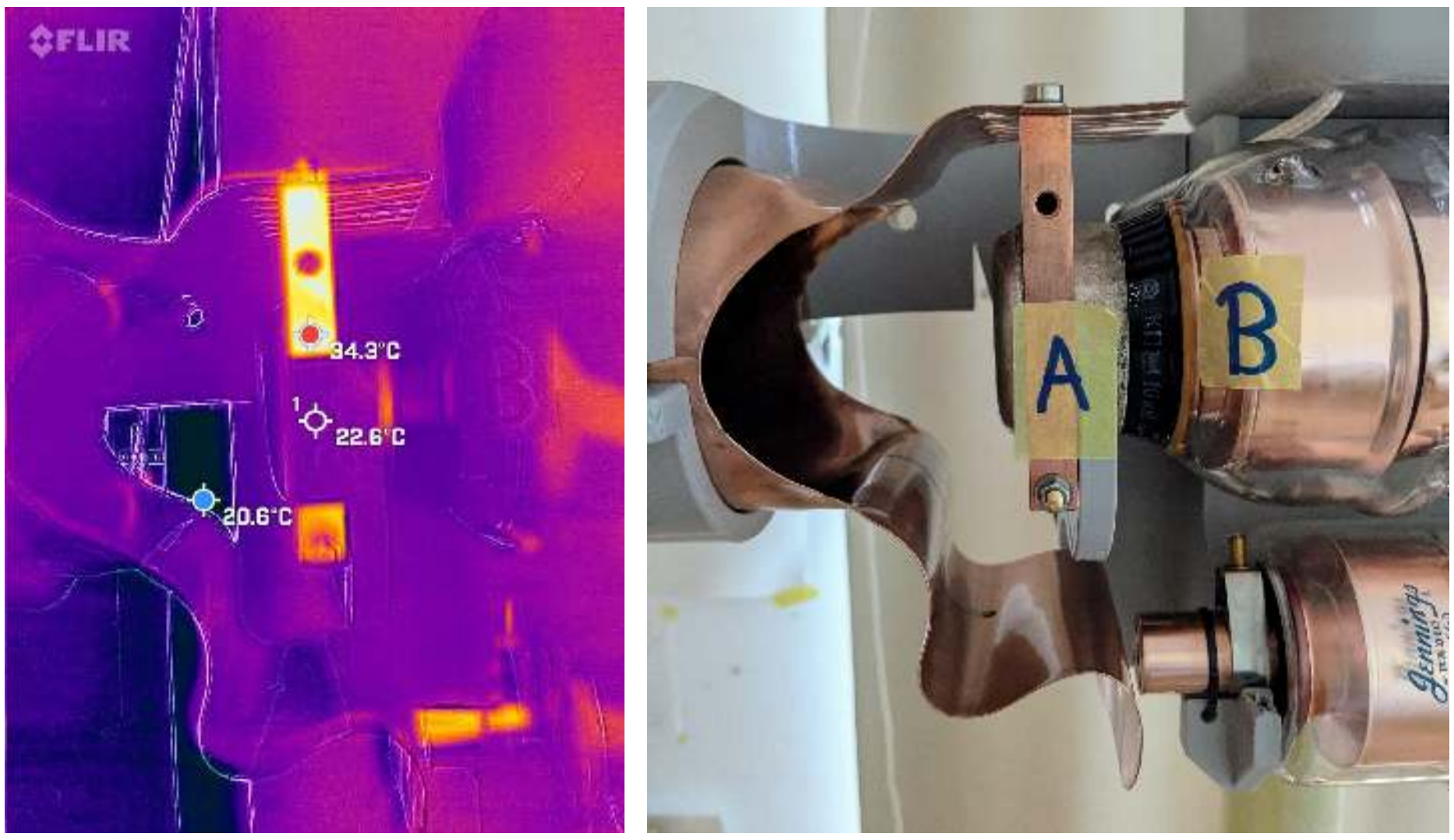

Figure 85: No antenna parts show measurable heating; temperatures are read on painter's tape (emissivity near 1), while shiny metal surfaces can appear hot due to reflected body radiation, so thermal images of reflective objects must be interpreted with care. In the left image, the copper plate surface appears hot, but this is only a reflection of my body heat. The tape labeled A shows the true temperature of the copper plate.

I heat the antenna using a heater with known power. I insert an LED strip into the left section of the main loop and let it run for 10 minutes. The LED power can be determined accurately from the measured current consumption: it is 5 W.

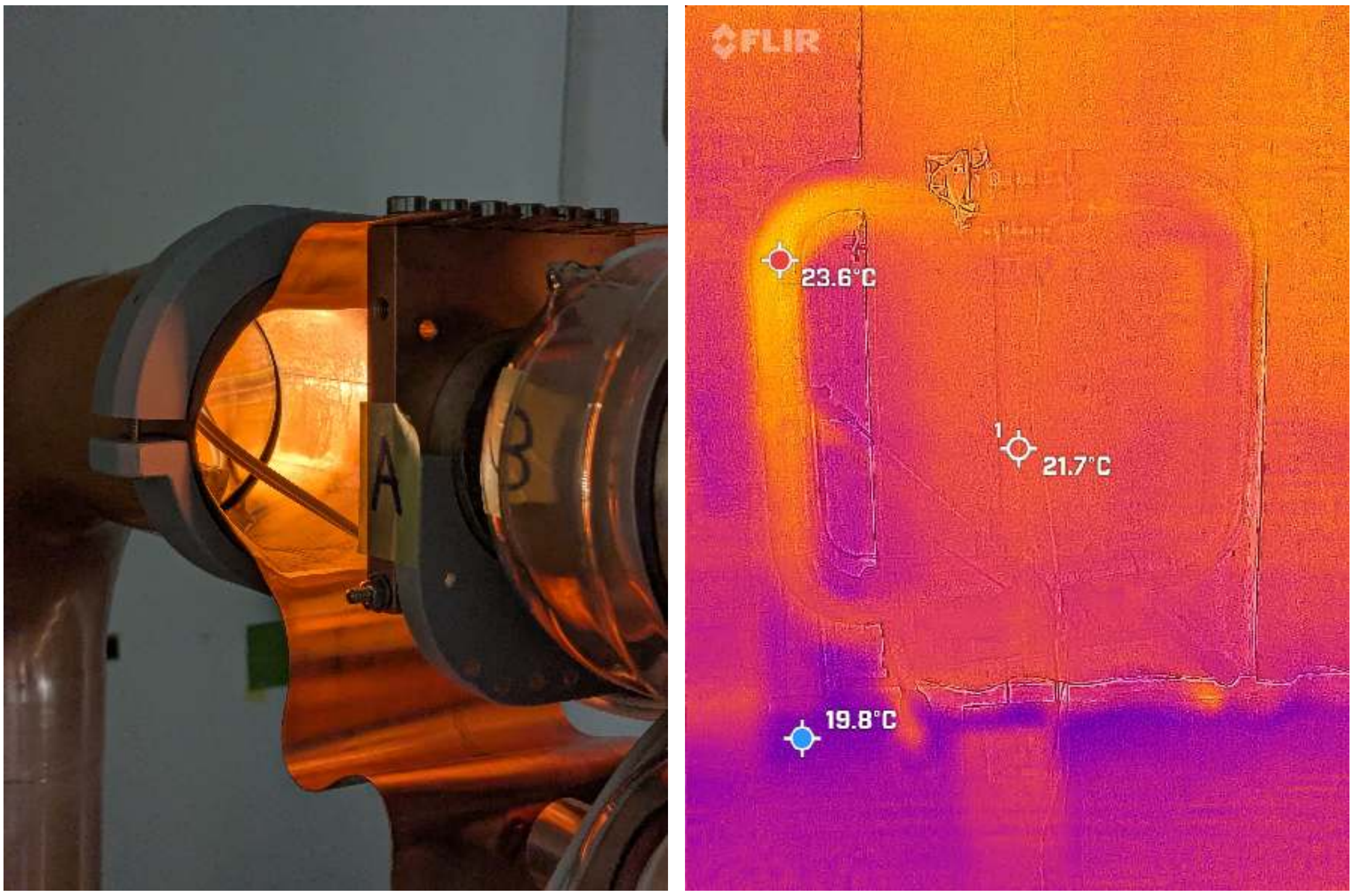

Figure 86: Left: view into the left tube section. The LEDs are lit and heat the tube. Right: thermal image of the antenna after 10 minutes with 5 W applied in the left tube section. The temperature rise is clearly visible.

The upper-left area is 24.3°C, and the upper-right area is 22.2°C; the temperature increase is therefore 2.1°C. Conclusion: 5 W of dissipated power is easy to detect. If I transmit with 100 W and cannot measure any heating of the antenna, I assume that the power dissipated in the antenna itself is clearly below 5 W.

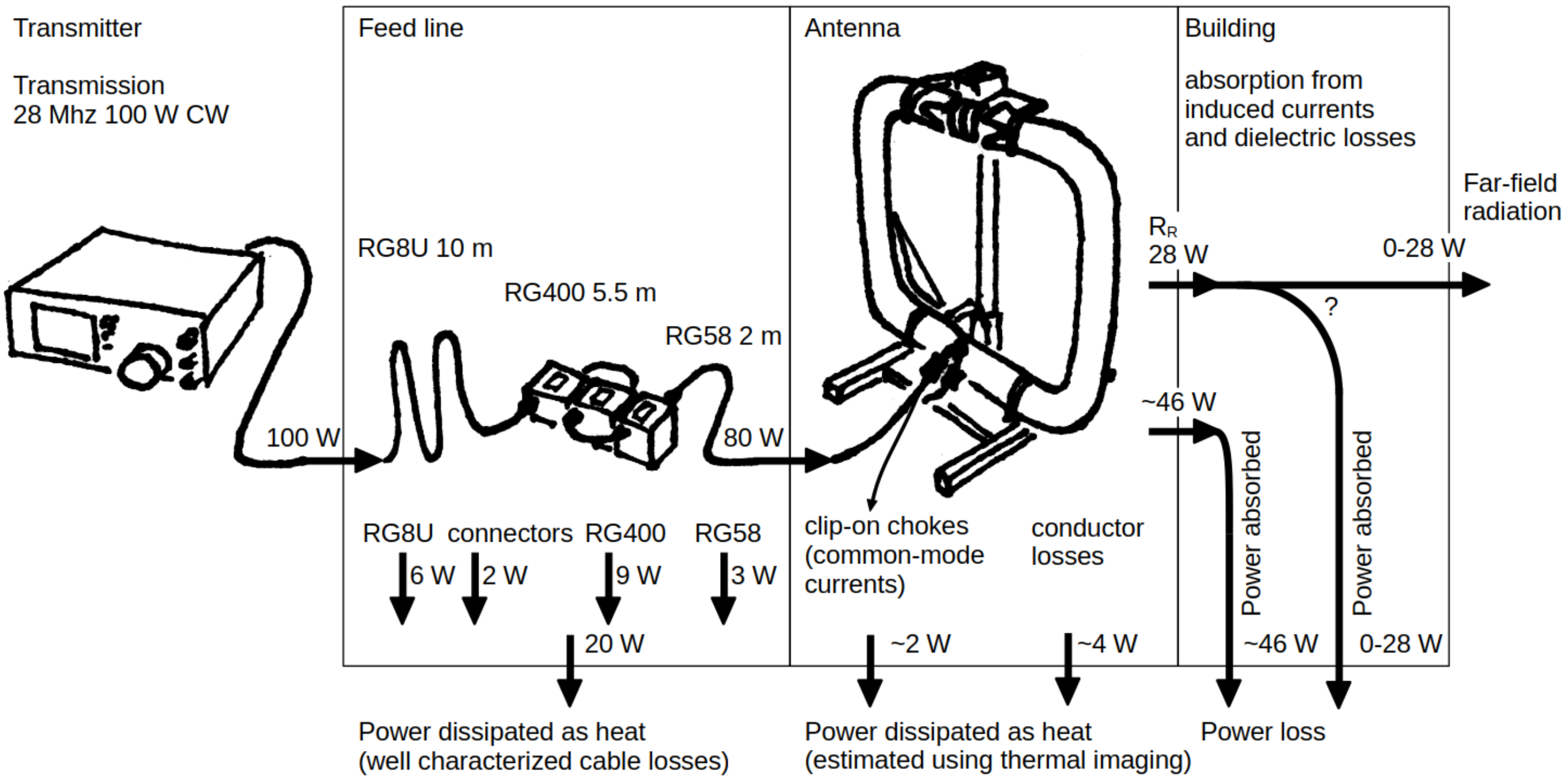

Figure 87: Summary of the heating experiment: power budget for 100 W transmit power.

Findings:

- I have measurable losses of 20 W in the feed cable. These are well characterized losses and I can see them with the thermal camera.
- The antenna does not heat up measurably. I estimate the power dissipated in the clip-on ferrites at 2 W and in the antenna itself at 4 W.
- In free space, the antenna would radiate 28 W based on the radiation resistance $R_{\mathrm{R}}$. However, not all of this power will reach the far field, because part of it is absorbed by the building.
- According to the power budget, the building absorbs between 46 W and up to 74 W of RF energy. However, it will not be the full 74 W, because radio operation works well, which suggests that a portion of the power does reach the far field. The H-field measurements in Section 3.7 suggest that the building does not significantly attenuate the H-field. I therefore venture to claim that the majority of the power attributed to $R_{\mathrm{R}}$ actually reaches the far field.
- In my indoor setup, the losses are dominated by the building. The antenna itself accounts for only a small fraction of the total losses.

## 3.13 Is antenna efficiency estimated correctly?

The term antenna efficiency refers to how much power is actually radiated compared to the power delivered to the antenna. However, this is strictly true only in free space, with nothing nearby. This is exactly what we can calculate and state using the standard formulas — correct and well defined. In typical applications on the ground, the earth alone is already in the way, and for that reason alone only a fraction of the power will reach the far field. Especially in the context of magnetic loop antennas, antenna efficiency must be interpreted with caution. It is a useful figure of merit when comparing different magnetic loops against each other. But for the question of how much power actually reaches the far field, the situation is far more complex.

## 3.14 Measurement uncertainty

Many factors influence the measurement results. Opening a window (with triple-glazed glass and an aluminum frame), slightly rotating the antenna, or routing a cable differently all affect the antenna behavior. When the antenna is outdoors at a height of 5 m and sways in the wind, the effect is even greater.

During measurements at 5 m height and with high Q values, the resonant-circuit trace in the Smith chart was no longer as circular as in textbook examples. I therefore adjusted the frequency slightly each time until the curve looked reasonable, measured the bandwidth, and then continued with the next measurement. Subsequent investigations revealed the likely cause: my NanoVNA V2 Plus4 (firmware version 20250121) exhibits significant measurement artifacts when measuring highly reactive magnetic loop antennas below approximately 5 MHz. The distorted Smith chart trace was therefore most probably an instrument artifact rather than a physical effect. Measurements in this frequency range should be interpreted with caution.

Even though the results are given with many digits, this should not obscure the fact that the measurement uncertainty is large.

## 3.15 Is the radiation resistance correct at higher frequencies?

The investigated antenna has a very large surface area because the tube diameter is 100 mm. Could this possibly be a combination in which the loop antenna also exhibits characteristics of a shortened dipole in the 10 m band? Additional radiation would result in a larger radiation resistance than the value I approximated using the King model. An underestimated radiation resistance could explain the small antenna efficiency determined at higher frequencies, e.g., outdoor in the 10 m band. I look forward to feedback from antenna experts.

# 4 The Journey Is the Reward

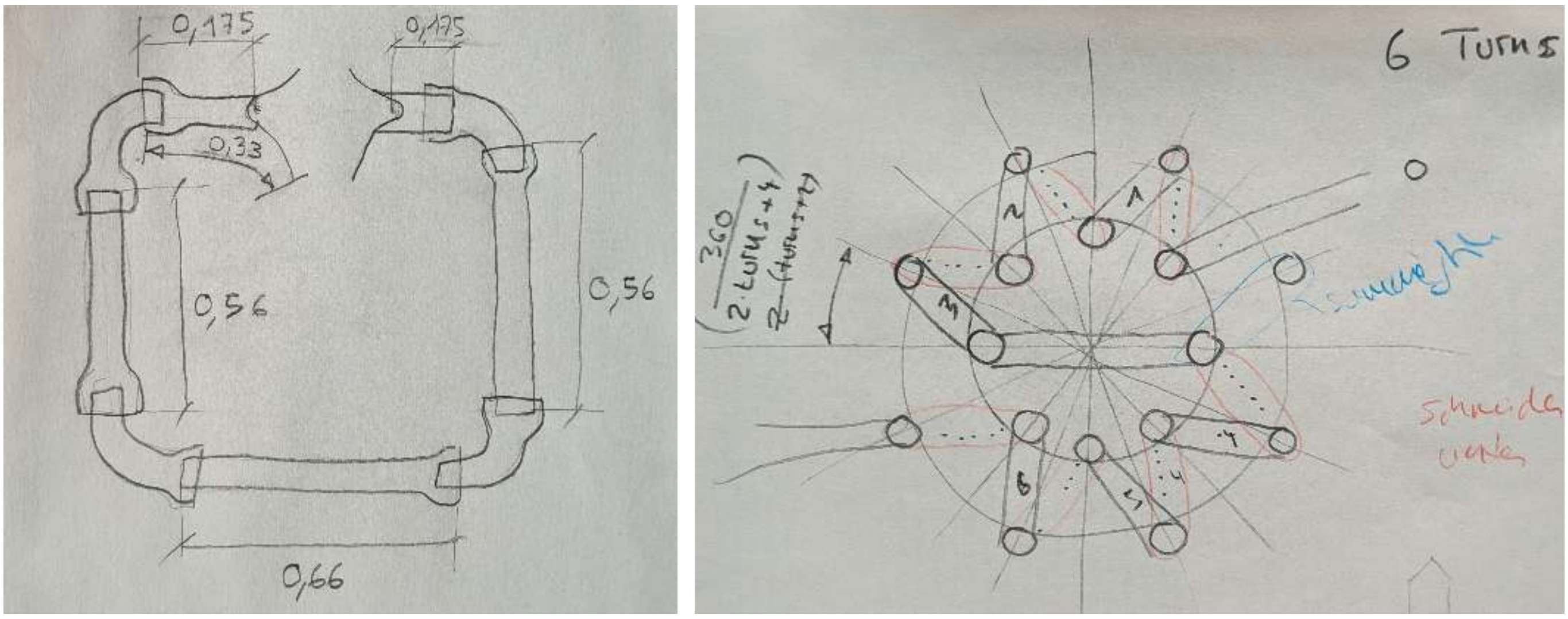

Figure 88: Not everything worked on the first attempt — it was a tough journey with many setbacks. But it was fun.

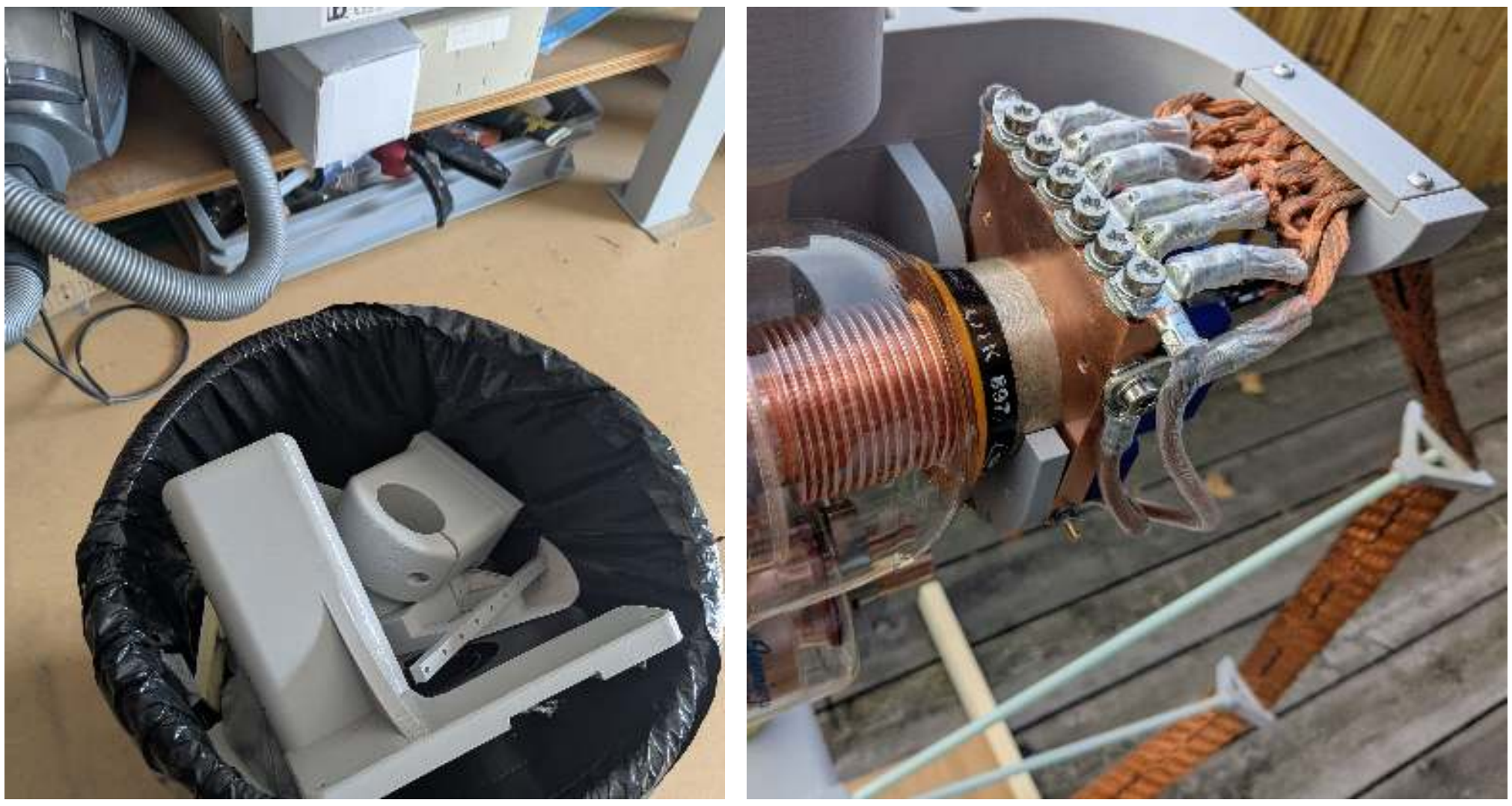

Figure 89: Many things ended up in the bin — or, in the case of this magnetic loop, in the scrap copper collection.

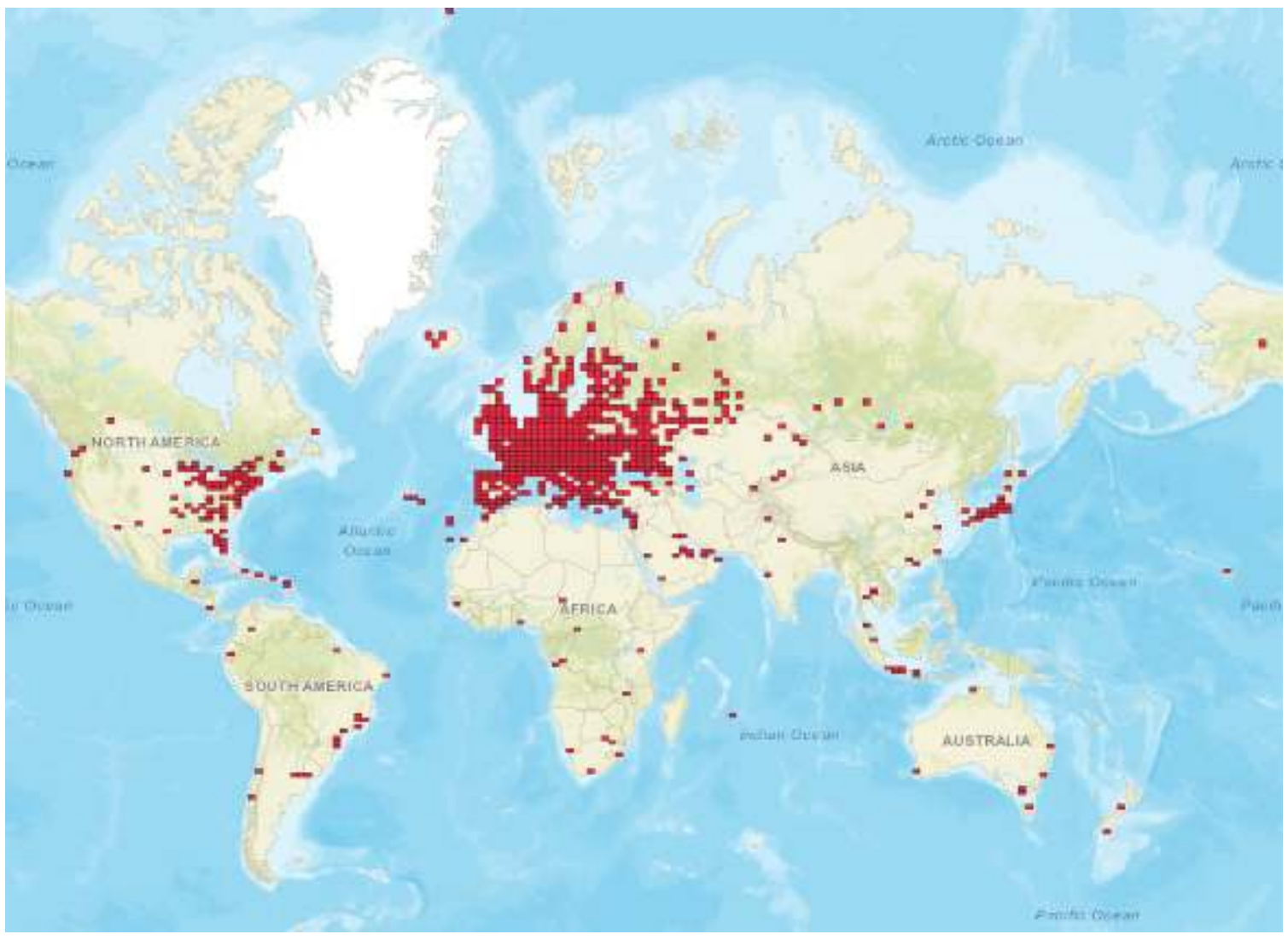

Figure 90: Map of digital-mode contacts made during the first months of operation. Source: GridTracker. I had many joyful hours on the air.

# 5 Acknowledgments

Many thanks to my brother Hans, who invested many days of work to help me develop software for automated measurement and tuning.

Many thanks to Peter Schär, who assisted me during the outdoor measurements — “a bit further to the left, wait, I need to measure again…”.

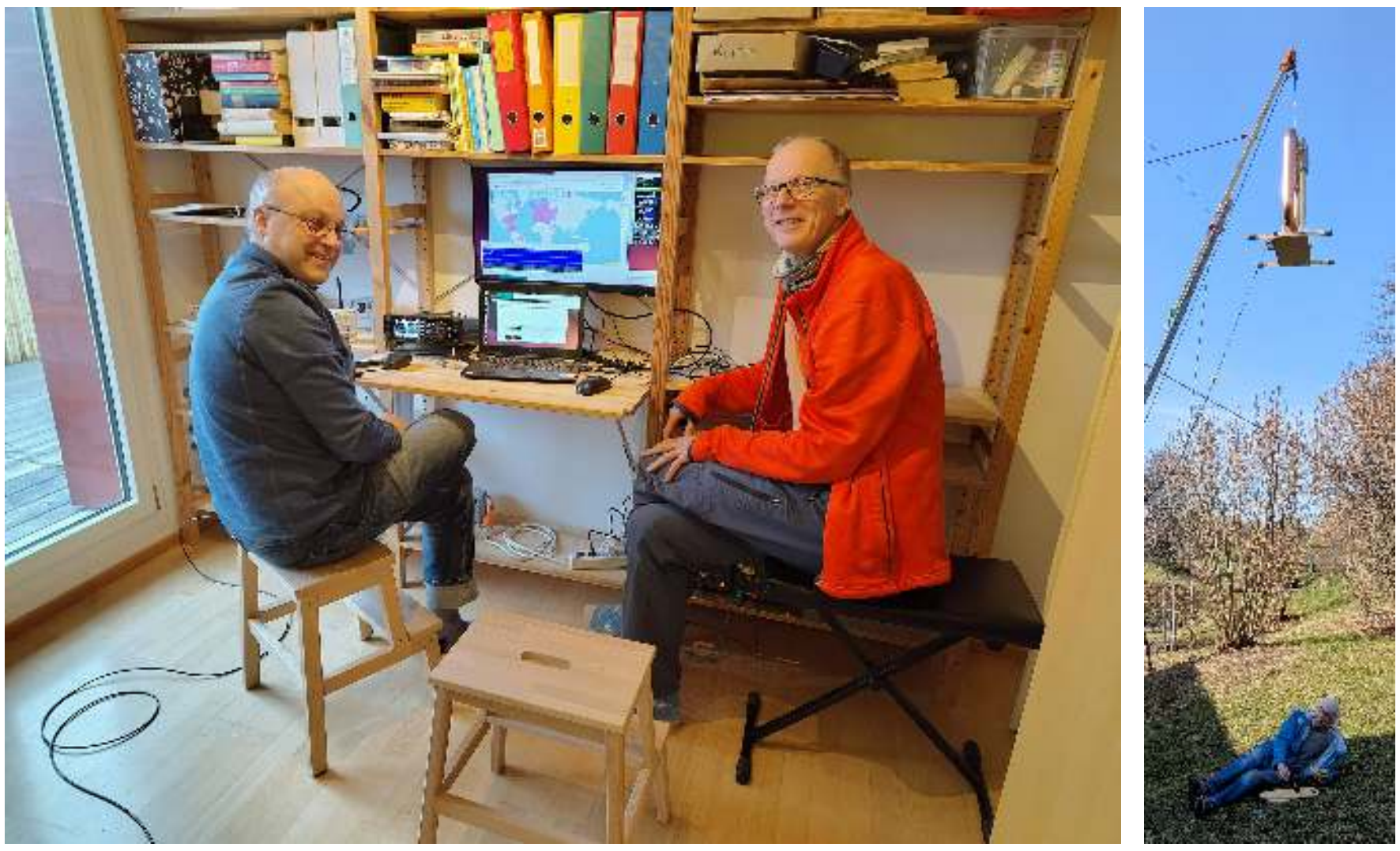

Figure 91: Left: Peter Märki with his brother Hans Märki; right: Peter Schär measuring antenna height above ground with a laser distance meter. We were concerned that the ladder might fail and the antenna could fall; despite this risk, he lay down beneath the antenna to take the measurement.

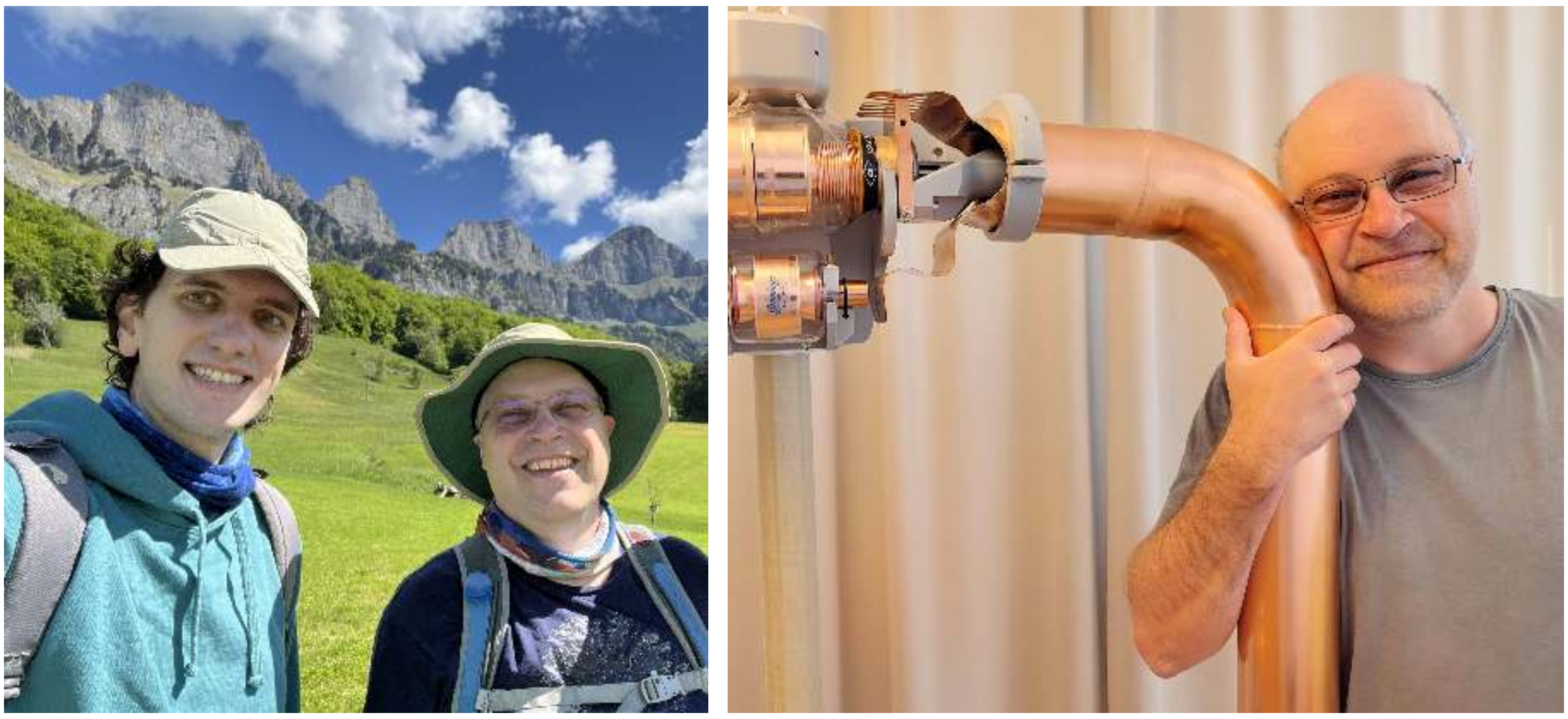

Figure 92: The authors: left, Markus Niese with Peter Märki; right, Loopie with Peter Märki.

Many thanks to the numerous other people with whom I had stimulating discussions and received valuable input.

# 7 Appendix

## 7.1 Derivation of the SWR 2.62 bandwidth

The unloaded resonant circuit can be represented by a series-equivalent RLC circuit. At the half-power frequencies, the magnitude of the series reactance equals the equivalent series resistance:

$$|X| = R \tag{14}$$

At resonance, the antenna is matched to the 50 Ω measurement system. Therefore, at the two half-power frequencies, the input impedance is

$$Z_{\text{in}} = 50\Omega(1 \pm j) \tag{15}$$

The reflection coefficient Γ at the band limit is given by:

$$\Gamma = \frac{Z_{\text{in}} - 50\Omega}{Z_{\text{in}} + 50\Omega} = \frac{\pm j50\Omega}{100\Omega \pm j50\Omega} = \frac{\pm j}{2 \pm j} \tag{16}$$

The magnitude of the reflection coefficient is the ratio of the magnitudes of the numerator and the denominator:

$$|\Gamma| = \frac{|\pm j|}{|2 \pm j|} \tag{17}$$

Using $|a + jb| = \sqrt{a^2 + b^2}$, we get:

$$|\pm j| = \sqrt{0^2 + 1^2} = 1 \tag{18}$$

$$|2 \pm j| = \sqrt{2^2 + 1^2} = \sqrt{5} \tag{19}$$

Thus:

$$|\Gamma| = \frac{1}{\sqrt{5}} \tag{20}$$

With this result, the SWR becomes:

$$\text{SWR} = \frac{1 + |\Gamma|}{1 - |\Gamma|} = \frac{\sqrt{5} + 1}{\sqrt{5} - 1} \approx 2.62 \tag{21}$$

The SWR = 2.62 points are therefore the half-power frequencies of the resonant circuit, and their separation defines $B_{\text{SWR 2.62}}$. For a series RLC resonant circuit, the half-power bandwidth equals $\frac{f_0}{Q_0}$, so:

$$Q_0 = \frac{f_0}{B_{\text{SWR 2.62}}} \tag{22}$$

## 7.2 Measurement at 3 dB Bandwidth

When operating the antenna (transmitting or receiving), the system bandwidth is determined by the loaded quality factor $Q_L$. In a perfectly matched system (critical coupling), the source impedance of the transceiver is transformed towards the antenna to match the antenna's total loss resistance $R_T$.

Consequently, the effective damping resistance in the loaded circuit $R_{\text{loaded}}$ is the sum of the antenna's total resistance $R_T$ and the transformed source resistance:

$$R_{\text{loaded}} = R_T + R_{\text{transceiver,transformed}} = R_T + R_T = 2R_T \tag{23}$$

The loaded quality factor $Q_L$ is then:

$$Q_L = \frac{X_L}{R_{\text{loaded}}} = \frac{X_L}{2R_T} \tag{24}$$

Since the 3 dB bandwidth is related to the quality factor we obtain:

$$\Delta f_{3\text{dB}} = \frac{f_0}{Q_L} = 2 \cdot \frac{f_0}{Q_0} \tag{25}$$

This shows that the bandwidth of the matched system is twice that of the unloaded antenna (where only $R_T$ determines the damping).

## 7.3 Effect of feeder loss on measured Q factor

I do not recommend measuring this type of antenna the way I did. It is better to include the feed-line in the VNA calibration. For my measurement approach, the corresponding correction factor calculation is given in Section 3.11.

Let $x = 2Q_0 \Delta f / f_0$. At the intrinsic SWR = 2.62 bandwidth, $x_{\text{real}} = 1$ (see Section 7.1); with feeder loss, the measured SWR = 2.62 point occurs at $x = x_{\text{meas}}$. Because the resonance bandwidth is narrow, the feeder attenuation $L$ (in dB) is assumed constant over the band, reducing the round-trip reflection coefficient by $10^{-\frac{L}{10}}$. The VNA bandwidth edge satisfies:

$$10^{-2\frac{L}{10}} \frac{(x_{\text{meas}})^2}{4 + (x_{\text{meas}})^2} = \frac{1}{5} \tag{26}$$

Since $B \propto \frac{x}{Q_0}$:

$$B_{\text{real}} = k \cdot B_{\text{meas}}, \quad k = \frac{x_{\text{real}}}{x_{\text{meas}}} = \frac{1}{2}\sqrt{5 \cdot 10^{-2\frac{L}{10}} - 1} \tag{27}$$

## 7.4 Terminology

"Quality factor" and "bandwidth" are referred to differently in the literature. To avoid confusion, it is important to clearly distinguish the terms. The following unambiguous designations are recommended:

- Intrinsic quality factor (or unloaded quality factor) of the antenna
- Loaded quality factor of the antenna (with the transceiver connected)
- Intrinsic bandwidth of the antenna $B_{\text{int}}$
- Bandwidth at SWR 2.62 of the antenna (defined by the frequencies at which SWR = 2.62)
- 3 dB bandwidth of the antenna with transceiver connected

For a resonant circuit without significant additional series reactance, $B_{\text{SWR 2.62}}$ is approximately equal to $B_{\text{int}}$.